# ENTREPRENEURIAL HIGHER EDUCATION

## Education, Knowledge & Wealth Creation

# Entrepreneurial Higher Education

## Education, Knowledge and Wealth Creation

**Rahmat Ullah**
Director ORIC
International Institute of Science, Arts & Technology, and
Executive Director, IRP

**Dr. Rashid Aftab**
Director,
Riphah Institute of Public Policy,
Islamabad

**Saeed Siyal**
Research Fellow,
School of Economics and Management,
Beijing University of Chemical Technology, Beijing China

**Kashif Zaheer**
Faculty Member,
Riphah Institute of Public Policy,
Islamabad

# Executive Summary

Pakistan took a great start after its independence by developing S&T infrastructure and using the S&T resources for economic development. The contribution of scientists like Dr. Saleemuzaman Siddique and Prof. Abdu Slalaam was very significant. The focus of earlier 20 years was economic development and industrialization using science and technology. Therefore, Pakistan has established six S&T councils covering all aspects of development like industrial efficiency, water resources, S&T policy, nuclear, agriculture and others. Pakistan was ahead of many counterparts in exports, economic indicators, and S&T establishments. Premier Chu En Lai visited BECO industries in 1956 and requested for training of Chinese engineers. Pakistan enjoyed the golden era of technology and economic development till 1970.

Pakistan continued establishing S&T organizations after 1970 till 2000 but could not use them for economic development due to political instability and policy inconsistency. Pakistan experienced nationalization that killed the industrial base of the country. The denationalization opened the market but was mostly dependent on raw materials export, semi-processing industry and general trade. Pakistan was unable to move into the high-tech industry and value-added production due to the very bitter experience of nationalization. This led to stagnation in export growth and investment in high tech production. The world moved to electronics, automobile, microelectronics, biotechnology, pharmaceuticals, mobile production, shipbuilding, and other high-tech industries. Pakistan was stuck in textile, and very low value assembly and formulation.

Pakistan moved to services sector like banking, real estate, and telecommunication after 2000 and totally ignored the high tech and value-added manufacturing. Pakistan also moved to higher education and started establishing universities, science labs, center of excellence and training of PhDs locally and from the international universities. The S&T sector having more than 300 labs, laboratories and R&D organizations are left orphans. This huge S&T resource was not revamped nor strengthened rather totally ignored. These R&D organizations become state liability and salaries paying machines having no productivity and no relevance to local industries. The vocational skills institutes having very poor conditions were also totally ignored and remained unproductive.

The era of higher education started in 2002 with the birth of Higher Education Commission HEC). HEC was the dawn of new hope and made success in terms of setting up new universities, increasing the number of publications, setting up new laboratories and centers of excellence. The mushroom production of graduates and PhDs started in Pakistan. The higher education sector grew to a large extent but was irrelevant to local industry and totally disconnected from the local market of Pakistan. The industry is totally dependent on foreign technologies and inputs for their production and expansion. Resultantly Pakistan is plunged into a vicious circle of four S&T liabilities. These S&T liabilities are 1) paying for technology imports from foreign markets, 2) paying for a semi-dead TEVT sector without productive skilled labor, 3) paying for more than 300 unproductive R&D organizations and 4) paying for huge science infrastructure in the universities irrelevant to local market needs. These four sectors are actually the backbone of industry and assets of the country producing economic returns. Pakistan has these S&T resources as liability and needs to make a serious turn around.

Pakistan is also confused about the role of higher education in the country. There is a school of thought that advocates for teaching and publications based on science regardless of relevance to market needs and demand. There is a school of thought that advocates for local relevance and economic contribution of the universities.

The increased political involvement, governance under bureaucratic system, very limited budget, infancy phase of university life, unsupportive culture and no demand of high-tech science puts greater challenge for higher education to produce any economic return.

This book presents detailed discussion on the role of higher education in terms of serving basic knowledge creation, teaching, and doing applied research for commercialization. The book presents an historical account on how this challenge was addressed earlier in education history, the cases of successful academic commercialization, the marriage between basic and applied science and how universities develop economies of the regions and countries. This book also discusses cultural and social challenges in research commercialization and pathways to break the status quo.

The lead author presents practical reflections and draws lessons from the last 17 years of working with more than 500 hundred universities, research labs, industries, S&T organizations, and policy institutes for research commercialization. The author has been involved in more than 100 academic projects for funding, development, and commercialization. The author has conducted training on applied research and been involved in policy dialogues. This localized experience along with active participation in the triple helix platform at international level helped the author understand the dynamic of research commercialization, social impediments and governance of technology development and diffusion in the society.

# Preface

Education which is life by itself is a path from Darkness to Light and the most powerful weapon of Entrepreneurial higher education is to work out the solution.

Entrepreneurship has become a vital force in driving innovation, job creation and economic growth. As such, higher education institutions have a critical role to play in developing the next generation of entrepreneurs. By providing students with the skills, knowledge, and experiences necessary to launch and grow successful ventures, universities can help to foster an entrepreneurial mindset and build a culture of innovation. This book is designed to provide a comprehensive overview of the current state of entrepreneurial higher education, exploring the various approaches and programs being used to develop entrepreneurial talent in Pakistan since 1947. It examines; the role played by Science & Technology Organizations, Higher Education Institutions and S&T Establishment.

This book offers a roadmap for creating and implementing effective entrepreneurial education through an integrated approach including; Research and Development (R&D) organizations, Higher Educational Institutions (HEIs), conducive Government policies/priorities and Industrial linkages. This book is an essential resource for anyone seeking to understand and harness the power of entrepreneurial education, entrepreneurial process and the role of entrepreneurship in the economy. This book explores the role of higher education and R&D organizations in fostering entrepreneurship, highlighting the challenges and opportunities for Higher Education Institutions as they strive to meet the needs of the entrepreneurial community. An effective entrepreneurship education is required for reshaping Pakistan's Higher education system considering the best practices from leading universities and organizations around the world. Higher education plays a critical role in preparing individuals for this dynamic and constantly evolving landscape, equipping them with the tools they need to turn their ideas into reality.

This book will provide you with valuable insights and practical advice for navigating the complex and ever-changing landscape of entrepreneurial higher education. With its focus on best practices and cutting-edge research, this book is a must-read for anyone looking to drive innovation and growth in the entrepreneurial community.

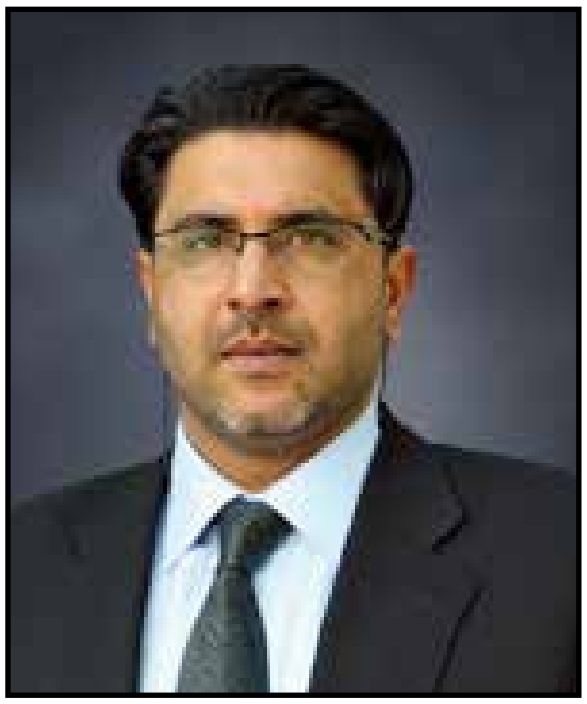

# Message

Dear all,

I am delighted to congratulate the authors on the publication of the book titled "Entrepreneurial Higher Education" Policy Perspective. This book discusses how the universities fit into the changing ecosystem of today's ever changing economy and future development. Sooner or later, the evolution of universities from institutions of theoetrical learning to creative hubs driving the students to thinks out of the box would be complete. But it's necessary to take the first important steps, to cultivate those entrepreneural mindsets and supporting them with actions that will bring about this much needed change.

Riphah International University has consistently supported the convergence of academic pursuits and research, further strengthening the entrepreneurial ecosystem. This book successfully shows the importance of Higher education in knowledge creation, teaching and applied research. I sincerely hope that the policy framework outlined in this book will be taken into account and integrated into our education policy. Given Pakistan's unique situation and the far-reaching impact of the business school, the use of this framework will enhance our competitiveness and promote economic development and growth.

Warm regards,

**Mr. Hassan Muhammad Khan**
Chancellor
Riphah International University

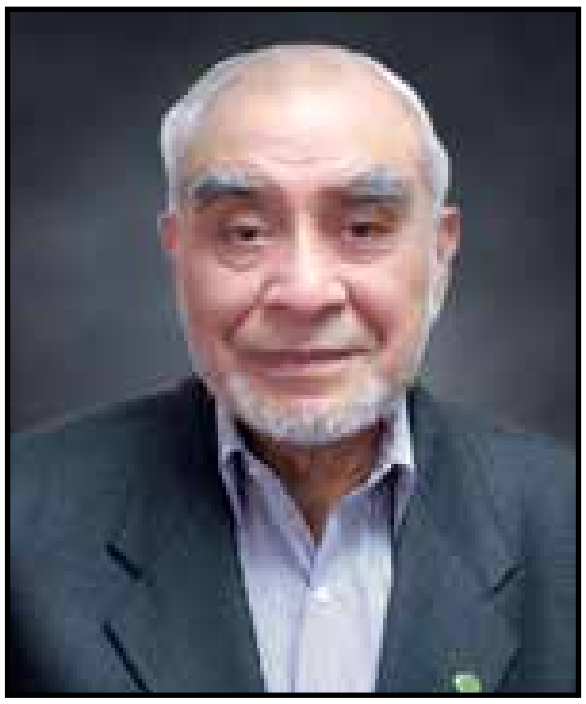

# Message

Assalamu Alaikum, Wa Rahmatullah

I am delighted that Riphah International University has partnered with the IRP to publish an important book called "Entrepreneurial Higher Education" Policy Perspective. This publication illuminates the evolving role of universities in the modern world and highlights the need for entrepreneurial strategies while maintaining the highest teaching and research standards. It examines how universities can effectively promote entrepreneurial ideas and activities among stakeholders, thereby making a valuable contribution to industrial innovation, job creation and overall economic growth.

In today's rapidly changing environment, universities are challenged to go beyond their traditional academic functions. They must excel not only in imparting knowledge and conducting fundamental research, but also in areas such as technology transfer, marketing execution, patent filing and creation of thriving spin-off companies.

The book introduces innovative policy initiatives and outlines a path forward for universities that need to develop entrepreneurial university architectures. It emphasizes the importance of cultivating a unique combination of traits that fosters an entrepreneurial mindset. By adopting these principles, Riphah International University aims to shape the future and make a significant contribution to the progress of society by fostering innovation, creativity and entrepreneurship.

Sincerely,

**Prof. Dr. Anis Ahmad**
Vice Chancellor,
Riphah International University

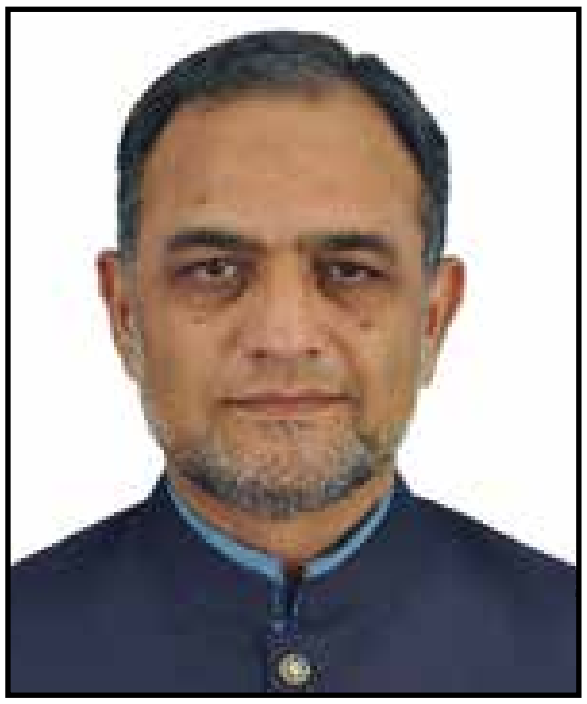

# Message

I have gone through the book" Entrepreneurial Higher Education" authored by Rahmat Ullah and others. I can say, without hesitation, that it's an excellent piece of work highly needed for the higher education sector. We must change from conventional education to entrepreneurial higher education, conventional universities to Entrepreneurial universities. In doing so, this book shall serve as handbook for all the stake holders and practitioners. Mr. Rahmat feels pain for the nation not progressing despite having all kinds of resources and he has been struggling to turn the stone. I believe, he would be successful one day, and we all are with him in this endeavour. This book is also going to contribute a lot in this regard. This book should be available to teachers and students at all educational institutions. It is also useful for policy makers and industrialist to understand the entrepreneurship eco system to perform their role effectively.

**Prof. Dr. Mahmood Saleem**
Vice Chancellor
MCUT, DG Khan

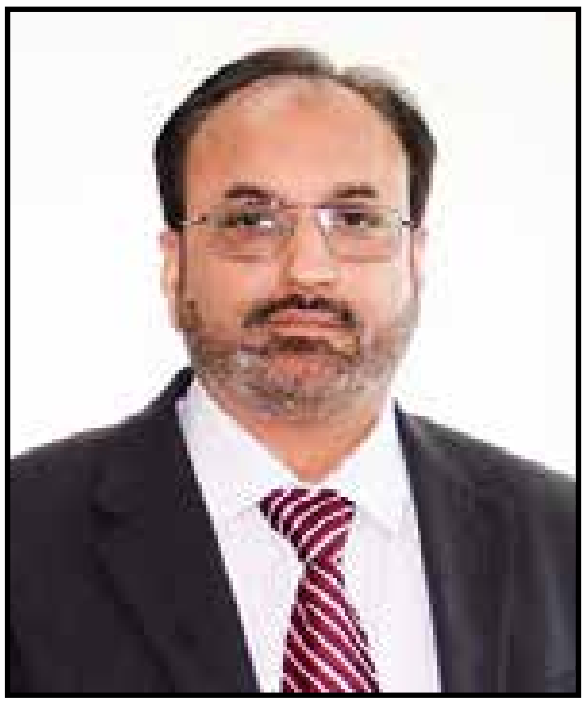

# Message

I recently had the opportunity to read your book, ENTREPRENEURIAL HIGHER EDUCATION: Education, Knowledge & wealth creation" and I must commend you on your comprehensive and insightful exploration of Pakistan's progress in the field of science and technology. Your book provides a detailed overview of the country's early successes in developing its S&T infrastructure and the contributions of renowned scientists such as Dr. Saleemuzaman Siddique and Prof. Abdu Slalaam.

I appreciate how you shed light on the initial establishment of S&T councils in various sectors, which positioned Pakistan ahead of many counterparts in terms of exports, economic indicators, and S&T establishments. However, you have also discussed the challenges that hindered Pakistan's progress in utilizing its S&T organizations for economic development, such as political instability and policy inconsistency.

Your book delves into the shift in focus towards the services sector in the 2000s, neglecting high-tech and value-added manufacturing. The emphasis on the importance of entrepreneurial higher education and its role in finding solutions is particularly compelling. It is evident that higher education institutions play a crucial role in nurturing the next generation of entrepreneurs and driving innovation, human resource development, and economic growth.

I also appreciate your focus on the challenges faced by the higher education sector, including disconnect between academia and industry, the need for demand-driven products/technologies, and the socio-cultural barriers in research commercialization. Your proposal to apply the triple helix model to establish Pakistan's presence in the global system is thought-provoking and could lead to significant advancements in research and innovation.

Furthermore, your emphasis on the Institute of Research Promotion (IRP) and its role in promoting applied research within the community is commendable. The collaborative efforts involving government departments, industries, universities, and policy institutes demonstrate a proactive approach towards fostering research and innovation. However, you rightly acknowledged the existing gap between efforts invested and desired outcomes, emphasizing the need for evaluation and improvement to achieve the shared vision of introducing innovative products to the market and contributing significantly to the GDP.

The comprehensive training program provided by the IRP is another remarkable aspect highlighted in your book. Equipping researchers with essential skills, promoting the integration of technology in research practices, and fostering collaboration between academia and industry through the University-Industry Partnership Program (UIP) showcase the commitment to practical approaches in applied research.

I also appreciate your recognition of policy barriers and the collaborative efforts between the IRP and the Pakistan Council of Science and Technology (PCST) in addressing these obstacles. The policy workshops conducted on a province-wide scale demonstrate a proactive approach to involve key stakeholders in shaping science and technology policies.

Overall, your book provides a valuable resource for understanding and harnessing the power of entrepreneurial education, research commercialization, and the role of entrepreneurship in the economy. Your dedication and skill as the authors are truly commendable, and I encourage you to continue your writing endeavours with the same level of excellence.

**Prof. Dr. Shahid Munir**
Chairperson
Punjab Higher Education Commission (PHEC)

# List of Contents

## List of Tables

## List of Figures

# List of Abbreviations

S&T – Science & Technology

R&D – Research & Development

HEC – Higher Education Commission

TEVT - Technical Education and Vocational Training

TTO - Technology Transfer Office

MIT - Massachusetts Institute of Technology

GDP – Gross Domestic Product

AUTM – Association of University Technology Managers

IITs - Indian Institutes of Technology

KIST – Korean Institute of Science and Technology

WW – World War

NRPU - National Research Program for Universities

AAAS - American Association for the Advancement of Science

CEO – Chief Executive Officer

GNP – Gross National Product

KAIST – Korea Advanced Institute of Science and Technology

AERI - Atomic Energy Research Institute

KAERI - Korea Atomic Energy Research Institute

GRI - Government-sponsored Research Institute

OSF – Oregon Shakespeare Festival

OSU - Oregon State University

ARPANET- The Advanced Research Projects Agency Network

TLO - Technology Licensing Office

HP - Hewlett-Packard

AAU - Association of American Universities

SEED - Stanford Institute for Innovation in Developing Economies

OLT - Office of Technology Licensing

IEEE - The Institute of Electrical and Electronics Engineers

FAO – Food and Agriculture Organization

IPR - Intellectual Property Rights

IPO – Intellectual Property Office

CIMP - Culture, Incentives, Mission, and Policy

ARP - Applied Research Innovation

ORIC – Office of Research Innovation and Commercialization

STM - Situational Teaching Methodology

ASTM - American Society for Testing and Materials

PESE – Personality, Environment, Scientific and Enterprising

CAT - Computerized Axial Tomography

EMI – Electronic and Musical Industry

IRP - Institute of Research Promotion

PASTIC - Pakistan Scientific and Technological Information Center

PSF – Pakistan Science Foundation

UIP - University-Industry Partnership

LCCI - Lahore Chamber of Commerce and Industry

RCCI – Rawalpindi Chamber of Commerce and Industry

KPCCI - Sarhad Chamber of Commerce & Industry

KCCI - Karachi Chamber of Commerce & Industry

FCCI – Faisalabad Chamber of Commerce & Industry

GCCI – Gujranwala Chamber of Commerce & Industry

PCST - Pakistan Council of Science and Technology

KTO - Knowledge Transfer Offices

IIM - Indian Institutes of Management

HEI – Higher Education Institution

# Chapter 01

## 1. The Introduction

### 1.1 Birth of the Concept; Entrepreneurial University

The universities have experienced an evolution by traveling from religious teaching to liberal arts teaching and then scientific discoveries from its inception till the mid-20th century. The universities almost took 1000 years to liberate themselves from influenced mode to an independent mode of thinking and reflection. The universities developed two missions of teaching and research mostly sponsored by the state and communities. The credit for today's science world goes to these early established universities as they trained humans, experimented with new ideas, and published the results for widespread sharing and exchange. In the late 19th century, Professor William Rogers of Virginia University found two missed components in this evolved two-mission university system. This system teaches content to the students and leaves them for the market to train them for practical skills and field orientation. Secondly, this system produces good research and leaves it for the market to exploit, develop products and modify it for the end-user. He thought of adding these two components to the university system later on called the third mission of the universities. He developed the MIT motto of men's et manus, which means mind and hand (Roberts, E. B., Murray, F., & Kim, J. D., 2019). He founded MIT to experiment with this third mission of knowledge exploitation and economic contribution. Skills-based teaching started supplying market-ready graduates from MIT. The professors started consultancies and created their own companies to exploit research. MIT set up its technology transfer office (TTO) to facilitate patent filing, patent licensing, and generating maximum revenues from faculty research. This emerged as an entrepreneurial university framework largely followed by the universities in the USA and the rest of the world (Etzkowitz, H., 2002). The entrepreneurial universities created regional economies around their research and faculty works. The universities developed a system to connect their technology producer with end-users and consumers of technology to speed up exploitation (Etzkowitz (2003). The entrepreneurial university needs a system to transfer created knowledge into patents and startups. An entrepreneurial culture will provide opportunities for students and faculty to exploit their ideas and improve life experiences.

Components of Entrepreneurial MIT

1. Professors were encouraged to spend one day in the industry and society under the 5-1 rule and do consultancy ventures.
2. Professors were offered revenue sharing from technology income and allowed to initiate for-profit businesses.
3. Faculty were given venture capital to start their research-based businesses.
4. Students were given skills, applied training, practical experience and trained fully for the market before they graduate.

Massachusetts Institute of Technology (MIT) identified the pathways of academic entrepreneurship that affected the local economy of Route 128, Bostan. MIT was also followed by Stanford to develop the Silicon Valley that affected the global economy and changed the course of technology history. There were green orchards and agricultural land before Stanford University. Stanford university professors and students were greatly inspired by MIT and converted agricultural land into Silicon Valley – the global innovation hub (Etzkowitz, H., 2002). The MIT born model also contributed to the development of economies of many countries and regions like Korea. According to Roberts, E. B., Murray, F., & Kim, J. D., (2019), there are currently 30000 living companies set up by MIT fellows with the employment of 4.6 million and USD 1.9 trillion in revenue. The economy of MIT fellows equals the GDP of the 10th largest economy in the world.

## 1.2 The Impact of MIT Model on US Economy

Here is the summary of the AUTM report 2017 related to the academic output for 23 years from 1996 to 2015. The USA entrepreneurial universities produced 11,000 new ventures, developed 200 drugs and vaccines in the market, disclosed 380,000 inventions, and got 80,000 patents granted in the last 25 years. The small companies and new startups received 70% patent licensing that fueled fresh energy in the economic system of the USA. This entrepreneurial academic system contributed USD 1.3 trillion to the industrial output of the USA, USD 591 billion to the GDP of the USA, and created almost 4.3 million new jobs. This is the economic impact of the third mission of entrepreneurial universities. The AUTM report 2017 also presented the performance of TTOs of universities by reporting statistics of technology output 2017. According to the AUTM 2017 report, 24998 new inventions were disclosed to TTOs, 15335 patents were filed, and 7459 final patents were granted to scientists and universities. The USA entrepreneurial higher education system gave 7849 licenses and contracts, supplied 755 innovative products to the market, and formed 1080 new startups. This entire development is the impact of 68.3 billion R&D expenditure made by USA universities.

The AUTM (Association of University Technology Managers) Report 2021 is an annual report that provides insight into the state of technology transfer activities at universities and research institutions. The report provides data on trends in the technology transfer industry, including the number of patents and licenses generated, the amount of funding received from licensing and the types of organizations licensing the technology.

One key finding of the 2021 report is that university technology transfer is an important source of innovation, with universities and research institutions generating a significant number of patents and licenses. Additionally, the report found that the total amount of funding received from licensing activities has increased in recent years, with universities and research institutions receiving a total of $2.9 billion in licensing revenue in 2020.

Another important finding of the report is that universities and research institutions are increasingly partnering with a diverse range of organizations, including startups, large corporations, and government agencies, to commercialize their technologies. The report also highlights the importance of international collaborations, with universities and research institutions licensing technology to organizations in countries around the world.

AUTM Report 2021 provides valuable insight into the state of technology transfer activities at universities and research institutions and highlights the importance of these activities in driving innovation and commercialization.

The 2021 report highlights key trends and insights in the technology transfer industry, including:

- Increase in licensing and startup activities: The report showed a continued increase in licensing and startup activities, with universities and research institutions generating more revenue from licensing and commercialization of their research.
- Growing interest in artificial intelligence and biotechnology: The report showed that there was a growing interest in artificial intelligence and biotechnology, with these areas attracting the highest number of licenses and startups in the past year.

- Importance of partnerships and collaborations: The report emphasized the importance of partnerships and collaborations between universities, research institutions, and industry for successful technology transfer.
- Increase in diversity and inclusion initiatives: The report also highlighted the increasing focus on diversity and inclusion initiatives, with universities and research institutions making efforts to promote diversity and equity in technology transfer.

Overall, the AUTM report provides valuable insights and information on the technology transfer landscape in the United States, helping stakeholders in the industry to better understand the trends and challenges facing the sector.

The AUTM (Association of University Technology Managers) Report 2022 provides insights into the latest trends and developments in technology transfer and commercialization activities within the academic research community.

In terms of new ventures, the report highlights an increase in the number of start-up companies emerging from universities and research institutions. These new ventures are leveraging cutting-edge research and technology to bring new products and services to market, fueling economic growth and creating jobs.

Patents and licensing activities also continued to grow in 2022, with universities and research institutions securing a record number of patents and executing more licensing agreements. This trend demonstrates the increasing importance of protecting and commercializing university-developed innovations, as well as the growing demand for cutting-edge technologies from industry partners.

The entrepreneurial impact of university research is also a key focus of the AUTM Report 2022. The report highlights the significant economic and societal impact that university-generated innovations are having, particularly in the areas of biotechnology, medical devices, and information technology.

Finally, the report provides an in-depth look at the activities of technology transfer offices (TTOs) across the country. TTOs play a critical role in facilitating the transfer of technology and knowledge from the academic research community to industry and the marketplace. The report reveals that TTOs are continuing to evolve and adopt new strategies and technologies to streamline their operations and improve their effectiveness in supporting the commercialization of university-generated innovations.

The report suggests that TTOs are continuously evolving and adapting new strategies to improve their effectiveness in supporting the commercialization of university-generated innovations. Overall, the AUTM Report 2022 provides valuable insights into the trends and developments in technology transfer and commercialization within the academic research community.

## 1.3 IITs the Asian MIT

According to Arab News (Khan, 2013), the Indian Institutes of Technology (IITs) are now ranked among top 10 institutions in science and technology in Asia. IIT maintains close collaboration with MIT for student exchange programs and other interventions. IIT supplies highly technical and competent graduates to top-ranked multinationals across the globe. The IIT graduates began to dominate at the top leadership positions of multinationals and technology companies also. IIT also receives research grants and funding from global organizations, multinational corporations, and institutions.

IIT followed the model and practices of MIT. MIT faculty was invited to help in faculty training, system design, and developing academic excellence and entrepreneurial culture at IIT. MIT graduates from India contributed significantly to the foundation of IITs, technology-led development, and the setting up of technology-based corporations in India. 73 unicorns out of 100 in 2021 have one co-founder from IITs (Patwardhan, 2022). This indicates IITs' contribution to the local economy of India (Bassett, R., 2016). MIT played a similar role to establish the Korean Institute of Science and Technology (KIST) and provided the foundation for industrialization in Korea. The MIT model was followed by KIST and replicated by other universities after observing its impact on the economic development of Korea.

# Chapter 02

## 2. Entrepreneurial Orientation in the Pakistan- A Debt to Retire

The entrepreneurial orientation advocates for optimum resource utilization and generating wealth from the available resources (Liu et al, 2014). The passive countries like Pakistan are blessed with significant resources but stay poor due to missed entrepreneurial approach of resource utilization. The absence of entrepreneurial orientation incapacitates and disables the society to gain higher returns from the given potentials and resources. The assets are turned into liability and strength turned into weakness. The countries mostly depend on external sources and are entrapped into begging, helplessness, and powerlessness.

Pakistan has a youth bulge, high investment in science capacity, abundant natural resources, and strategic geographical location. Pakistan is increasing its foreign debt, unemployment, food import, energy import, and stagnant export. Pakistan has observed increased investment in science capacity and food import despite being an agricultural country due to absence of entrepreneurial orientation that can turn liability into assets generating wealth. Pakistan is blessed with the highest rate of youth in the world (Shabbir et a, 2018). These God gifted assets can turn into great opportunities for leapfrog progress. Pakistan needs to train its youth on entrepreneurial orientation and provide an enabling environment for youth to grow and progress. Pakistan has great potential to outbeat many countries by developing entrepreneurial orientation in youth.

## 2.1 The Entrepreneurial Orientation

There are nations in the world who turned into ashes in WWII and during the British era of ruling. They rose from the ashes and emerged as leading nations in the world now. They managed to throw away their luggage of historical misery and stood tall in the leagues of rising nations within a few decades. The example includes Turkey, Malaysia, Singapore, South Korea, China, Japan, and others (Jarausch, 2015). In contrast, the few nations stayed poor in spite of more human and natural resources. They are unable to develop entrepreneurial capability that harness resource utilization and wealth generation from the available resources. They are unable to incorporate entrepreneurial approaches in their education, governance system, policies, and political system. The entrepreneurial approach advocates for wealth generation and value creation whereas passive nations relied on external debt and depended on foreign aids to fuel their empty bellies. The passive nations believe in consuming the resources given by others and mentally trained to live on the ventilators. The nations with entrepreneurial orientation believe in generating value and wealth from the resources and ensure minimum possible dependence on foreign aids and support.

## 2.2 The Passive Nation and Wealth Generation

The major difference between passive nations and entrepreneurial nations is the ability to generate value and wealth from the available resources. Few nations with very limited natural resources like Singapore and Japan emerged as leading nations in the world. They developed their capacity to generate high value and wealth from the very limited resources. The per capita income of these generations ranges from $50000 to $100000. The few countries like Pakistan in spite of abundant natural resources and youth bulge, produces $5000 to $10000 per capita income. The passive nations have 10 times less value generation capacity than the entrepreneurial nation. Pakistan has more resources than Dubai, but Dubai is the global financial, logistics and trade hub now due to their entrepreneurial approach of generating high value and wealth from the available resources (Stephens, 2008).

The passive nations are characterized by the following:

- Depend on foreign aids.
- Wait for external help and support.
- Believe in an external role in their bad and good luck.
- Trained to consume wealth.
- Generate wealth through corruption and unfair means.
- Very low in value addition
- Contained on limited gains.
- Prefer secured jobs over risky entrepreneurship.
- Blame others for their incapabilities.
- Unable to create high value from available resources.
- Unable to generate wealth from innovation and value addition.

**Table 1: Per Capita GDP of Korea, Singapore, and Pakistan from 1970- 2020**

| GDP per Capita (Current US$) | | | |
|---|---|---|---|
| **Year** | **Pakistan** | **UAE** | **Singapore** |
| 2020 | 1,322 | 31721 | 60,729 |
| 2010 | 911 | 23087 | 47223 |
| 2000 | 531 | 12257 | 23825 |
| 1990 | 346 | 6610 | 11861 |
| 1980 | 293 | 1715 | 4928 |
| 1970 | 169 | 279 | 925 |
| Source: https://data.worldbank.org/indicator/NY.GDP.PCAP.CD?end=2021&locations=SG&most_recent_year_desc=false&start=1960&view=chart | | | |

## 2.3 The Case of S&T Resource in Pakistan

Thanks to visionary leadership, Pakistan was able to spend billions on developing its science capacity under higher education programs. Pakistan has spent more than 100 billion PKR per year consistently over the last two decades to develop and maintain its S&T infrastructure in higher education. The huge amount is spent on parallel S&T infrastructure available under the S & T ministry of Pakistan. Pakistan got PhDs trained from advanced countries, established universities, setup labs and laboratories and gave around 100 billion for research through NRPU and other research grants and projects. This S&T resource has turned into liability as more money is needed to maintain this infrastructure.

While Pakistan was spending around 100 billion on research, the import of Pakistan kept increasing and export of Pakistan stayed stagnant. Why spending on S&T resources could not reduce Pakistan imports and increase exports is the prominent question? This explains the passive and non-entrepreneurial orientation phenomenon in the country. Pakistan as a nation is the passive country that has mostly lived on foreign aid throughout the history. Pakistan lacks entrepreneurial orientation that can generate value and wealth from the resources. Pakistan developed S&T resources under its passive approach of spending money. Due to lack of entrepreneurial approach, Pakistan was unable to produce value and wealth from the S&T resources. The entire nation needs to get trained on entrepreneurial orientation and shift from consuming resources to generating value and wealth from resources.

## 2.4 The Entrepreneurial Orientation Framework

The majority of Pakistani people follow Islam as a religion that advocates for balance in life. The Quran mentions charity, offering food to the poor and helping orphans as the most favored things in the eyes of The Allah Almighty. The people need to generate wealth even extra from their living means to spend on charity and food for the hungry. Similarly, the Quran mentions Zakat along with the prayer most of the time to teach the importance of Zakat in the eyes of The Allah Almighty. The people need to generate extra wealth to pay the zakat. The Islam presents five basic pillars of Islam as The Belief, The Prayer, The Fasting, The Pilgrim, and The Zakat. The last two pillars demand wealth creation and saving money to spend on The Pilgrim and The Zakat. 40% of Islamic foundation belongs to wealth creation and wealth spending. The followers of Islam were supposed to be the most entrepreneurial nation generating value and wealth from the resources and spending to pay zakat. This nation turned into debt beggars instead of zakat givers having 10 times less per capita income from the advanced countries. The development of entrepreneurial orientation in society can turn the passive nation into the entrepreneurial nation ranked high in the generation of value and wealth, giving Zakaat to others.

### 2.4.1 Believe in High Valued Life

The individual human lives a life for a very short period like a passenger. Should we make this world a dirty place just because our tour to this world is very short? Should we not try to make this world a better place just because we will not live there for a longer period? Islam teaches the prayer asking "O, God give us betterment of this world and betterment of the hereafter". Therefore, making this world a better living and leaving a better world behind is equally important. The people in the entrepreneurial society need to make efforts to create a high valued life for themselves and for others too. Most innovations come from the advanced world that improve human living experiences and add value to the life. The passive nations like Pakistan mostly consume these valuable innovations. The people need to be entrepreneurial, believing in high valued life and creating value through innovations for the human being.

The service to human beings is highly appreciated in the teachings of Islam. That service is only possible if you believe in the high quality of this life and therefore you will serve humans to live a better life. Even the service to animals is highly regarded and appraised. According to the teachings of Islam, you are advised not to waste water even if you are at the bank of a river (Metwally, 1997). Therefore, the importance of this life and its quality must be part of people's faith and they should try to improve this life for human beings.

### 2.4.2 Believe in Wealth Generation

The difference between per capita income between the advanced entrepreneurial world and poor passive world clearly shows the fundamental belief system and view towards wealth generation. The passive nation thinks wealth generation is near to sin and against the wishes of God. The balanced life as advocated by Islam demands wealth creation through right means and spending for right purposes like Hajj, Zakaat, charity, food for the hunger, and shelter for the orphans. Therefore, wealth generation becomes the act of sawaab, and serves the wishes of The God.

As an entrepreneurial nation, the people of Pakistan need to believe in wealth generation through right means of value addition and innovation. The wealth generation through corruption and unfair means is not a healthy sign as it does not have residual value nor spillover effects to improve the lives of others. The people of Pakistan must create wealth through high value-added enterprises that provide employment and livelihoods in the entire value chain of the business. The wealth creation becomes an act of Sawaab when spent on the less privileged people to reduce the inequality gap and improve the life of people who are left behind.

### 2.4.3 Believe in Self Dignity and Honor

The rise of a nation depends on how the nation thinks and behaves in terms of self-respect, dignity, and honor. The people with high esteem tend to preserve their honor and strive hard to gain back if lost due to some circumstances. The people who value themselves low and carry low self-esteem, do not struggle for their honor and dignity. These kinds of nations are obsessed for centuries and trapped in various kinds of slavery. They become the victim of frog syndrome and unable to realize the meaning of dignity and honor. In order to become an entrepreneurial nation, the people of Pakistan must live and struggle for a high level of dignity and honor. This dignified spirit will drive them to become self-sufficient in their needs and produce high value goods and services. This spirit will end their dependence on foreign aid, and they will generate enough value and wealth to live an honored life.

### 2.4.4 Believe in Sharing and Honor for Contract

The passive societies tend to share less and exchange less due to internal fears of losing value of something unique they have. This puts a limit to the impact of unique value and ends with the creator. The passive societies do not have a legal system nor social culture to protect people's unique value and invention. Therefore, passive societies tend to share and exchange less that significantly reduce the spillover effects of unique value created by the individual or companies.

The entrepreneurial nations have a higher level of sharing and exchange culture that provides a high level of spillover effect in the downstream development. The academic publishing in the form of journals is the best example of sharing knowledge so the maximum people can benefit from the academic research. Similarly, the advanced nations have created strong intellectual property laws and enforcement to ensure protection of each unique value and innovation. These two instruments of exchange and sharing created a huge impact on the overall culture of sharing and exchange. This sharing culture provides a competitive edge to these nations for fast growth and development. This sharing accelerates the wealth creation process as one innovation touches the minds of millions, and all benefit and make advancement in it. The five big intellectual property offices belong to five big nations such as the USA, EU, Japan, China, and South Korea. These nations also have strong justice systems and honor for contracts among the parties. Pakistan needs to develop sharing and exchange culture to grow as an entrepreneurial nation. The people of Pakistan also need to change their habits and views about exchange and shift to a higher level of sharing exchange culture. This will have a big spillover effect and create fast growth in the mainstream.

## 2.5 Orientation of Creating High Value

The passive nations like Pakistan tend to rely mostly on the trade of raw materials produced from the natural sources. They do not have capacity nor high will to convert their natural resources into value added fine products and services. They earn very less from the trade of natural resources. They are more into real estate businesses or asset-based life. They tend to preserve gold and lands as their social safety net. They are unable to enter the knowledge economy and high value production.

In contrast, the entrepreneurial nation has a higher capacity to create value and earn more. Therefore, they import raw materials from poor countries and sell back high value products to the same countries earning high from value creation. This is the serious orientation problem with the passive countries like Pakistan. There is a need of reorientation from value-less businesses to value creating business and enterprises. The people of Pakistan need to be oriented about the benefits of high value creation. The indigenous value creation will make Pakistan an entrepreneurial country earning much higher revenue.

## 2.6 Orientation of Generating Wealth from Value

More than 100 countries share their borders with the sea but around 30 countries are known as maritime nations. A few countries exploit the sea borders and generate wealth as compared to others. China, Japan, USA, Singapore, Germany, Greece, and Norway lead the rest of the world as maritime nations. This explains the phenomenon of ability to commercialize and capacity to generate wealth through high value resources. These countries are able to generate more wealth from their ports as compared to other nations. The similar examples go to other sectors and areas. The countries can generate wealth from the resources they do not even have due to their capacity of wealth generation.

The passive countries have very poor orientation of wealth generation in terms of value creation and value marketing. The entrepreneurial nations have an extreme level of orientation and capacity to create value and market value. Pakistan mostly exports raw materials and semi processed goods without creating much value. The indigenous unique value-added products are also not marketed to the world to cash the global market. Pakistan holds a unique value in terms of geographical indication in Basmati rice. Pakistan was unable to commercialize this value in the global market and lost the economic game. Indian basmati dominates the global market along with other 100 plus varieties. The people of Pakistan need to be trained in commercial orientation on how to generate high wealth from created value.

## 2.7 The Driving Entrepreneurial Orientation

The entrepreneurial orientation has to be developed in the society of Pakistan to foster the growth and catch up with the advanced nations. This challenge needs to be taken up by each segment of society. There are four main institutions that can drive the entrepreneurial orientation in society.

**Education Sector**: The institutions in education from primary to tertiary need to plan, design, and execute programs that develop entrepreneurial orientation in the society. The education sector needs to include entrepreneurial orientation in their current subjects and curriculum. There should be special awards and incentives for teaching entrepreneurial orientation. The students who exhibit entrepreneurial orientation must be facilitated and mentored to initiate their ventures and lead a great entrepreneurial life.

**Media Sector**: The media can play a significant role in developing entrepreneurial orientation in the society. The media in the advanced world dedicate 20-30% of time for social and education including science, technology, and business. The media of Pakistan must dare to initiate such programs and activities that promote entrepreneurship and creativity. The media can influence a large segment of society by presenting and appreciating the people who exhibit a high level of entrepreneurship. The media also gives space to the people who have content and can teach about entrepreneurial orientation. The various media products need to be created on the subject of entrepreneurial orientation.

**Industry Bodies**: The industry bodies have the largest network of business communities. These chamber and business associations can influence the business community to learn entrepreneurial orientation and practice high value creation in their businesses. The industry bodies can launch training programs and awareness seminars on high value creation and wealth generation through value added products and services. The industry bodies can connect local industries with global technology companies to experience value creation and commercialization processes.

**Public Sector**: The government has the key responsibility of promoting entrepreneurship in the society. The govt. can enact such laws and make such policies that incentivize value creation and risk taking in the business. The government must protect and incentivize high risk entrepreneurial initiatives by offering relief in taxes, duties, regulations, and procedural matters. The over regulations and bureaucratic procedures are the biggest enemy of innovation and entrepreneurial initiatives. The government can provide an enabling environment and ease of doing business by offering one window solution in Govt related procedures.

The state facilitation will drive entrepreneurial orientation in the society and enable people to create value and generate wealth by commercializing created value.

# Chapter 03

## 3. The Case Studies of Entrepreneurial Universities

## 3.1 Case Study 01- MIT as Entrepreneurial University

### 3.1.1 Background

The US has a long history of origins and development of research and entrepreneurial institutions in its states and around the world. Four scientists (William Barton Rogers, Karl Compton, Vannevar Bush and Frederick Terman) are well known for their key role in establishing a contemporary and entrepreneurial institution (Massachusetts Institute of Technology (MIT) which combines teaching and research in the economic development of the region by creation of high-tech spin-off firms. Although all these scientists showed remarkable contributions, Professor William Barton Rogers continued to work until he succeeded in establishing MIT and got the honor of being its founder. Professor Roger was a geologist at the University of Virginia. He envisioned a science-based institution dedicated to developing its regional industries. His vision was much greater than the available resources and hardly met 30 percent of the resources required to establish such an institution. His excellence continued striving and MIT assumed to be a classical college with a specific common curriculum which largely consisted of contents of a polytechnic education. However, MIT's foundation was also based on the ideal training in liberal arts and practical disciplines to create leaders rather than technicians who used to work for Harvard graduates.

MIT unified several academic formats such as classical teaching college, polytechnic engineering school, land grant and research university into an exceptional configuration. All these academic paradigms showed inspiration for numerous traits of the development of MIT as a technological university with a strong base in science and liberal arts. In the $19^{th}$ century, academia in the US was divided into streams of pure science and technology-based institutions along with diverse cultures and missions. Majority of the well-known institutions in the US failed to compete in both the streams. On the other hand, MIT successfully developed its strength in both the streams of science and technology. Soon after continuous developments, MIT created its distinctive identity by building close links with industry, the existing and new firms which evolved from the research and consulting of its faculty. The educational design of MIT put up the technological education and led it to top position in the arts and sciences. However, several decades passed in realizing Roger's nascent entrepreneurial design after the foundation of MIT in 1862. It took over twenty years from concept initiation to funding and decades to fully realize and understand the original idea to establish MIT. Later, MIT influenced the academic development of the US and transferred its entrepreneurial university model to several other universities to Stanford and other institutions across the globe. This case study discusses the foundation of MIT with the idea of a top ranked entrepreneurial university which could transform US academia in terms of science and technology and diffuse it to other schools in US and around the world to enable them to develop their regional economies.

### 3.1.2 The Founding of MIT: A Science-Based Institute of Technology

MIT is a distinctive entrepreneurial university having strong engineering, science, technology, and humanities streams. The founder of MIT, Professor Roger's intentions of a stronger MIT made it to the fact that it proved to be broader and stronger in technology among its competitors. Roger presented an innovative concept for a distinctive technological institution based on science that links to industry and is useful for the society and economy. MIT was established to be broader, and purpose built as compared to the current engineering schools in the US which include Rensselaer Polytechnic Institute and West Point, with their respective specializations in the civilian and military branches of civil engineering. Professor Roger didn't find any institution practically linked to industry in the areas of science, technology, arts of construction, manufactures or agriculture which motivated him to establish such an institution which could be useful for this purpose (Rogers, William. 1954).

Science and technology have been believed to be consistent and equally supportive which serve the common purpose of rationalizing the process of production in the industries of Boston along with creation of new industries through scientific discoveries. Thus, he realized that his ideas could not come into reality in rural area, Professor Rogers shifted from Virginia to Boston where his brother Henry had been doing business. Henry introduced Rogers to the people who could potentially support his project to establish MIT. Boston was the hub for educational, industrial, and technological streams in the US. It was a perfect and fertile ground to implant the notion of a contemporary technological university. To do so, Rogers hired experts from the manufacturers, merchants and intelligentsia in Boston who assisted him to gain private and state funds in addition to the share of the federal government's land grant to the commonwealth of Massachusetts.

Professor Rogers found an interactive and receptive audience among Boston's industrialists for his innovative concept of technological university. Thus, MIT was established with assistance from the share of about 30 percent of the land grant of Massachusetts. Each state was already provided land to establish the higher education institutes to assist the industrial and agriculture development of the states under the Morrill Act. Rogers highlighted the higher level of industrial development and requested the government to release significant funds and land to contribute to the establishment of non-agricultural schools. The outcome is widely credited to the ideas of Rogers and his lobbying activities that discovered fertile soil having a sound base in terms of industries and technology.

### 3.1.3 The Confluence of Academic Streams

In the second half of the 19th century, the state amalgamated two formats of organizations: research institutions and the teaching colleges allocating the first in the agriculture and later in the industrial regions. The combination of both emphasized practical results but was based on basic analyses. The academic institutions performing both roles were considered as a base for an agricultural innovation system which has evolved since the 19th century. This effective and innovative model started spreading in some interesting phases from agriculture to the industry. The change also occurred by means of synthesis with the model of higher education university which was established as an independent and differing format to the system of land grant.

The establishment of MIT is considered as an interplay in between the academic model designed by the international polytechnic movement and the US collegiate and university designs which include the land grants and research university designs. Teaching colleges that were based on the traditional unitary classical curriculum (such as Harvard and Columbia) expanded into the universities of general purpose consisting graduate schools pertaining to arts and sciences along with separate technical schools and colleges. New universities including Johns Hopkins and University of Chicago were established on the notion of pure research-based universities with no any technical majors. The land grant universities were established to advance the agricultural and mechanical arts just after the federal law of the year 1862.

Each of the designed academic models served their specific purpose. For instance, the teaching colleges are meant to train their students in a common body of knowledge which qualify them to the learned professional areas. Harvard college, MIT's Cambridge neighbor established in the year 1636 demonstrates the classical teaching colleges in their origin. Hence, the second stream is known as the "land grant" university that trained students in the areas of agriculture and mechanical which constitute the major industry of the nation till the end of $19^{th}$ century. The third stream relates to the polytechnic institute imported from Europe that emphasized on training the students in the field of engineering to bring industrial revolution. The fourth stream pertains to the research university that is based on the tenets of investigation in the humanities and sciences. All the four academic streams were never separate. They were always interrelated with each other. Numerous research universities have been established based on classical and polytechnic colleges which were affiliated with the research universities.

### 3.1.4 The Origins of the Land Grant System

The concept of the land grant system originated in agriculture to involve the universities in terms of economical and industrial perspectives. The very first station to conduct agricultural experiments was established in Connecticut state in 1816 when agriculture was the only leading industry in the US and was supported by the majority of the people. Despite being the leading industry, it also needed research to improve their crops and production. Hence, the farmers and public pressurized the government to establish the institutions which could conduct research on agriculture. After the establishment of such research institutions, it started replicating in each state.

Later, it was realized that establishing research institutions is not enough and a complete solution to their problems. Conducting research in the isolated setup didn't prove to be effective due to the lack of practice. This required the transmission of that knowledge to the colleges that may train the farmers and their children to put the knowledge into practice. People schooled in scientific agriculture filled the loop in between experimentation and putting it into practice in the land. Hence, the land grant universities transformed the US agriculture and led it to be the global leader in terms of agriculture progression and research.

### 3.1.5 The Rise of the Polytechnic Movement

The rise of polytechnic institutions provided an additional building block for the development of MIT in addition to the land grant movement. Supporters of the polytechnic model had laid out a concept of link between science and practice. They emphasized that science and technology are interconnected and equally supported to serve a common purpose by rationalizing the production processes of the current industries and establishment of newer ones by utilizing the scientific discoveries. To maintain the division of labor among the scientists and engineers, the scientists prioritized developing the advanced and effective physical laws and assisting them in utilizing the current laws to standardize the industrial activities. The engineers were to restructure craft practices into an efficient body of knowledge by means of relevant science.

MIT focused connections with industry instead of agriculture as compared to the typical land grant schools in the region. One of the closest analogues may be the Georgia Tech which was founded as a part of the New South movement initiated after the civil war for renovating an agriculture economy into the technological and industrial lines. Similarly, other industrial regions also attempted to establish polytechnic universities but all of them failed to do so on the tenets of institute plan. The local elites supported the rising industrial base to uplift the introduction of an innovative model of higher education that uses science to standardize the processes of industries and required the universities to train a new technical and managerial class of the students and rising leaders.

### 3.1.6 The Evolution of the Classical College into the Entrepreneurial University

The splitting of the academic world into two tenets: pure and practical strains marked a line in higher education institutions. In contrast to the former in which MIT and its affiliates struggled for getting legitimacy for educational design. The universities which formed from other grounds are modeling them on the tenets of MIT. For example, millions of pounds were committed by the British government in support of the recent collaboration of Cambridge University and MIT. The collaboration was meant to impart older foundations with the luster of comparative newcomers. Practical topics started being introduced into the current curriculum in the colleges to create leaders in the relevant fields in the 19$^{th}$ century. This made colleges to advance their teaching and it started dispersing to hundreds of other colleges in the region and country wide. Faculty members also started introducing and recommending new and practical topics to be added into the curriculum as an addition or in the form of another course in geology and chemistry. Changes also started occurring directly from the leadership and founder of the new university proposing different and innovative ideas for the courses and programs to be offered or taught at the university. Thus, the academic entrepreneurs had the wish to design new subjects from the foundations to make their university distinguished from the traditional ones. Hence, agricultural coursework started being offered at Cornell and urban themes, commercial studies were introduced at the New York University. New England which was home for the first agricultural experiment in the United Nations was also the location of MIT. It was home for the teaching colleges including Harvard and Yale which soon went through the very first ever "academic revolution" by means of the introduction of research into institutions along with teaching. The revolution started diffusing extensively to the current institutions and new ones such as University of Chicago, Clark University and Johns Hopkins were also established to serve this purpose.

### 3.1.7 The Great University

In the 19th century, the model for basic or pure research was proposed by Henry Rowland, a physicist at Johns Hopkins University, in his Presidential address to the American Association for the Advancement of Science (AAAS). Rowland's model suggested to start from the science which drives curiosity and goes to applied research which eventually benefits for the long term. This model got accepted and followed as the institutional ideology for major universities which were being established in the late 19$^{th}$ century by the great industrial fortunes. They believed that this ideology seemed to be an effective fiction which provided a way forward to protect the universities from anticipated and dreaded interference from their respective funders. The academic research programs were rarely intervened due to any external issues related to political instability or any other social issues. Even the issues of the universities concerning political or social sciences didn't interfere with the research activities of the universities. Any preventive actions were taken in the form of anticipatory strikes inside the institution without affecting its research operations. Several examples of such issues are found in the literature, such as the reappointment of the president Henry in the University of Chicago (Storr, Richard. 1966). This would save the goodwill and market worth of the universities in the eyes of public and industrialists who invest and give their fortunes to the research universities. Thus, it brought the concept of academic autonomy and diffused it in the beliefs of public, faculty and industrialists which engrained the development of research universities.

### 3.1.8 The Development of Entrepreneurial Culture at MIT

MIT transformed into a research university due to the development and establishment of research laboratories within the industries and corporations which were run by its engineering graduates. Indeed, several MIT professors such as Wills Whitney have been assisting industries to establish laboratories within the industries. Whitney's developed laboratory became the model for other technology-based companies in the US to develop corporate laboratories with the consultancy of professors of MIT and other leading research universities (Wise, George. 1985). Later, a question on academic research was raised regarding its support in terms of finances. A report generated by the Smithsonian Institution raised a question for academic science and research. It argued that pure scientific research can not be supported directly from the finances raised by its discoveries and findings as the industrial research does (Carty, J.J. 1917). One point of view was that schools need to pay professors from the tuition fees income and endowment funds for supporting the research laboratories of universities. The other suggestion was that professors can do consulting for the industrial firms to get financial support for their basic and applied research conducted in the universities. Thus, the research activities jointly supported by MIT and industries started emerging in chemistry and other fields in the early $20^{th}$ century. Moreover, a certain impetus was also seen from the alumni of MIT and other leading research universities to tend towards basic and applied research to solve the issues pertaining to industry and society.

Several research laboratories were established in MIT with the joint support by school funding and professors consultancy income in the areas of chemistry and other sciences. These laboratories conducted such emerging research that served as role models of contemporary entrepreneurial science which moves back and forth in between the academic, corporate, and business world to benefit academia, society, corporate and economy of the regions and states.

## 3.2 Case Study 02- Dr. Langer Lab as Entrepreneurial Research Center

### 3.2.1 Background

Professor Robert Samuel Langer studied chemical engineering in Cornell University due to his interest in the field. In the final year of his undergraduate degree, he picked chemistry as the major as he was only good in this subject. After his bachelors, he joined MIT for further studies which could have an impact on society. Although his friends joined oil companies with highly paid salaries, he didn't do so. He was criticized for this to ignore job and going for doing something which could transform the world. All his colleagues discouraged him from doing what he wanted to do. Finally, he heard about a professor called Judah Folkman who hires unusual people which motivated Langer to write to him. Judah Folkman was a medical researcher at Harvard University and a surgeon at Boston Children's Hospital. Folkman's current investigation was to stop the blood vessels forming in the tumor. He tried to block them from growing, resulting in starvation of tumors and saving their lives. After getting to know about Langer, Folkman recalled that all the scientists with excellent medical and biological background have not yet solved the problem. So, a scientist with different expertise and background may be useful to solve this problem and thus he appointed Langer in his lab. Langer started working on the issue and finally came up with a solution to stop the growth of blood vessels to form tumors.

### 3.2.2 Academic Accomplishments

Robert Samuel Langer is an American chemical engineer, scientist, entrepreneur, inventor and one of the twelve Institute Professors at the Massachusetts Institute of Technology. He is a former Germeshausen Professor in Chemical and Biomedical Engineering at the Chemical Engineering Department of MIT. He is also a faculty member of the Harvard–MIT Program in Health Sciences and Technology and the Koch Institute for Integrative Cancer Research. He was the only professor who advised his students to do something big which can change the world instead of adding a slight value. He is a pioneer in the areas of controlled release drug delivery and tissue engineering. His services and dedication led his lab to be one of the most productive research facilities across the globe. He joined MIT as Assistant Professor of Nutritional Biochemistry in 1978. Dr Langer has written over 1,250 articles and has more than 1000 patents. He is one the highly cited researchers in history with over 376,000 citations. He is a well-known and widely cited scientist in the areas of biotechnology, drug delivery and tissue engineering.

### 3.2.3 Industrial Accomplishments

Dr Langer's patents have been licensed or sublicensed to over 250 pharmaceutical, chemical, biotechnology, and medical device companies. Dr Langer is honored with over 220 major awards. He is one of 5 living individuals to have received both the United States National Medal of Science (2006) and the United States National Medal of Technology and Innovation (2011). Langer has been playing a founding role for several companies. He has over 20 in partnership with one of the venture capital firms 'Polaris Partners'. Professor Langer is in the Advisory Board of numerous organizations including Patient Innovations which is a multinational non-profit free venue for the patients of any disease, Xconomists, tech news and media companies and others.

### 3.2.4 Five-Dimensional Approach

Langer proposed a five-pronged approach which accelerates the pace of discoveries and ensures their transformation from academia into products in the real world. This approach starts with a focus on 1)the high-impact ideas which can create a difference, 2)a systematic process to cross the typical "valley of death" among the research and commercial development in academia and industry, 3)methods to facilitate multidisciplinary collaboration to create an impact, 4)practices to make the continuous turnover of practitioners and researchers and limiting the funding duration, and 5)the leadership style which could balance the freedom and be supportive in all circumstances. His approach and dedication are creating potential for wealth generation in academia and industry. Professor Langer's lab has an annual budget of $17.3 million at MIT. His lab is one of the labs having the highest budget and creating impact on society. Langer's lab has produced over 40 companies that are working independently or as a part of other companies making an estimated market value of $23 billion. His patents are used by over 300 companies having an impact on over 4.7 billion lives. The final product of his lab are the people who have scored to more than 900 researchers earning graduate degrees, postdocs, and making a distinguished career in academia, industry, business, and venture capital.

### 3.2.5 The Impact

Professor Langer always followed the mantras of considering the potential impact on society, not the money while taking projects. He believed that if something different with potential impact is created, it will automatically attract customers and create wealth. The same is in terms of research to hit the obstacle as ambitious research always does. To the belief of Langer, an invention shows the impact by the number of people it helps. The enterprises emerging from his lab pertaining to life sciences have benefited nearly 5 billion people. For instance, one of the products produced by his lab is a wafer implanted in the human brain to deliver the chemotherapy. Another is produced to cure type 1 diabetes. Looking at such potential projects, foundations, companies, scientists from other labs and government agencies have joined Langer's lab to support up to 63% of the lab's annual budget. The main motive of his lab is to meet the medical and scientific challenges by applying directly or expanding the available science either at his lab or in collaboration with other relevant labs.

## 3.3 Case Study 03- IIT as Entrepreneurial Institution and Transformation of India

### 3.3.1 Background

The concept of Indian Institute of Technology (IIT) originated before the independence of India and after the end of 2nd world war in the year 1945 (Bassett, 2016) when sir Ardeshir Dalal who was a Parsi administrator and one of the successful business tycoons of that time proposed that it's the technology that can brighten the future of India and bring prosperity rather than capital. He emphasized establishing the council of scientific and industrial research and conceptualized numerous institutions to train and educate the workforce (Francis, 2011). Moreover, Jogendra Singh, who was a member of the Executive Council, department of education and the Indian educationist had set up a committee to build higher technical institutions which could bring industrial developments in the post war era in India. These conceptualizations have been known as the base for foundation of the very first IIT in India. The conceptualization of Dalal caught the intention of the chief minister of that time Dr. B.C Roy who showed interest to execute it. A meeting consisting of 22 members was held following the lines of the Massachusetts Institute of Technology (MIT) USA recommended that at least four higher technical institutions in all the regions along with some affiliated institutions to develop the workforce and industry of the country (Jayaram, 2011). The report emphasized the establishment of such institutions on priority basis. The committee revealed that such institutions will not only increase the technology undergrads with high standards as MIT and other world class institutions but will also engage them in research and development of the industry and academia. It was recommended to establish the first institute in west Bengal due to the high concentration of industries there.

Soon after the independence, the country was poor, with hungry and homeless people in majority. In this condition, the leadership of the country followed the trends in their neighborhood who believed that.

*"To plan for 1 year, plant rice. To plan for 20 years, plant trees. To plan for 100 years, teach children."*

Following this recommendation, the government decided to establish the first Indian Institute of Technology (IIT) in Hijli, Kharagpur which is in the eastern part of India in the year 1950. The first established IIT was funded and regulated by the ministry of human resources and development of India which were later given the powers to provide instructions and facilitate other institutions regarding research and development of workforce and industry under the institute of technology act 1961. Thus, the IIT was given the authority to hold exams and award academic and honorary degrees, get fundings and extend cooperation with academic and technical institutions in India and across the globe. In addition, the exchange of fellowships and scholarships were also included in the functions of IIT. The foundations of the very first IIT started working and showed remarkable progress within its vicinity and around the country by training, educating, and developing the workforce and industry which produced the need for extending the network of IIT and forming the other IITs across the country and certain education and training institutions to be affiliated with IIT. Thus, the foundation of IITs and affiliated institutions transformed academia, industry and science and technology in India which enlisted the country among the big economies of the world. Today, India is one of the countries having the best technology as compared to others. The CEOs and directors of the world's largest technology companies are Indians who graduated from IIT or MIT and other technology institutions.

The current case study unfolds the establishment of IITs and the way they operated and transformed the country into one of the largest technological advanced countries in the world. It reveals the way IITs were seeded to develop, educate, and train the workforce which could work for the development of the industries and technology advancements in the country. This study discusses how IITs uplifted the country's technology and economy, transformed it from the hungry, poor, and homeless people to the developing and technological advanced workforce that enjoys all the luxuries of life within the country and across the globe. The rapid expansion of the IIT network locally and globally led the government to support IITs in serving the purpose and work for the advancement of technology, support and facilitate them. The policies pertaining to the technology advancements, education and training of the workforce were revised and made easy for the academia and industry. The establishment and development of the multiple IITs resulted in the economic and technological development of the country and strengthened their cooperation with academia, industry and research and development institutions within and outside India.

### 3.3.2 Foundation of Indian Institute of Technology (IIT)

After the 2$^{nd}$ world war, India was in a very critical condition and struggling for independence. The untiring struggles led the country to get independence, but the country faced hunger, poverty and the majority of the public had no homes to live in. The economy of the country was far behind to uplift the state and bring prosperity for the common people. Thus, few people including politicians, businessmen and educationists such as Sir Adresh, Jogendra Singh, Dr. B.C Roy and others revealed that the capital can't develop the country until we don't work for advancing the technology to educate and train our workforce to develop the industry and technology usage in the country. They proposed that it's the only technology that can transform the country and diminish hunger and poverty from the country. They emphasized on the establishment of the technology institutes following the MIT model and affiliating several technical institutions to it for training and educating the workforce and making it trained and technical to improve the performance of industries in the country. It was necessary to raise the standard of the education so that it could produce the undergraduates and postgraduates competent to the graduates of MIT and other world class technology institutions.

Considering the need for technology and the conception of the technology institutions, the government succeeded in establishing the first Indian Institute of Technology (IIT) in the year 1950 in Kharagpur due to the reason that there were several industries in the surrounding. Hence, the Indian government approached the US government and MIT to assist them in developing IIT on the model of MIT and signed a ten-year agreement with MIT and nine other US institutions of technology to execute it. The sole purpose of establishing IITs was to transform India and make it self-sufficient in terms of technology and technical education to develop their workforce and industry. In addition, several higher education institutes were also affiliated with it to train and educate the workforce. The IIT started functioning and showing outstanding performance which led the government to establish IITs in other regions of the country and making the number of IITs to 16 in the country along with several affiliated institutions. The IITs were funded and managed by the ministry of human resources and development and were given powers to make several decisions on their own without any external influence. IITs consist of four major administrative bodies which include board of governors, senate, finance committee and the committee for building and works. The board of governors consisted of chairman, director, one official nominated by the sitting government's IIT zone, four experts in the field, and two professors serving in the institute. The management of the IITs are designed in such a way to maintain transparency and meritocracy in the institutes.

The network of IITs expanded due to the increasing focus of Indians on engineering and technology which led to the incredible success in India and US. The increasing tendency of Indians to MIT created the circumstances and a motivation and thus MIT played a vital role in redesigning the technology and technical education in India. The first Indian succeeded to get into MIT in the year 1882 which revealed in the "Kesari" newspaper that MIT can probably offer several contributions to India and transform its technology to new heights.

### 3.3.3 MIT Collaboration

The foundation of the very first IIT was only possible because of the collaboration with MIT. When the government of India paid attention to the recommendations given by the educationists and businessmen of the time to establish IT institutions and affiliated colleges. They emphasized on IT which could only change the fate of India and transform industrialization. The government of India contacted the US government and requested for MIT collaboration to assist them in planning to establish IIT. MIT sent their team to survey the country and identify possible ways of establishing the institution on the model of MIT. Thus, MIT and the Indian government signed an agreement to design a module on the lines of MIT for the formation of an IT institution to boost up the performance of industry and economy of the country. Moreover, IIT-MIT collaboration is not limited to the establishment of IIT. It was just a start to build IIT, later their collaboration expanded to all areas of development ranging from education, skill development, workforce education and training to strengthen industry, boost up the economy in all possible ways. Recently, IIT-MIT collaboration has been tied up in the form of "Sandhi (relating to science and heritage) and Shantiniketan projects focusing on key planning issues including heritage, culture, archaeology, and ecology. IIT-MIT collaboration has also worked on improving agricultural practices throughout the country to improve productivity.

### 3.3.4 IIT and Silicone Valley

IIT graduates have contributed largely into Silicon Valley as numerous IITians rushed into Silicon Valley in Northern California starting from the 1970s. Soon, IITians made a breakthrough and reached to the leading and powerful positions in the domain of technology and media. Being highly skilled and talented, IITians have got the privilege of being pioneers of majority of the American big companies pertaining to technology. Recently, an Indian replaced the CEO of Microsoft due to having a better profile in the field of IT. There are several other examples of IITians making history to the highest positions. Intel Pentium, Android, Chrome and apps at Google, electronic design automation company Daisy Systems, pioneering web service, Hotmail, Adobe, google are some of the huge companies pioneered, excelled, and managed by the IITians. Thus, IITians have greatly penetrated Silicon Valley in all domains.

The great contribution of Indians in general and IITians in particular have led one of India's most populous cities to transform into the Silicon Valley of India. In the year 1980, the city started rising to an IT hub by unfolding its potential in terms of IT and engineering. Bangalore is known as the home for several high caliber academic and research institutions along with the headquarters of numerous global corporations constituting more than 50% of the IT industry of India. Moreover, there are several other major industries in Bangalore such as aircraft and aerospace manufacturing, electronics, biotechnology, and machine making. These specifications have made the city an outsourcing center of India and across the globe. Several international corporations sign contracts for their IT projects in India. Recently, hundreds of IT experts have started returning to their country and serve after getting experience in various multinational corporations in the US and other countries.

### 3.3.5 IIT and Global Tech Joints

IITs are not behind in shaping the technology around the globe despite being less known in the western world. Majority of the founders and leaders are the IITians including Adobe, Palo Alto Networks, VMware, Sun Microsystems, and Match Group. A well-known journal 'Wall Street Journal' highlighted them as a unicorn nursery which ranked 4th globally after Harvard, Stanford and the University of California which produce unicorns by their graduates. Delhivery, Flipkart, Zomato, Meesho, Swiggy, M-Fine, and upGrad are also some of the unicorns produced by IITians. This trend of producing successful tech leaders reveals brighter regarding the digital future of India and its name across the world. India started establishing IITs out of desperation after the 2nd world war as the country's infrastructure was very poor with no basics to meet the people's fit for survival. The country lacked all the tools needed for the development which include literacy, electricity, industries, factories, and hospitals to meet the needs of people. Thus, the country was in dire need for engineers which pushed it to strive for the establishment of IITs. Soon after the successful operationalization all over India, the IITs started driving innovation beyond the borders. One of the IIT alumni co-founded Sun Microsystems that led to the development of Java which is believed to be the heart of the computer world. IITians have also excelled in the field of robotics as one of the alumni recently created salad preparing robotics firm.

### 3.3.6 IIT and Indian Unicorn

Over the last few years, several IITs have started transforming the startup ecosystem of India and producing unicorns. They started raising the value of startups over a million US dollars. Few examples of recent unicorns include flipkart, Zomato, ShopClues, and Snapdeel which are created by different IITs across India having worth of more than $1bn. IIT graduates face various barriers of entrepreneurship at the time of initiating startups which tend the investors to bet on their ideas more

easily as compared to the graduates of other schools. It is all possible due to the distinguishing qualities of the IITs which polish and train their graduates to succeed in establishing and developing startups. The students are facilitated with the trained and qualified faculty which prepare rigorous study programs in competitive culture, give them confidence to start again and again despite failing, encourage them to take a realistic approach that connects to society, let them identify the problems persisting in society and find solutions. These qualities of the IITs create a strong ecosystem supporting the startups and entrepreneurs which in turn creates the environment promoting entrepreneurship and unicorns. Furthermore, some of the IIT graduates choose to join Silicon Valley for pursuing higher degrees and get indulged into a better entrepreneurial ecosystem to be able to create billion-dollar startups. This would enable them to be job providers rather than job seekers.

### 3.3.7 IIT and Indian Transformation

The establishment of IITs India was in a very critical stage after the 2nd world war in 1945 and soon got independence within two years but the situation started getting worse due to poverty, hunger, and homelessness. This made some educationists, businessmen and politicians to think of possible ways of bringing prosperity in the country. Professor Ross Basset from North Carolina State University has mentioned in his book "The Technological Indian-2016" that the majority of the Indians graduating from MIT from its founding to 2000 came from the families of high-tech professionals, political families, and businessmen. The author once visited India and found a huge impact of MIT on the infrastructure, art galleries and several companies such TCS, Datamatics and numerous brands. Further, he found all those associated with the MIT graduates. He found 850 Indians having studied engineering form the MIT who came back to their country and served their homeland. Moreover, India decided to develop their education and training system in general and engineering and technology education in specific and make it compatible with MIT. This developed India and took it to new heights in terms of education, technology, industrialization and so on.

IITs have been emphasizing on scientific culture and abstractions following the MIT model which focuses more on doing actual engineering in addition to solving problems related to science, technology and educating and training the workforce to develop industry (Cech, E. 2014). IITs have developed MIT oriented systems which produced TCS, Bangalore (center of high-tech industries) and the experts who can successfully build and run companies in the US environment. Many of the IITians have successfully developed global companies within India which adds to the economy of the country. In the last 20-30 years, Indian IIT and MIT graduates have been doing things in India, US and other countries and are working on the top positions in the world leading technology companies including Google, Fortune 500, IBM, FedEx, Microsoft, Barclays, Twitter, etc. Indian companies are producing several technologically advanced products being sold in the local and global market (Bandyopadhyay & Banerjee, 2003). IIT graduates have transformed the IT industry of India and led it to one of the industries generating a revenue of more than $100 billion and employing a huge workforce of about 3 million.

Besides expanding IT and industrialization in the country by educating and training the workforce, IITs have also been contributing a major part in start-ups in the country in the form of small business and unicorns. Unicorns are the start-ups that are worth more than $1 billion and create a huge impact in the line of start-ups. Recently, the IIT graduates have made a history to give their 100$^{th}$ unicorn to India in the form of Neo-banking fintech portal in the year 2017 (Chugwani, 2022). This unicorn offers several services such as business banking, payments transfer, and management of the expenses to the small and midsize business within the country. Currently, India has 107 unicorns around the country which are initiated by the IIT graduates. All the unicorns have raised more than $94 Bn as funding and are worth $343 Bn, adding a lot to the economy of the country.

The establishment of Indian Institute of Technology and all its campuses across the country along with several affiliated institutions have collectively played a major role in transforming India in terms of technology, industrialization, educating and training the workforce to produce technologically sophisticated products and the economy of the country. The graduates of IITs are doing wonders within the country and globally and are excelling in the field of Information Technology. All this could be possible with the foundation of IIT and rapid expansion across the country along with delegation of the powers necessary for the fair and smooth operationalization of the institutes without any external interference. The country started producing quality products and human resources which are the demand of the current local and global market.

## 3.4 Case Study 04- KIST as Entrepreneurial Institution and Transformation of Korea

### 3.4.1 Background

Korea has a long history of being known as a stagnant agrarian society which had no such plans for industrial, economic, and technological developments. After untiring hard work in the early 1960s, Korea became successful in launching its first industrialization efforts. Before this, Korea was listed in the poor and large population developing countries with very poor resources and production base having a small domestic market. During that time, Korea was facing a very difficult time due to its large population and poor resources which made their life difficult and uneasy. In 1961, the GNP of Korea was only $2.3 billion which equaled approximately $87 per capita, and the main sources of this GNP were primary sectors of the country. The share of manufacturing sectors in that GNP was hardly 15% due to its poor technology, performance, and below average quality of products. The international trade ratio of Korea with other countries was also very low and at the initial stage with an average export of up to $55 million and import of about $390 million which was huge as compared to export of the country (Chung, S., 2011). The import was higher than the export due to the reason that Korean industries were not able to produce quality products as required in the local and international market and those didn't meet the requirements of the international market and developed nations. Besides, the local people also started preferring quality products and services available in the international market which made the Korean government and companies to import the high-quality products to meet the customer demands. This scenario clearly tells that Korea has faced a tough time in the past due to its weaknesses and lack of resources along with poor technology.

Considering the economic degradation of the country, General Park visited the US and requested the president of the United States to assist the country to cope with the economic downfall. After a detailed discussion, they came up with a joint statement on establishing a research center that could raise the industrialization, education and science and technology in the country. Soon after the joint statement, both the countries started planning for its execution. Korea developed a plan and advisory board in preparation for the execution (Kim, G. B. (1990). The US president called upon the team of scientists from MIT to build a team who could visit Korea and conduct the feasibility of the execution of the same. MIT is one of oldest, prestigious, and high ranked research institutions of the world founded in 1865 in Cambridge which has been playing a key role in the development of modern technology and science across the world. A team of scientists headed by Dr. Donald F. Hornig visited Korea in July 1965 and conducted on-site surveys to initiate the execution plan. The team concluded that the current education and training system of Korea is incapable of transforming Korea and its industrialization to uplift the economy of the country. Following the MIT model, a new entrepreneurial research institution needs to be established in the country which could educate and mentor the scientists who could bring innovations in science and technology and transform Korea by bringing industrial, economic, and technological developments. Dr Choi led this project in collaboration with the scientists from MIT and worked for the execution of a joint agreement to establish such an entrepreneurial institution having the ability to produce scientists in the field of science and technology.

Very soon after a few years, their persistent hard work started paying off and they succeeded in establishing research institutions Korea Institute of Science and Technology (KIST) and Korea Advanced Institute of Science and Technology (KAIST) with the untiring efforts of their scientists and scholars (Shin, T., Hong, S., & Kang, J., 2012). After the establishment, KIST conducted its first survey and reported a demand of 16 industries in the country to boost up the economy and meet the production needs of the country. These research institutions started looking for highly qualified scholars and scientist in the areas of economic development, technology advancement, start-ups, innovation, and others, which helped the country in establishing world prominence in such important technology areas which include semiconductors, LCD, telecommunication equipment, manufacturing, automobiles, shipbuilding, and international trade to increase export. These research institutions, specifically the KIST changed the fate of Korea and transformed it into one of the largest economies which competes no less than others in the line of trading countries. Today, Korea has emerged from nowhere to the 13th largest economy and one of the key international players in the global economy.

This case study discusses how the foundation of the Korean Institute of Science and Technology (KIST) changed the fate of Korea and transformed it into an industrial and one of the largest economies of the world. KIST being an entrepreneurial institution was seeded to develop the capacity of research and development along with the advancement in S&T due to the expansion of industrial needs (Korea Science, 2010). This study unfolds how KIST transformed the lower performance labor-intensive industry into the S & T industry of Korea. It boosted the R&D capacity of academic institutions and initiated entrepreneurial activities. The rapid development of KIST called the attention of the government to revise the policy focusing on industrialization, R&D capacity, innovation policies to increase the dynamism among innovation units and the Korean innovation system. The development and advancement of KIST proved that economic development was not possible without development of S&T. Moreover, it strengthened the university-industry relations to advance the use of S&T.

### 3.4.2 Foundation of Korea Institute of Science and Technology (KIST)

The foundation of KIST was only possible due to the hard work and expertise of Dr Choi, Hyung Sup (Nov 2, 1920 – May 29, 2004) on February 10, 1966, with economic aid from the United States. Dr Choi belonged to a small city in Kyungnam

Province of Korea (Shin, T., Hong, S., & Kang, J., 2012). He moved to Japan to study mining and metallurgy at Waseda University. Soon after independence from Japanese Colonialism, he came back to his homeland to serve his nation and started teaching mining and metallurgy at Kyungsung University, which was the predecessor of Seoul National University. Later he proceeded for higher studies in the field of Metallurgy to the United States where he studied Masters from the University of Notre Dame and doctoral degrees at Minnesota University. After completing his studies in the US, he started his career as a researcher in the department of Interior before coming back to Korea. Upon his return to Korea in 1961, he joined Atomic Energy Research Institute (AERI) founded in early 1959 which he developed and led to Korea Atomic Energy Research Institute (KAERI).

He brought several advancements in the institution which enabled him to be the Director of KAERI within a year and a concurrent director general of the mining division in the Ministry of Trade and Industry of Korea. His hard work and strive for excellence continued for years and he developed KIST in covering diverse research domains under the Korean government-sponsored research institute (GRI). Thus, he became the first president of KIST. After a few years, the government was impressed by his outstanding leadership in establishing and developing KIST and offered him to be the Minister of Science and Technology in recognition of his excellent achievements at KIST. During his tenure of around eight years, he established and strengthened the foundations of Korean science and technology in several domains which include but not limited to semiconductors, LCD, telecommunication equipment, manufacturing, automobiles, shipbuilding, and international trade to increase export. He brought revolutions in the science and technology in multiple domains of the marketplace along with production and manufacturing which increased the export and minimized import of the country. He attracted local scientists and researchers to serve KIST, especially the ones who had been to US and other countries for higher studies. Dr Choi actively contributed during service and after retirement in systemizing the developments in Korean Science and Technology and diffusing the successful science and technology model to the international community by means of consultations.

Dr Choi's love for his nation made him extend his services to the national level that he started contributing to the formation of the industrial policies for several projects and long-term plans. His plans such as "Long-term Plan for Energy Supply" (1967), "Basic Direction for Industrialization of the Machine Industry" (1969), and "A Feasibility Study for an Integrated Steel and Iron Mill" (1969-1975) have been recognized to have a profound impact on the success of Korea. In association with KIST and minister for Science and Technology, Dr Choi extended his services and formed a long-term plan for Korean S&T, technical colleges, and research and development wings. He also emphasized on mentoring the scientists and engineers and promoting public awareness regarding the importance of science and technology policies of the country.

### 3.4.3 Long-Term Plan for Korean S&T

Dr Choi found no specific systematic policies for the long-term plan for Korean S&T and felt the dire need for that to sustain the development of science and technology in the country. For this, he set up three basic directions and principles for Korean science and technology policy with a long-term view. Those directions include:

- Building a robust foundation for science and technology development
- The development of strategic industrial technology
- The promotion of public awareness in science and technology.

With this approach, firstly, Dr Choi built an effective system to develop S&T human resources. In the second direction, he developed strategies to make S&T available for the economic development of the country. In the third approach, he created such circumstances for the S&T policies to get the support of the public (Lee, D. H., Bae, Z. T., & Lee, J., 1991). Further, numerous projects were initiated to produce grand frameworks for the development of S&T and the national innovation system. For which, some legal frameworks were also developed that proposed new laws and regulations in terms of S&T operationalization and development. His remarkable contributions made him involved in developing the first and second five-year plans for the development of S&T. In terms of international cooperation, he developed a five-year plan for international technology cooperation with developing and developed nations (1972-1976). All the efforts taken under the KIST established groundwork for the current developments in S&T in Korea.

### 3.4.4 Technical Colleges

Dr Choi realized and emphasized that human resources are believed to be the key components in the development of science and technology. Under the umbrella of KIST, Dr Choi conducted an analytical survey regarding the supply and demand of human resources for the successful industrialization across the country. The survey revealed that with only 5% of high-quality researchers in S&T, 10-15% technicians and 80-85% low skilled craftsmen are needed. The findings were quite critical which disappointed Dr Choi, and he took serious measures in response. As a result, he developed a 15-year plan for educating and training the S&T human resources for industrialization in diverse domains across the country. In this plan, he introduced the concept of establishing Technical Colleges to educate and mentor the qualified proficient technicians in all areas to develop industrialization and improve their performance. Before these policies and regulations, the Koreans used to value and respect only highly qualified scholars. The technicians were treated as a low category employee in the country which created a barrier for the people to choose this line of work. Dr Choi's policies introduced the "Act of National Technician Qualification System" which emphasized on uplifting the status of technicians and providing them all the incentives and benefits they deserved.

### 3.4.5 Research and Development

Being an expert in S&T development, Dr Choi understood well the significance of research and development to promote science and technology. He emphasized that R & D is as important as capital investment in the development of science and technology, and it is in dire need for long term investment. To cope with the situation, he developed numerous research institutions supported by the government focusing on diverse industrial technologies such as chemical, electronics and materials. Research complex was created to draw synergy effects and link all the research institutions closely. To promote industrial technology development, Dr Choi established the "Promotion Act for Development of Advanced Technology' (Promotion Act) in 1972. This act proposed all the rules and regulations for tax credits and public-private partnerships across the country.

### 3.4.6 Scientists and Engineers

Dr Choi prioritized on producing and mentoring high quality scientists and engineers to develop and sustain S&T. He revealed that there are only 5% resources majored in S&T which was a very low ratio to support the S&T development as other staff didn't have the basic skills to perform their duties and be the assets of the institutions. Consequently, he started replacing staff from junior to senior levels with the staff having skills and background of S&T. He also facilitated the staff with basic knowledge to learn the required skills in S&T. He directed the staff to develop long-term plans and policies by consulting with scientists and engineers in the relevant areas. He always welcomed researchers including professors in S&T schools to discuss the problems being encountered and find the best possible solutions. This way, it raised the morale of researchers and scientists and people got interested in pursuing careers in S&T.

### 3.4.7 Public Awareness

Dr Choi always preferred campaigns for public awareness of S&T and created the "Campaign for National Science Awareness". This changed the attitude of the public from ignoring and looking down on S&T which was a barrier in creating scientists and engineers. For this, he founded a special unit named "Division for Fostering S&T Climate" which developed diverse policy measures to promote awareness of the public on S&T. This campaign was nationalized by the government with the initiatives of 'One Skill Per Person', 'Science in Daily Lives,' and 'Technological Independence'. In addition, the national science museum was also established for distributing books to the public and facilitating free classes on practical S&T.

The formation of the Korean Institute of Science and Technology transformed Korea from a poor economy with minimum resources and huge population to one of the biggest global economies of the world which has all the resources of their own with high exports around the world. All this was only possible because of the enablement and continuous development in the KIST which was initiated from a very few domains but covered all the important domains required for the development of the country. Dr Choi being the founder of KIST and rendering his services to uplift the S&T of Korea is remembered as the founder of contemporary S&T in Korea.

## 3.5 Case Study 05- Oregon State University (OSU) as Entrepreneurial University

Etzkowitz,. (2013) stated that Oregon State University (OSU) developed the economy of festival clusters in Ashland, Oregon, USA through the Oregon Shakespeare Festival (OSF) initiated by Professor Angus Bowmer back in 1935. Entrepreneurial professors of social studies and arts working in the entrepreneurial university created the economy of billions (Etzkowitz,. (2015).

### 3.5.1 The Initiation

Angus Bowmer joined the arts faculty as a drama instructor and developed an idea to celebrate 4th July through some festivals in 1935. Bowmer was a truly entrepreneurial scientist of arts. He connected faculty, students, and local towns' people to conduct festivals based on Shakespeare plays. He was an instructor, actor, director, and outreach mobilizer too. They named it Oregon Shakespeare Festival (OSF) and conducted it on the college premises for two years. After realizing its potential, Bowmer arranged for a group of people from a local town to manage OSF regularly. They got it registered as a non-profit independent organization to conduct annual events of OSF.

### 3.5.2 The State Support

Bowmer approached local state authorities and sold his idea of OSF. He secured a lot of support along with numerous venues and places for conducting festivals. He made Govt people part of his project and projected how the festival will serve the development goals of the local community. The local authorities played an active role and contributed to making OSF a success.

### 3.5.3 Academic Support

Bowmer convinced the president of his college for expanded support. OSU provided faculty, students, funding, and administration support to manage these annual events. Faculty designed plays and played various roles. The students volunteered, did research, and performed management work.

Bowmer later enrolled with Dr. Bailey of Stanford University for his Ph.D. He convinced Dr. Bailey for extended support, active participation, and mentorship for OSF. President OSU also invited Dr. Bailey for teaching and research to strengthen the activities of festival plays. The journey now reached from OSU only to OSU-Stanford collaboration to support OSF. The faculty of both institutions contributed actively to the planning and execution of theatrical festivals based on Shakespeare plays.

### 3.5.4 The Growth and Impact

OSF grew to a large number of interconnected festivals in Ashland running for eight months and attracting 20 million visitors annually. A large number of other festivals started making Ashland a big cluster of festivals. Prof. Bowmer and Dr. Bailey of OSU and Stanford gave birth to civic entrepreneurship in collaboration with state support. The growth of festivals attracted tourism, the restaurant industry, the hotel industry, hospitality services, and numerous related economic activities. Professors and students of OSU worked for OSF and other festivals. The practitioners in OSF and other festivals teach as professors of practice, train students, and sponsor research at OSU. The Government of Ashland provides many incentives and services to attract investment in the festival industry. Entrepreneurial professors of social studies working in the entrepreneurial university created an economy of billions.

## 3.6 Case Study 06- Stanford as Entrepreneurial University: Historical Perspectives and Developments

MIT (Massachusetts Institute of Technology) and Stanford University are two of the most prominent and innovative institutions of higher education in the United States. Despite being located on opposite sides of the country, the two universities have a long history of collaboration and shared achievements in various fields. Here are some examples of their shared history and collaboration:

ARPANET: In the 1960s, both MIT and Stanford were involved in the development of ARPANET, the precursor to the modern internet. Researchers at MIT's Lincoln Laboratory helped develop the first packet-switched network, while researchers at Stanford's Artificial Intelligence Laboratory helped develop the first host-to-host protocols.

Entrepreneurship: Both MIT and Stanford are known for their entrepreneurial spirit and have helped launch many successful companies. For example, MIT's Technology Licensing Office has helped launch over 150 companies, including Akamai Technologies and Zipcar, while Stanford's Office of Technology Licensing has helped launch over 70 companies, including Google and Hewlett-Packard.

Open Courseware: In 2001, MIT launched Open Courseware, a free and open online platform for sharing educational materials. Stanford followed suit in 2011 with the launch of Stanford Online, a similar platform. Both universities have since shared course materials and collaborated on online learning initiatives.

Research collaborations: MIT and Stanford researchers have collaborated on numerous research projects, including in the fields of artificial intelligence, robotics, and biotechnology. For example, in 2015, researchers from both universities collaborated on the development of a new type of battery that could potentially revolutionize energy storage.

Joint programs: MIT and Stanford have also established joint programs to promote collaboration and exchange between students and faculty. For example, the MIT-Stanford Joint Program in Policy, Energy, and the Environment is a collaborative program that brings together researchers from both institutions to address global energy and environmental challenges.

Overall, the shared history and collaboration between MIT and Stanford highlight the importance of interdisciplinary and cross-institutional cooperation in advancing innovation and knowledge.

### 3.6.1 Faculty and Students Exchange between MIT and Stanford

MIT and Stanford have a long-standing relationship when it comes to faculty and student exchange. Both universities have strong academic programs and research centers, making them prime destinations for scholars and students seeking to expand their knowledge and experience. Here are some examples of faculty and student exchange between MIT and Stanford:

Faculty Exchange: Faculty members from both universities frequently visit each other's campuses to participate in collaborative research projects, deliver lectures, and share their expertise. For instance, in 2020, MIT Professor Andrew Lo spent a year at Stanford's Graduate School of Business as a visiting professor, while Stanford Professor Alexei Efros spent a year at MIT's Computer Science and Artificial Intelligence Laboratory.

Fellowship Programs: Both MIT and Stanford offer a variety of fellowship programs that allow students and faculty to study and conduct research at the other institution. For example, the Stanford Graduate Fellowship program offers MIT students the opportunity to pursue advanced degrees at Stanford, while the MIT Energy Initiative offers fellowships for Stanford students interested in energy research at MIT.

Research Collaborations: Many research projects at MIT and Stanford involve collaboration between faculty and students from both institutions. These collaborations enable researchers to access a broader range of expertise and resources than they would be able to on their own. For example, in 2019, MIT and Stanford researchers collaborated on a project to develop a new type of carbon capture technology.

Student Exchange: MIT and Stanford students can participate in exchange programs that allow them to study at the other institution for a semester or a year. For instance, the MIT International Science and Technology Initiatives program offers a semester exchange program for MIT and Stanford students interested in international development.

Overall, the exchange of faculty and students between MIT and Stanford fosters a collaborative spirit that drives innovation and discovery. While it's difficult to quantify the number of MIT fellows who have joined Stanford, it's safe to say that the exchange of ideas and expertise has benefited both institutions greatly over the years.

### 3.6.2 Pioneer Faculty from MIT who Joined Stanford.

There have been several notable professors who have moved from MIT to Stanford over the years. Here are a few examples of earlier professors from MIT who joined Stanford:

Frederick Terman: Frederick Terman is often credited with transforming Stanford into a world-class research university. He was a graduate of Stanford and later went on to earn a PhD in electrical engineering from MIT. Terman returned to Stanford as a professor in 1925 and later became Dean of the School of Engineering. He is widely recognized for his role in promoting the commercialization of technology and for his efforts to establish Silicon Valley as a hub for innovation and entrepreneurship.

William Shockley: William Shockley was a Nobel Prize-winning physicist who earned his PhD from MIT in 1936. He later joined Bell Labs and worked on the development of the first transistor. In 1956, Shockley left Bell Labs to join the faculty at Stanford, where he continued his research on semiconductor technology. He is known for his controversial views on race and intelligence, which overshadowed his scientific contributions in later years.

Richard Taylor: Richard Taylor was a renowned physicist who earned his PhD from Stanford and later joined the faculty at MIT. He was awarded the Nobel Prize in Physics in 1990 for his work on understanding the structure of the proton. In 1994, Taylor returned to Stanford to become the James and Anna Marie Spilker Professor in the School of Humanities and Sciences.

Thomas Kailath: Thomas Kailath is a renowned electrical engineer who earned his PhD from MIT in 1961. He later joined the faculty at Stanford and has made numerous contributions to the field of signal processing and information theory. Kailath is a member of the National Academy of Sciences and the National Academy of Engineering and has received numerous awards for his research contributions.

These are just a few examples of the many professors who have moved between MIT and Stanford over the years, contributing to the advancement of science and technology at both institutions.

### 3.6.3 The Stanford and Start of Silicon Valley

Stanford faculty and students played a significant role in the development of Silicon Valley, which has become a hub for technology and entrepreneurship in the United States. Here are some of the ways in which they contributed to the region's growth:

**Research and Innovation:** Stanford has long been a center of innovation in science and engineering, with researchers and faculty members pursuing groundbreaking work in fields such as computer science, electrical engineering, and materials science. Many of these researchers and faculty members have gone on to found startups based on their research and have helped to establish Silicon Valley as a center for technology innovation.

**Entrepreneurship Education:** Stanford has also been a leader in entrepreneurship education, with the establishment of the Stanford Graduate School of Business in 1925 and the Stanford Technology Ventures Program in 1999. These programs have provided training and mentorship to generations of entrepreneurs, helping to foster a culture of innovation and risk-taking in the region.

**Venture Capital:** Stanford faculty and alumni have played a significant role in the growth of the venture capital industry, which has provided funding and support to countless startups in Silicon Valley. In the 1960s and 1970s, several Stanford faculty members and alumni founded venture capital firms, including Ilya Strebulaev, who helped to finance many emerging technology companies, and Jennifer Carolan, who helped to finance Genentech and Amazon.

**Collaborative Culture:** Stanford has a collaborative culture that encourages interdisciplinary research and the exchange of ideas between researchers and faculty members. This culture has helped to foster a spirit of innovation and entrepreneurship and has helped to bring together people from different fields to work on new and innovative projects.

In summary, Stanford faculty and students have played a key role in the development of Silicon Valley, contributing to the region's growth through their research and innovation, entrepreneurship education, venture capital investments, and collaborative culture. Their contributions have helped to establish Silicon Valley as a global center for technology and innovation.

**Earlier startups and spinoff of Stanford started in Silicon Valley:**
Stanford has a long history of spawning successful startups and spinoffs that have gone on to become some of the most successful companies in Silicon Valley. Here are some examples of earlier startups and spinoffs from Stanford:

**The Birth of Silicon and Shockley Semiconductor Laboratory:** Dr. William Shockly PhD from MIT and Professor at Stanford is credited for co-inventing silicon based semiconductors and starting production. He was awarded the Nobel Prize along with his co-inventors. He pioneered with his company to produce transistors and revolutionize the computing industry. The eight employees of his company called eight traitors left his company and seeded the foundation of silicon valley.

- Fairchild Semiconductor Company was founded by these Traitors Eight
- Intel was founded by Robert Noyce and Gordon Moore, two of the members of the "Traitorous Eight" and joined by Sheldon Roberts third traitor
- Julius Blank and Victor Grinich co-founded Amelco, which specialized in manufacturing diodes and other electronic components.
- Jean Hoerni established Union Carbide's Semiconductor Division, where he developed the planar process, a technique crucial to the production of integrated circuits.
- Eugene Kleiner, along with Arthur Rock, went on to become one of the pioneering venture capitalists in Silicon Valley who played a significant role in the growth of numerous technology companies.
- Jay Last founded his own company, called J. Last Engineering, which focused on integrated circuit design and manufacturing.

**Hewlett-Packard (HP):** HP was co-founded by Bill Hewlett and Dave Packard with their professor in 1939, while they were still students at Stanford. The company grew to become one of the largest technology companies in the world, producing a wide range of products including computers, printers, and scanners.

**Google:** Google was founded in 1998 by Larry Page and Sergey Brin while they were PhD students in computer science at Stanford. The company began as a research project focused on developing a better search engine and has since grown to become one of the world's most valuable companies, with products ranging from search to cloud computing to self-driving cars.

**Sun Microsystems:** Sun Microsystems was founded in 1982 by Vinod Khosla, Andy Bechtolsheim, and Scott McNealy, all of whom were Stanford alumni. The company produced a range of computer hardware and software products, including the popular Java programming language, and was eventually acquired by Oracle Corporation in 2010.

**Cisco Systems:** Cisco Systems was founded in 1984 by Leonard Bosack and Sandy Lerner, both of whom were working as computer science staff members at Stanford at the time. The company produced a range of networking equipment and software products and has become one of the largest technology companies in the world.

**PayPal:** PayPal was founded in 1998 by a team of entrepreneurs including Peter Thiel, Max Levchin, and Elon Musk, all of whom had ties to Stanford. The company developed an online payment system that allowed users to send and receive money electronically and was eventually acquired by eBay in 2002.

These are just a few examples of the many successful startups and spinoffs that have emerged from Stanford over the years, highlighting the important role that the university has played in the growth and development of Silicon Valley.

### 3.6.5 Economic Impact of Stanford

Stanford University has had a significant economic impact on both the region of Silicon Valley and the United States as a whole. Here are some keyways in which Stanford has contributed to the economy:

Job Creation: Stanford is one of the largest employers in the region, with more than 13,000 faculty and staff members. In addition, the university has spawned countless startups and spinoffs that have created tens of thousands of jobs in the region and beyond. According to a 2017 study by the Bay Area Council Economic Institute, companies founded by Stanford alumni have created 5.4 million jobs and generated $2.7 trillion in annual revenue.

Innovation and Entrepreneurship: Stanford has a long history of producing successful entrepreneurs and startups, many of which have become major players in the technology industry. In addition to creating jobs, these startups have also generated significant economic value through their products and services. A 2018 study by the Stanford Institute for Economic Policy Research found that companies founded by Stanford alumni have a combined valuation of $2.7 trillion and have generated $8.3 trillion in revenue.

Research and Development: Stanford is one of the leading research universities in the world, with faculty members and researchers pursuing groundbreaking work in fields ranging from computer science to medicine to renewable energy. This research has led to numerous technological advances and innovations, many of which have had significant economic impact. According to a 2017 report by the Association of American Universities, research at Stanford and other AAU universities contributed $71.2 billion to the U.S. economy in 2015 alone.

Philanthropy and Giving: Stanford has a long tradition of philanthropy and giving back to the community, with many alumni and donors contributing significant amounts of money to support education, research, and other causes. In 2019, Stanford received a record $1.3 billion in philanthropic gifts and donations, which will help to support the university's ongoing research and educational activities.

In summary, Stanford University has had a significant economic impact on both the region of Silicon Valley and the United States as a whole, through job creation, innovation and entrepreneurship, research and development, and philanthropy and giving. Its contributions have helped to drive economic growth, create new industries, and improve the lives of people around the world.

### 3.6.6 Economic Impact of Silicon Valley

Silicon Valley, a region in Northern California that is home to many of the world's leading technology companies and startups, has had a profound economic impact on the United States and the world. Here are some keyways in which Silicon Valley has contributed to the economy:

Job Creation: Silicon Valley is a major hub for technology and innovation and has created countless jobs over the years. According to a 2021 report by the Silicon Valley Institute for Regional Studies, the region had a total employment of over 1.5 million people in 2020, with the technology sector accounting for approximately one-third of those jobs.

Innovation and Entrepreneurship: Silicon Valley has long been known as a center of innovation and entrepreneurship, with many of the world's leading technology companies and startups based in the region. These companies have created new products and services that have transformed industries and disrupted traditional business models, generating significant economic value in the process.

Investment and Funding: Silicon Valley has attracted significant investment and funding over the years, with venture capital firms and other investors pouring billions of dollars into the region's startups and technology companies. According to a 2021 report by Pitchbook, Silicon Valley accounted for over 40% of all venture capital invested in the United States in 2020.

Research and Development: Silicon Valley is home to many of the world's leading research institutions and universities, including Stanford University and the University of California, Berkeley. These institutions have produced groundbreaking research and innovations in fields such as computer science, biotechnology, and renewable energy, driving technological progress and economic growth.

Philanthropy and Giving: Silicon Valley is also home to many philanthropists and donors who have contributed significant amounts of money to support education, research, and other causes. In 2020, Silicon Valley Community Foundation, one of the region's largest philanthropic organizations, gave over $2 billion in grants to support a range of charitable activities.

In summary, Silicon Valley has had a profound economic impact on the United States and the world, through job creation, innovation and entrepreneurship, investment and funding, research and development, and philanthropy and giving. Its contributions have helped to drive technological progress, create new industries, and improve the lives of people around the world.

### 3.6.7 Contribution of Stanford to other Countries outside of USA

Stanford University has made significant contributions to countries outside of the United States through its global engagement efforts, which include research collaborations, educational partnerships, and outreach programs. Here are some examples of Stanford's contributions to other countries:

Research Collaborations: Stanford faculty members and researchers collaborate with colleagues in countries around the world on a wide range of research projects. These collaborations have led to groundbreaking discoveries and innovations in fields such as medicine, engineering, and environmental science. For example, Stanford researchers have worked with colleagues in China to develop new therapies for cancer and Alzheimer's disease and have collaborated with partners in India to improve access to clean water and sanitation.

Educational Partnerships: Stanford has established partnerships with universities and educational institutions in countries around the world to support the development of global leaders and innovators. These partnerships include joint research projects, student and faculty exchanges, and joint degree programs. For example, Stanford has partnered with the National University of Singapore to establish a joint research institute focused on energy and the environment and has collaborated with Tsinghua University in China to establish a joint degree program in engineering.

Outreach Programs: Stanford also has outreach programs that aim to address global challenges and support development in countries around the world. These programs include initiatives focused on health, education, and economic development, among others. For example, the Stanford Center for Health Education has developed training programs for healthcare workers in countries such as Rwanda and Kenya, while the Stanford Institute for Innovation in Developing Economies (SEED) supports entrepreneurs and small businesses in developing countries.

In summary, Stanford University has made significant contributions to countries outside of the United States through its global engagement efforts, including research collaborations, educational partnerships, and outreach programs. These efforts aim to address global challenges, support development, and promote collaboration and innovation across borders.

### 3.6.8 Contribution of Stanford in the Corporations and Large Companies of USA

Stanford University has made significant contributions to the corporations and large companies of the United States, particularly in the technology sector. Here are some ways in which Stanford has contributed:

Entrepreneurship and Innovation: Stanford has a long history of fostering entrepreneurship and innovation and has played a key role in the development of many of the largest and most successful technology companies in the United States. For example, founders of companies such as Google, Yahoo, Hewlett-Packard, and Sun Microsystems were all Stanford alumni.

Research and Development: Stanford faculty members and researchers have made significant contributions to the research and development efforts of many corporations and large companies in the United States. Stanford researchers have conducted groundbreaking research in fields such as artificial intelligence, biotechnology, and renewable energy, which have contributed to the development of new products and services for companies across multiple industries.

Talent Pipeline: Stanford has a strong reputation for producing talented and innovative graduates who are highly sought after by corporations and large companies in the United States. Many Stanford graduates go on to work for leading companies in industries such as technology, finance, and consulting, bringing with them the skills and knowledge gained during their time at Stanford.

Corporate Partnerships: Stanford has established partnerships with many corporations and large companies in the United States, collaborating on research projects, sponsoring events and programs, and providing internships and other career opportunities for Stanford students and graduates. These partnerships have provided valuable resources and expertise to corporations and large companies, while also providing students and faculty with opportunities to gain real-world experience and insights.

In summary, Stanford University has made significant contributions to the corporations and large companies of the United States through its emphasis on entrepreneurship and innovation, research and development, talent pipeline, and corporate partnerships. These contributions have helped to drive technological progress and economic growth, while also providing students and faculty with valuable opportunities to engage with the business community and make a positive impact on society.

### 3.6.9 The Influence of Stanford on Entrepreneurship and Technology

Stanford University has had a significant influence on universities across the United States in the areas of academic entrepreneurship and technology. Here are some ways in which Stanford has influenced other universities:

Emphasis on Entrepreneurship: Stanford has a long-standing emphasis on entrepreneurship, innovation, and commercialization of research. Through initiatives like the Stanford Technology Ventures Program and the Stanford StartX accelerator, Stanford has developed a culture that encourages students and faculty to pursue entrepreneurial opportunities and turn their research into new products and services. Other universities have followed suit by developing their own entrepreneurship programs and incubators, and by promoting entrepreneurship as a viable career path for students and faculty.

Collaboration with Industry: Stanford has established strong partnerships with companies in the technology sector, providing opportunities for collaboration on research projects, internships for students, and funding for research. This has helped to bridge the gap between academia and industry and has encouraged other universities to develop similar partnerships with companies in their own regions.

Technology Transfer: Stanford has a well-established technology transfer program that helps to bring new technologies and innovations to the market. Through this program, Stanford has licensed many of its technologies to companies in a variety of industries, creating new opportunities for commercialization and economic growth. Other universities have followed Stanford's lead by establishing their own technology transfer offices and programs.

Educational Programs: Stanford offers a range of educational programs in entrepreneurship and technology, including undergraduate and graduate degree programs, as well as short courses and workshops. By providing students with access to cutting-edge research and industry expertise, Stanford has helped to develop a new generation of entrepreneurs and innovators. Other universities have introduced similar programs and courses, helping to spread the culture of entrepreneurship and technology across the country.

In summary, Stanford University has had a significant influence on universities across the United States in the areas of academic entrepreneurship and technology. Through its emphasis on entrepreneurship, collaboration with industry, technology transfer, and educational programs, Stanford has helped to create a culture that encourages innovation and commercialization of research and has inspired other universities to follow its lead.

### 3.6.10 USA Universities before and after Stanford

Before Stanford University, academic entrepreneurship was not a well-established concept in the United States. Universities were primarily focused on research and education, with little emphasis on commercializing their research or turning their ideas into products or services. There were few resources available to support entrepreneurship, and there was little collaboration between universities and industry.

However, Stanford University helped to change this situation by pioneering new approaches to academic entrepreneurship. In the 1970s and 1980s, Stanford faculty members such as Burton McMurtry, James Gibbons, and James Lattin started to develop new courses and programs that focused on entrepreneurship and innovation. They also created new partnerships with industry, providing opportunities for students and faculty to work on real-world problems and turn their research into new products and services.

Over time, other universities began to take notice of Stanford's success and started to develop their own programs in entrepreneurship and innovation. They started to establish technology transfer offices, incubators, and other resources to support entrepreneurial activity on campus. Today, universities across the United States have robust programs in entrepreneurship and innovation, and there are many resources available to support students and faculty in turning their ideas into successful companies.

In summary, Stanford University played a critical role in transforming the landscape of academic entrepreneurship in the United States. Its pioneering efforts helped to establish new approaches to entrepreneurship education, fostered closer collaboration between universities and industry, and inspired other universities to develop similar programs and resources. As a result, academic entrepreneurship has become a well-established concept in the United States, with universities playing a key role in driving economic growth and technological innovation.

### 3.6.11 Earning of Stanford from Startups, Technology Licensing, Patent Royalty, and Faculty Consultancy

Stanford University is known for its significant contributions to the world of entrepreneurship and technology, and as a result, it has generated a significant amount of revenue from startups, technology licensing, patent royalties, and faculty consultancy. Here is a breakdown of some of the financial benefits Stanford has gained from these sources:

**Startups:** Stanford has a rich history of supporting entrepreneurship, and many successful companies have been founded by Stanford faculty, students, and alumni. According to a report published by Stanford in 2018, companies founded by Stanford alumni have generated more than $2.7 trillion in annual revenue and have created 5.4 million jobs since the 1930s. Stanford invests in the startup and earns equity that contributes in the total tech revenue of the university.

**Technology Licensing:** Stanford has a robust technology licensing program, which allows the university to license its intellectual property to companies for commercial use. According to its 2020 annual report, Stanford received $102 million in licensing

revenue in the fiscal year 2019-2020.

**Patent Royalties:** Stanford holds a significant number of patents, which generate royalties when licensed to companies. In the fiscal year 2019-2020, Stanford received $71 million in patent royalties, according to its annual report.

**Faculty Consultancy:** Stanford faculty members are often sought after for their expertise in various fields, and they frequently provide consultancy services to companies. According to its annual report, Stanford earned $13 million in revenue from faculty consultancy services in the fiscal year 2019-2020.

In summary, Stanford University has generated a significant amount of revenue from startups, technology licensing, patent royalties, and faculty consultancy. The revenue generated from these sources not only benefits the university, but also helps to drive economic growth and innovation, creating new jobs and opportunities for people around the world,

### 3.6.12 Annual Patents, Startups and Technology Licenses by the Stanford

Stanford University is a leading institution in the world of entrepreneurship and technology, and as such, it has a significant number of patents, startups, and technology licenses. Here is a breakdown of the numbers for each of these categories annually:

**Patents:** Stanford is a prolific inventor and has been awarded many patents over the years. According to the university's Office of Technology Licensing (OTL), Stanford researchers were granted 244 U.S. patents in the fiscal year 2020-2021.

**Startups:** Stanford has a rich history of supporting entrepreneurship, and many successful companies have been founded by Stanford faculty, students, and alumni. According to a report published by Stanford in 2018, more than 51,000 companies have been founded by Stanford alumni since the 1930s. In the fiscal year 2020-2021, Stanford reported that it had 41 new startups created based on university technology.

**Technology Licenses:** Stanford has a robust technology licensing program, which allows the university to license its intellectual property to companies for commercial use. According to the OTL, Stanford had 465 active licenses in the fiscal year 2020-2021, generating $102 million in licensing revenue.

In summary, Stanford University continues to be a major player in the world of entrepreneurship and technology, with a significant number of patents, startups, and technology licenses. The university's commitment to innovation and entrepreneurship has helped to drive economic growth and create new opportunities for people around the world.

## 3.7 Case Study 07-Cases of Various Entrepreneurial Scientists

The field of science is full of opportunities, and there are numerous examples of scientists who have turned their research findings into successful businesses. In this case study, we will explore the stories of three local entrepreneurial scientists and two international scientists who have successfully commercialized their scientific research.

### 3.7.1 Case: 1. Dr. G. Sarwar Markhand, Pakistan

Dr. G. Sarwar Markhand developed tissue culture technology for date palm growers. He established the Date Palm Research Institute, which provided high-quality, disease-free, and rapidly growing date palm trees to local farmers. The plants were uniform in age, had a very low mortality rate, and could produce fruit in the third year. Dr. Markhand sold these plants at a subsidized rate, benefiting the rural community and the economy. DPRI sells these plants at a subsidized rate of Rs.1000, benefiting the local farmers by reducing the wait time and mortality rates. Dr. Markhand has a Ph.D. in molecular genetics from the UK, won a Commonwealth scholarship, published over twenty research papers, and supervised MPhil and PhDs.

### 3.7.2 Case: 2. Dr. Waheed Noor, Pakistan

Dr. Waheed Noor, an assistant professor at the University of Balochistan, developed application software for the university and industry to improve process automation. He saved at least 15 million through in-house development, which also contributed to revenue generation. His team won the project of developing an internal examination system for the university, which saved millions and prevented corruption. They are also working on automating student matters through a system that records, tracks, updates, and accesses all information. Dr. Noor obtained his Ph.D. in machine learning from the Asian Institute of Technology, Thailand, in 2013, and is an approved HEC supervisor and technical reviewer of IEEE journals. His team won an FAO project and delivered it successfully, saving the organization a lot of money. In-house development provided IT services to the public and private sectors, generating revenue.

### 3.7.3 Case 3: Dr. Muhammad Zubair, Pakistan

Dr. Muhammad Zubair, an assistant professor at Khushhal Khan Khatak University Karak, developed the olive growing industry in Karak, Pakistan. The district had high potential but lacked growth. Dr. Zubair provided olive seedlings, technical support, and marketing channels, increasing olive yield from 10 kg to 500 kg per tree, and farmers' income from Rs. 5,000 to Rs. 80,000 per acre. He also introduced a new olive oil extraction technology that reduces extraction time from hours to minutes, retaining quality and taste. Dr. Zubair obtained his Ph.D. from Dalian University of Technology, China. Dr. Muhammad Zubair developed the olive growing industry in Karak District, which had high potential but had not yet been established. He introduced new techniques and varieties, trained farmers, and established nurseries. His efforts led to the establishment of a thriving olive industry, providing economic benefits to the local community.

### 3.7.4 Case 4: Dr. Abul Hussam, USA

Naturally occurring arsenic in nature is found to be polluting groundwater at higher concentrations in especially in the areas with deep tube-wells. The presence of arsenic pollution is a menace in Bangladesh where 61 districts out of 64 have crossed the permissible limit and has caused chronic arsenic poisoning to about 77 million people. Environmental activists, government and academic institutions have been putting efforts on developing an easy to use and market a cost-efficient technology for safe water. Adversity of the situation was addressed remarkably by Dr. Abul Hussam, a Bangladeshi chemist at George Mason University in the United States. He developed a cost effective, simple and zero energy input system for arsenic removal from water. From 2001 to 2010, about a million of Bangladeshi have been benefiting from this "SONO" filtration system. The licensed NGO has commercially produced about 160,000 SONO filters up till 2010 being used in Bangladesh as well as in India and Nepal (Wipo Case Studies, 2629).

### 3.7.5 Case 5: Dr. Maurice Iwu. Nigeria

The pain of the nation was felt by a patriot named Dr. Maurice Iwu, who in 1992 developed a non-profit, non-governmental platform (NGO) as Bioresources Development and Conservation Programme (BDCP). Its role is to collaborate local and foreign partners on traditional health practices, medicinal plant varieties and their effective utilization. By doing so, it aims to proliferate and support Nigeria's biological and human resources. In the global pharmaceutical market, hundreds of these are plant derived. Of which, 75% of these herbs are from tropical forests in Africa and South America. The discoveries and commercialization of herbal medicines originated from the knowledge and information obtained from Traditional Health Practitioners (Wipo Case Studies, 3229).

These five entrepreneurial scientists, Dr. G. Sarwar Markhand, Dr. Waheed Noor, Dr. Muhammad Zubair, Dr Hussam, Dr. Maurice Iwu, demonstrate that there is no one-size-fits-all approach to commercializing scientific research. However, successful scientists turned entrepreneurs share common traits such as identifying gaps in the market, utilizing their scientific expertise, and having a strong drive for success. The commercialization of scientific research is essential in driving innovation and solving real-world problems, and these entrepreneurs serve as inspiration to others in the field.

## 3.8 S&T Reflections- IRP Case Study

We started the Institute of Research Promotion (IRP) as an informal community platform to promote applied research in Pakistan. The government departments, industries, chambers of commerce, business associations, universities, policy institutes and international S&T community actively participated in the IRP platform. The numerous types of diversified activities and interventions are planned and executed through the IRP platform. IRP has become a shared history of S&T stakeholders in Pakistan. All the partners tried maximum to promote applied research and produce economic returns from university research. However, concrete outcomes in terms of innovative products in the market, revenue from technology sale, new startups and significant contribution to the GDP is not achieved. The networking, mobilization, collaborative working and increased understanding of dynamics of university-industry joint working is achieved substantially.

### 3.8.1 Training in Applied Research

IRP started with an extensive training program on research methods, publishing, data analysis, writing and winning grants, commercial research and policies and governance of applied research. The Software tools were introduced and promoted to increase use of technology in research conduct. Around 500 training sessions and workshops were conducted from 2005 to 2019 in more than 100 universities across Pakistan. The series of workshops were conducted all over Pakistan on how to collaborate with industry and develop industry needed technology in collaboration with PASTIC (PSF), Ministry of Science and Technology, Government of Pakistan. The industry persons (from SRC Pvt. LTD. and Others) having rich experience of applied R&D used to train faculty on technology development for industry. The industry used to offer applied research projects along with financial and non-financial support to the interested faculty.

### 3.8.2 UIP Symposium

We initiated the University-Industry Partnership Program (UIP) with PASTIC (PSF) under the visionary leadership of that time. The UIP idea was to invite academia to display their technologies in leading chambers of Pakistan for industrial connectivity. The UIP symposiums were organized in various leading chambers like LCCI, RCCI, KPCCI, KCCI, FCCI, GCCI, and others. The university faculty actively participated in the symposium and demonstrated their technologies. The industry participated and noticed the academic works.

### 3.8.3 Innovation Summit

The symposium work was further advanced, and the next version was planned in the form of an innovation summit at the bigger scale. The Punjab university hosted the first innovation summit followed by leading universities of the other provinces as UET Peshawar, Karachi University, and University of Balochistan. This innovation summit was actively organized in four provinces for almost seven to eight years starting from 2012 till COVID pandemic. More than 100 universities, industries and development organizations co-organized these summits in their respective provinces. Around 600 organizations including universities, policy institutes, development agencies, international development agencies, industries, trade bodies, R&D labs and S&T organizations actively participated and contributed to the summits.

University of Management and Technology contributed significantly in the summit drive under the leadership and patronage of Dr. Hasan Sohaib Murad and Mr. Abid H K Shirwani. The summits were a huge R&D and innovation drive across Pakistan that connected S&T stakeholders and promoted collaborations.

### 3.8.4 Policy Workshops

The policies behind the university-industry interaction were identified as primary hurdles for applied research. The series of policy workshops were planned and executed in collaboration with the Pakistan Council of Science and Technology (PCST). PCST is the premier body for S&T policies and hosts a national commission for S&T headed by the prime minister of Pakistan. The S&T stakeholders were invited in all the provinces of Pakistan in the series of workshops to explore policies needed to promote problem solving research in Pakistan. The stakeholders' consultative workshops helped to understand root causes of university-industry gaps and highlighted the policy proposals to address this gap.

### 3.8.5 Technology Portal

The technology portal was conceived after observing the rise of global technology portals like Yet2, Innocentive and others. The portal was launched in collaboration with Lahore Chamber of Commerce and Industries to promote academic research in industry circles. The portal was linked with the LCCI website and universities were approached to add their potential research into the portal. More than 1000 research records were added. The portal was actively marketed in the universities and industries.

Unfortunately, the research in the portal was perceived as irrelevant to industry needs and demands. The strategy was revised, and industry problems were collected and shared on the portal. This achieved some success as a lot of university projects were initiated based on industry needs and problems shared on this portal.

### 3.8.6 R&D Projects

Based upon the multiple interventions, numerous collaborations took place and joint projects were initiated. These projects were submitted for grants from national funding agencies. Many projects won funding too. IRP identified around 500 potential industry projects that were picked by academic scientists. IRP then facilitated academic faculty to connect with industry, plan projects jointly, and execute the projects as per industry needs and requirements.

Most collaborative projects failed due to poor orientation of faculty towards industry environment, less availability of time for the projects, weak capacity to solve industrial problems, absence of economic viability, and non-supportive attitude of industry people.

# Chapter 04

## 4. The Policy Framework for Entrepreneurial Education

### 4.1 The Policy Debate for Higher Education

The policymakers of Pakistan need to address the confusion regarding the outputs and outcomes of higher education in Pakistan. The answers to these questions may lead to the development of strategic direction, realignment, defined goals, and achieving the impact of higher education on the industry and society. Here are fundamentals 10 policy questions regarding the higher education policy of Pakistan:

1. Is higher education creating more job opportunities through new businesses or increasing unemployment?
2. Are higher education training graduates according to industry needs?
3. Are higher education policies, incentives, and assessments flexible and supportive for basic research, applied research, and both?
4. What is the outcome of investment made in training human resources, and setting up scientific labs and laboratories in the universities?
5. What should be the national priority policy for primary, secondary, technical, professional, and higher education?
6. What impact does Pakistan need from higher education?
7. What are the needs of Pakistan and the industry specifically for human resource training?
8. Should higher education focus on basic research or applied or both or both?
9. Should investment in higher education be linked with economic returns like an increase in export, GDP, business activities, and wealth creation.
10. Should universities contribute to socioeconomic development beyond teaching and research?

Pakistan is constrained with scarcity of resources and needs to design its buildup and catch up in technology. There are fundamental steps in terms of people training culture, system, policies, and rate of technology diffusion in the society as prerequisite of knowledge economy. Pakistan needs to set its priorities and gradual developmental pathway to grow as an innovative country.

**Technology Growth Model**

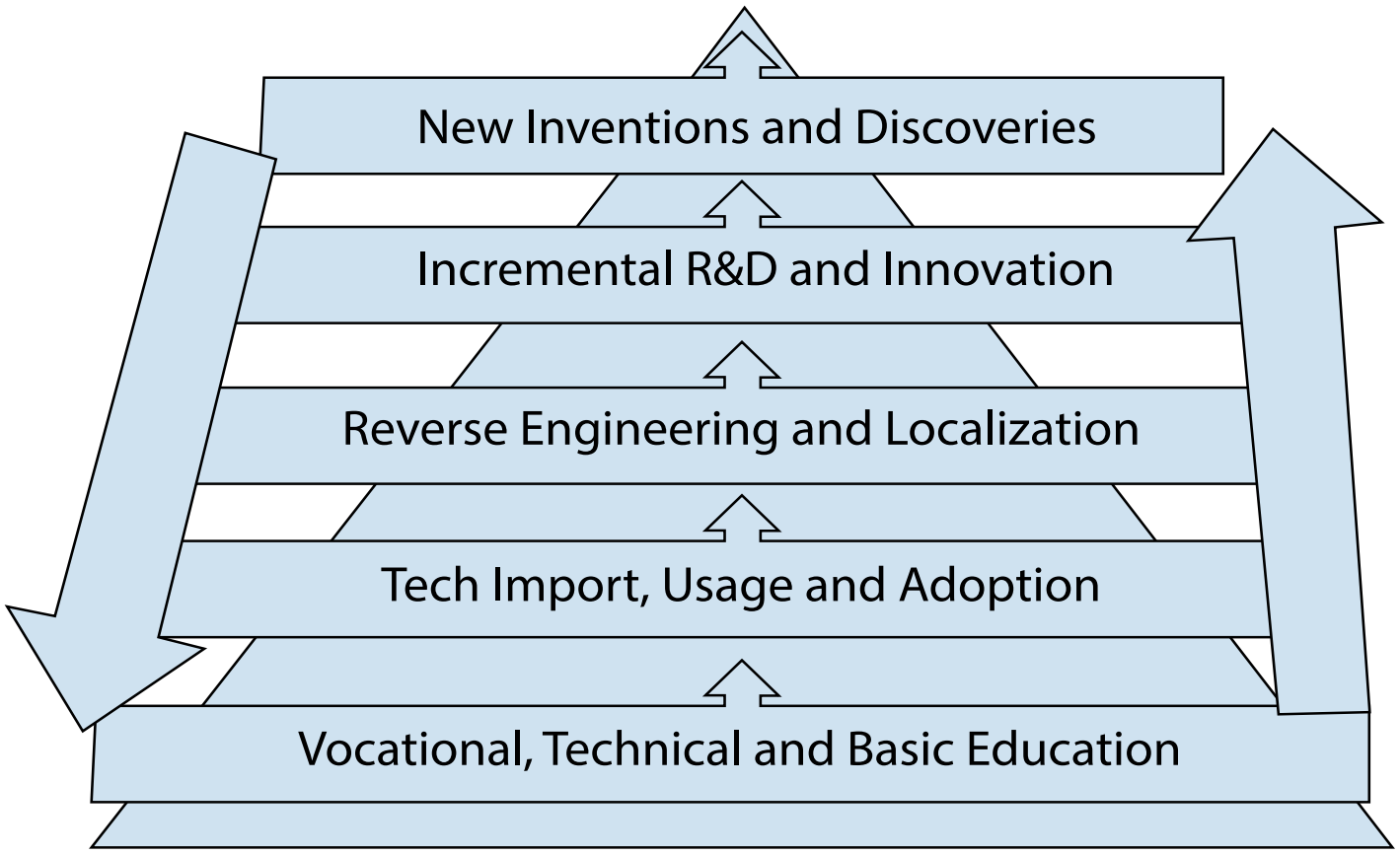

## 4.2 The Death of Industrial Demand for R&D

Pakistan experienced a glorious period of industrialization and innovation back in the 1950-1960s. Pakistan did nationalization in the 1970s and killed the industrial competitiveness of the country. Pakistan tried hard after that in terms of denationalization and attracting foreign investment but could not achieve the desired level. The countries in the region surpassed Pakistan in terms of their industrial base, export, and innovative technologies. Pakistan is stuck in trade, repackaging, and export of non-value-added products. Real estate investment, trade, and services are preferred over manufacturing in Pakistan due to a serious trust deficit between the state and private sectors. A gulf of misunderstanding and negative perception between the state and private sector killed the collaborative efforts of industrial innovation and competitiveness. Therefore, the industry of Pakistan focuses less on R & D intensive products and prefers not to invest in high-value-added sectors. This leads to very less demand for innovation and technology from the local R&D institutions and universities.

## 4.3 The Four R&D Liabilities of Pakistan

Pakistan invests huge amounts of taxpayers' money in four sectors as: 1) defense-related R&D, 2) R&D labs and laboratories of various federal and provincial ministries, 3) higher education laboratories and centers of excellence, and 4) technology imports by Pakistan that consume huge foreign reserves. Pakistan and the world in terms of investment in research and development behave very differently.

**Table 2: The Four R&D Liabilities of Pakistan**

| Areas | Pakistan Practices | The World Practices |
|---|---|---|
| **R&D Organizations** | Pakistan invests to maintain more than 300 R&D labs and laboratories run under the various provincial and federal ministries without economic return. Contrarily, Pakistan spends billions on technical testing done by international labs like SGS. | *The world earns billions from R&D labs in terms of testing, analytical business, new product development, and development of new processes innovated by these labs* |
| **Defense R&D** | Pakistan invests huge amounts in defense-related R&D and successfully developed technologies to ensure national security without significant commercial gains. | *The world invests in defense R&D,ensures national security but also offers technologies to the private sector, and generates a new economy of trillions from defense offshoots.* |
| **Universities** | Pakistan has been investing huge amounts in higher education to strengthen R&D resources in the universities like labs and laboratories and research grants. Pakistan is unable to produce significant economic returns. | *The world earns billions from R&D labs in terms of testing, analytical business, new product development, and development of new processes innovated by these labs. Royalty and Startup income of universities is growing* |
| **Imports** | Pakistan spends huge foreign reserves on imports of new technologies, new plants, and equipment. Pakistan's imports stand at around USD 60 billion and exports around USD 25 Billion which reflects serious negligence of R&D planning. | *The world manages its import substitution and export increase through investment in R&D of the universities. The examples of Korea, Singapore, Taiwan, Israel, and China are very evident, converting the trade deficit into trade surplus.* |

## 4.4 The R&D Turn Around - From Liability to the Asset

The serious question is how long Pakistan can afford to pay for the non-productive R&D sectors from the taxpayers' money. Pakistan needs to revisit, review, restructure and redesign its R&D sector that can produce significant economic returns. Pakistan also needs a serious audit and review of how Pakistan is indulged in this trap of R&D liabilities. There are five basic questions:

- Can Pakistan develop a national consensus for R&D development?
- What should be the legal and policy framework of the R&D sector of Pakistan?
- What should be the governance structure to drive smart and viable R&D?
- How can the R & D sector produce economic returns in the short, medium, and long terms?
- What are the social changes needed to develop the S&T -led economy in Pakistan?
- How to combine fundamental research, applied research and technology commercialization through a system and governance framework?

**Table 3: Five factors to make a turnaround in innovation outcome and generate wealth instead of consuming wealth**

| Present Situation of R&D | Revised Situation of R&D |
|---|---|
| 1. Consumes taxes and state money. | 1. Generate taxes and state money. |
| 2. Innovation is high risk. | 2. Innovation is derisked by state and society. |
| 3. Totally invested by Govt from taxpayer money | 3. Invested by Govt, Industry and Philanthropy |
| 4. Appreciation and acceptance for only academic research. | 4. Appreciation for clinical, technical, vocational, applied, and allied practices. |
| 5. Uniformity and single subject focused | 5. Diversity, holistic, interdisciplinary, and connected with market. |

## 4.5 Policy Proposals for Industrial Demand for S&T

The industry demand for innovation and technology from local universities is very critical for the innovation ecosystem. The absence of this demand can lead to wastage of university resources and cause serious R&D failure. The same is in practice in Pakistan since the development of higher education. The industry demand needs to be incentivized through various measures like:

- R&D expenses on local innovative products to be tax-deductible.
- The duty-free import of inputs for local R&D and technology piloting
- 50% duty reduction on plant and machinery for locally developed R&D product
- Special technology zones for R&D-led enterprises (already initiated)
- Tax credit for newly developed R&D-based products
- Interest-free loans for new innovative projects and plan
- Technology grants for high-risk technology ventures like green projects
- Technology funding to industry for import substitution and export enhancing projects.

There is world over practice of providing technology funding to the industry like the recent USD 280 billion Chip and Science Act of USA. This technology funding for industry is still missed in Pakistan and needs to be initiated as a step one for university-industry connectivity.

## 4.6 Policy Proposals for Social Drive for S&T

The social norms and culture of a society provide incubation for S&T endeavors to pop up and grow. The innovation-friendly social and civic culture determines the future growth and prosperity of the society. The social norms of Pakistan are very less conducive to innovation and technology commercialization. The following norms and values need to be promoted to make entrepreneurial higher education effective in the country.

- The scientific thinking and logic to be promoted from class one to higher education.
- The wealth creation and economic prosperity through knowledge need to be promoted.

- Risk-taking and encouragement for failure need to be appreciated.
- Media should spend 30% of time airing science-related contents.
- Innovation should be considered as an act of high value and dignity in the society
- Venture capital should be made available for innovative ideas and projects.
- Business should be highly valued and appreciated in the society.
- R&D people should be treated very special and of high dignity.
- Students need to be inspired for new ideas and critical thinking
- Symbolic leadership position to be given to the scientists.
- The scientists should be ranked as people of high status.

## 4.7 Policy Proposals for Intellectual Property (IPR)

Intellectual property rights are the mother of innovation in society. IPR inspires people to do innovation, protects them from infringement, and ensures the reward of innovation for the long term. The absence of IPR kills the spirit and usefulness of new inventions. The weak IPR in Pakistan is responsible for poor innovation culture. The strong IPR will play the foundation role for entrepreneurial universities to grow. The industrial revolution of the UK is based on intellectual property rights. The people started investing in new technologies when other innovators were given royal protection in terms of patent grants for certain periods. The USA technology revolution is also based on intellectual property. The US lawmakers converted a patent grant by the royal into the legal right of a citizen to claim intellectual property just like other properties. The enacted law protected the intellectual outputs of the citizens just like land and cars. This inspired people of the USA to do maximum innovation, get protection and earn royalties by selling IPRs.

This IPR law and policy created the economy of trillions. Pakistan is very weak in IPR and needs serious policy revisions.

IP organization needs to be strengthened and empowered to reduce patent grant time from 06 years to 02 years.

- The awareness in society needs to be increased about IPRs.

- The IP course needs to be introduced in primary, middle, secondary, TEVT and higher education.
- IP awareness contents like drama, movies, and plays to be made and propagated.
- IP awareness literature to be produced and publicized.
- IP-related collaborations to be made with the top five IPOs of the world.

## 4.8 Policy Proposal for University Governance

The entrepreneurial university needs to provide a barrier-free environment and incentives for entrepreneurial activities by the faculty and students. The universities of Pakistan present high barriers and very few incentives for entrepreneurial activities. Here are five fundamental barriers that prevent entrepreneurial working in the universities of Pakistan:

1. Very poor belief in commercialization and wealth creation
2. Highly overloaded with in-class teaching and learning and disconnected from the market.
3. No prior training in applied works and exposure to market working.
4. A lot of bureaucratic approvals for administrative, procurements, and mobility works.
5. Industry engagement is not highly appreciated, incentivized not made part of the university system.

Prof Bill Kays, Dean of Engineering at Stanford University said that a person without industrial consultancy engagement is not worth considering for teaching engineering subjects at Stanford. Prof Karl Compton (President of MIT in 1930) allowed the faculty of MIT to spend 20% time outside for consultancy and industrial engagement (Feldman, M., & Desrochers, P., 2003).

**Table 4: Policy Proposal for University Governance**

| Traditional University | Entrepreneurial University |
|---|---|
| Uni-Discipline Studies | Cross Discipline Studies |
| Regulations Driven | Mission Driven |
| Isolated Working Culture | Collaborative Working Culture |
| Striving for Compliance | Striving for Impact and Excellence |
| Knowledge Inspired Research | Problem Inspired Research |
| Publication is Primary and Usage is Secondary Objective | Usage is Primary and Publication is Secondary Objective |
| Concept Memorization | Concept Application |
| In Class Learning from Old Resources | Lab and Market Learning of Real World |
| Pride in Theories | Pride in Application and Usage |
| Knowledge is Ultimate Outcome | Development and Wealth Creation is Ultimate Outcome |

We proposed a governance framework for entrepreneurial universities within the context of Pakistan. The framework consists of culture, incentives, mission, and policy (CIMP) to drive the impact of a university on the regional and national economy.

## 4.9 CIMP Framework for Entrepreneurial University

**Culture**: Culture is the strong element of human living, especially in Asian societies like Pakistan. The people value their culture the most and carry it ahead of anything. The culture includes norms, values, customs, and what is appreciated and not appreciated in the society. The universities aim to transform into entrepreneurial universities and need to start with building a new culture of entrepreneurship. The following components will reflect an entrepreneurial culture in the university.

- Failure is accepted as an instrument of learning.
- Outside mobility and engagement are highly encouraged
- The top leadership is involved in entrepreneurial activities.
- The people of practices are involved in the academic programs and activities.
- The piloting and experimentation are highly supported.

**The Incentives**: Humans are driven by incentives and rewards. The people move towards higher incentives and rewards and make adjustments accordingly. Therefore, it is very important to incentivize entrepreneurial activities in academia like:

- Financial incentives to engage with industry and society.
- Non-financial rewards to contribute to the socio-economic development.
- Known income sharing if faculty earns from outside market.
- Empowerment and autonomy for funded projects won by the faculty.
- Workloads waive-off for faculty having industry contracts and research grants.

**The Mission:** The mission drives the organization towards planned destinations and guides about the pathways to reaching the goals. The mission serves as the soul of the organization. The entrepreneurial universities must include entrepreneurship in their mission of the university and consider the followings:

- The mission includes a commitment to entrepreneurship and research impact.
- The mission guides how the university will contribute to the regional and national economy
- The mission awareness is given to faculty and students to inspire them for entrepreneurship.
- The mission impact assessment is conducted to gauge the impact of entrepreneurship activities of the universities.
- The mission achievement awards and rewards are given to appreciate the implementation of entrepreneurship plans.

**The Policies:** The policies provide the foundation and build the trust of the stakeholders in the system. The policies bind everyone and set clear roles and responsibilities for people to plan and work. The policies also secure the interest of people and drive their inspiration. The entrepreneurial university needs to have a set of policies designed and consulted. The following policy aspects need to be considered:

- The IP policy defines ownership, rewards, and roles in intellectual property.
- The commercialization policy to define financial sharing of startups, consultancy, and other faculty contracts yielding income.
- The implementation of policies, review, and revision to accommodate changing requirements.

- Tracking the impact of policies in terms of financial and non-financial returns and sustainability
- Tracking the impact of policies in terms of contribution to the regional and national socio-economic development

## 4.10 Policy Proposals for Entrepreneurial Teaching and Assessment

The assessment of students guides the entire system of the university. The current assessment is mostly based on root learning, reproduction of memorized content, and presentation of contents based on PowerPoint slides. The application of learning, conceptual clearance, problem-solving, practical orientation, real-life extermination, and market relevance is quite absent in the higher education system. The in-class teaching covers 90% of learning and a very short part includes outside learning.

The entrepreneurial university needs to consider the followings:

- In class, learning should not exceed 50%
- The application and skills-based learning in a real environment should be 30%
- The curriculum, community engagement, and activity-based learning should be 20%
- The students must be assessed on skills as well as concepts.
- The students must be assessed on civic and personality skills.

The students need to be trained in applied environment having market needed skills and personality.

## 4.11 Policy Proposals for Faculty Assessment and Appraisal

The assessment and appraisal of faculty guide them on what to do and what not to do. The faculty's annual assessment and appraisal will decide either faculty should do publications only or convert the publications into saleable technology. The following needs to be considered to achieve commercial outcomes of university research.

- Faculty assessment includes 1) teaching, 2) publication, 3) administration, 4) community engagement, and 5) commercial activities.
- Faculty assessment includes technology funding, research grants, and earnings from commercial works.
- The faculty having own companies to be given high value as model faculty and significant weight in the appraisal.
- The faculty hiring includes a high weight for prior commercial experience and training.
- The entrepreneurial university needs to have 20%-30% faculty of practices and they are assessed based on practical experiences and growth.
- The faculty assessment should not be —one rule fits for all‖ and present flexibility to choose any path for growth

Universities need to liberate their assessment and offer choices and career pathways to faculty. Faculty may choose any pathway to excel and achieve growth-milestones.

## 4.12 Faculty Career Pathways

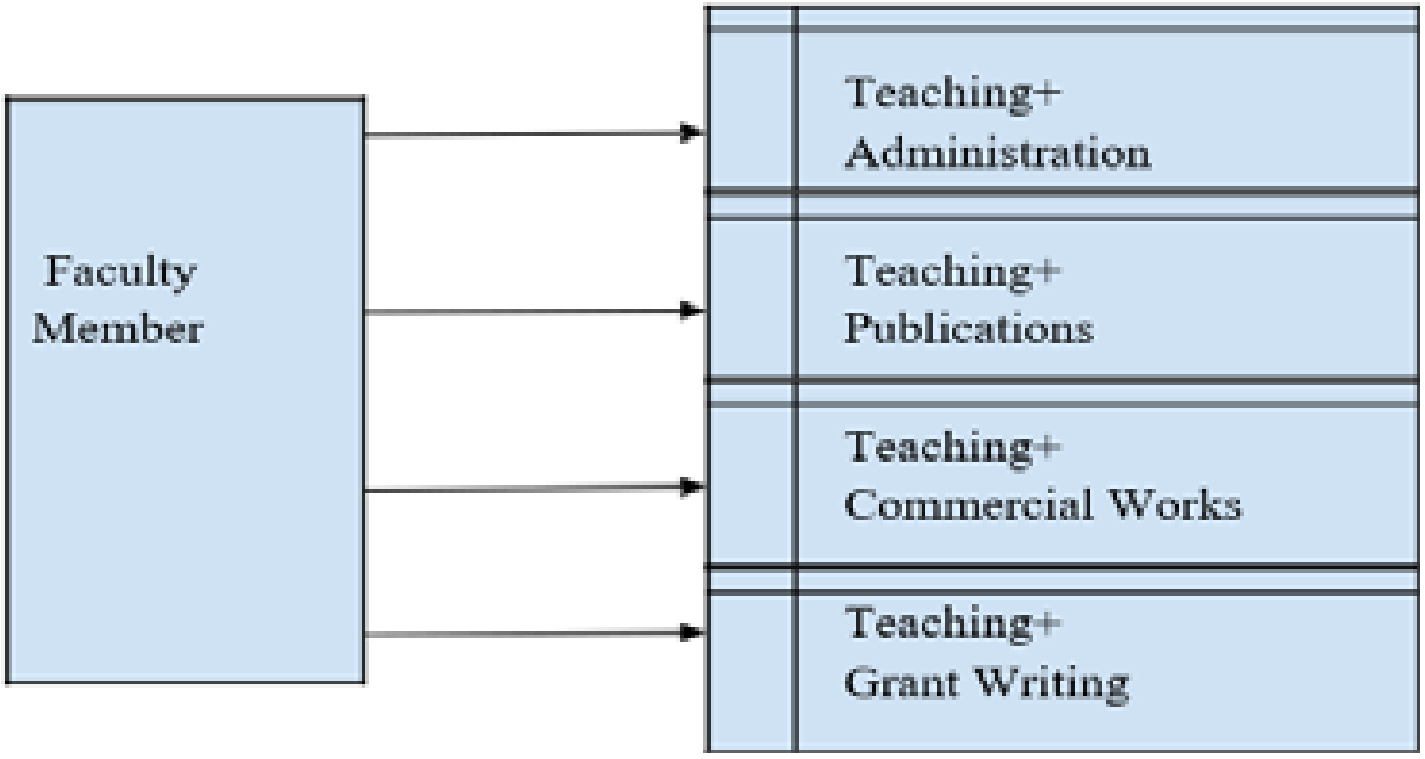

**Figure 2: Faculty Career Pathways**

## 4.13 Policy Proposals for University Assessment

The Pakistan education ministry and policymakers need to develop a university ranking system that responds to the local needs of Pakistan. Pakistan is at a very developing stage and needs a very different response to higher education as compared to universities in the advanced world. The advanced world needs new knowledge creation and assessment of universities in this direction. Pakistan needs to substitute imports and increase exports and expects universities to respond to this applied nature of the challenge. The universities need to partner with the industry for reverse engineering of imported products and technologies and contribute to wealth creation. This wealth creation, when it reaches a surplus level, can be spent on basic research and knowledge creation to respond to the challenge of the knowledge economy. A nation with hunger, poverty, and mountains of debt, first needs to create wealth through the utilization of available knowledge, followed by knowledge creation.

The universities need to adopt, use and diffuse available knowledge first to create wealth and need to be assessed on knowledge diffusion along with knowledge creation.

CIMP Framework needs to be used to assess the universities and drive them to grow as entrepreneurial universities.

- The Culture of the university supports entrepreneurship.
- There are significant Incentives that motivate faculty for entrepreneurship.
- The Mission of the university includes entrepreneurship and emphasizes on the university's impact on the regional and national economy
- The Polices drive university faculty and students toward entrepreneurship.

## 4.14 Policy Proposals for Technology Funding Program

Pakistan has been providing research grants for the last many decades but mostly in the domain of basic research. Recently HEC has initiated technology development and provided funding for applied projects of industrial needs and economic importance. Pakistan needs at least 10 such technology development programs for various needs of the country. The policymakers may consider the following to drive technology development in the industry.

- The technology funding needs to be given for technology development, technology purchase, licensing, piloting, and commercialization also.
- The technology funding needs to be given to academia, R&D institutions, non-profit organizations, startups, industry, and public sector departments, aiming to bring innovation to the market.
- Technology funding needs to be planned for 10 priority sectors like chemicals, food, agriculture, construction, ICT, healthcare, pharmaceuticals, textile, sustainability, and energy.
- The technology funding also needs to be planned specifically for import substitution and export increase connected with import/export data.
- Technology funding needs to be given to commercialize defense R&D offshoots and diffuse military-spared technologies in the civil sector.
- The industry contribution like export development fund to be given for technology development in the industry.

# Chapter 05

## 5. Applied Research Program- ARP

The Applied Research Program – ARP gives choice to faculty to opt for an application-oriented research track having interest to interact and serve society. The program presents comprehensive planning to create an enabling environment around these faculty members who aim to enjoy a hybrid role of serving science and society/industry also.

ARP program is sustainable as it invites institutions to invest initially and earn good amounts and resources in returns from research grants, contract research and technology licensing. ARP program may lead to more economic sustainability of institutions by increasing the share of research-based revenue along with contribution in the development of society and industry.

This policy document proposes a program to promote applied research in Pakistan and create wealth from faculty research. The commercialization of research increases financial and non-financial gains for faculty, students, university, industry, and society. ARP will increase research grants and funding many folds. The program can be implemented by a department, a faculty, a school, and the entire university also.

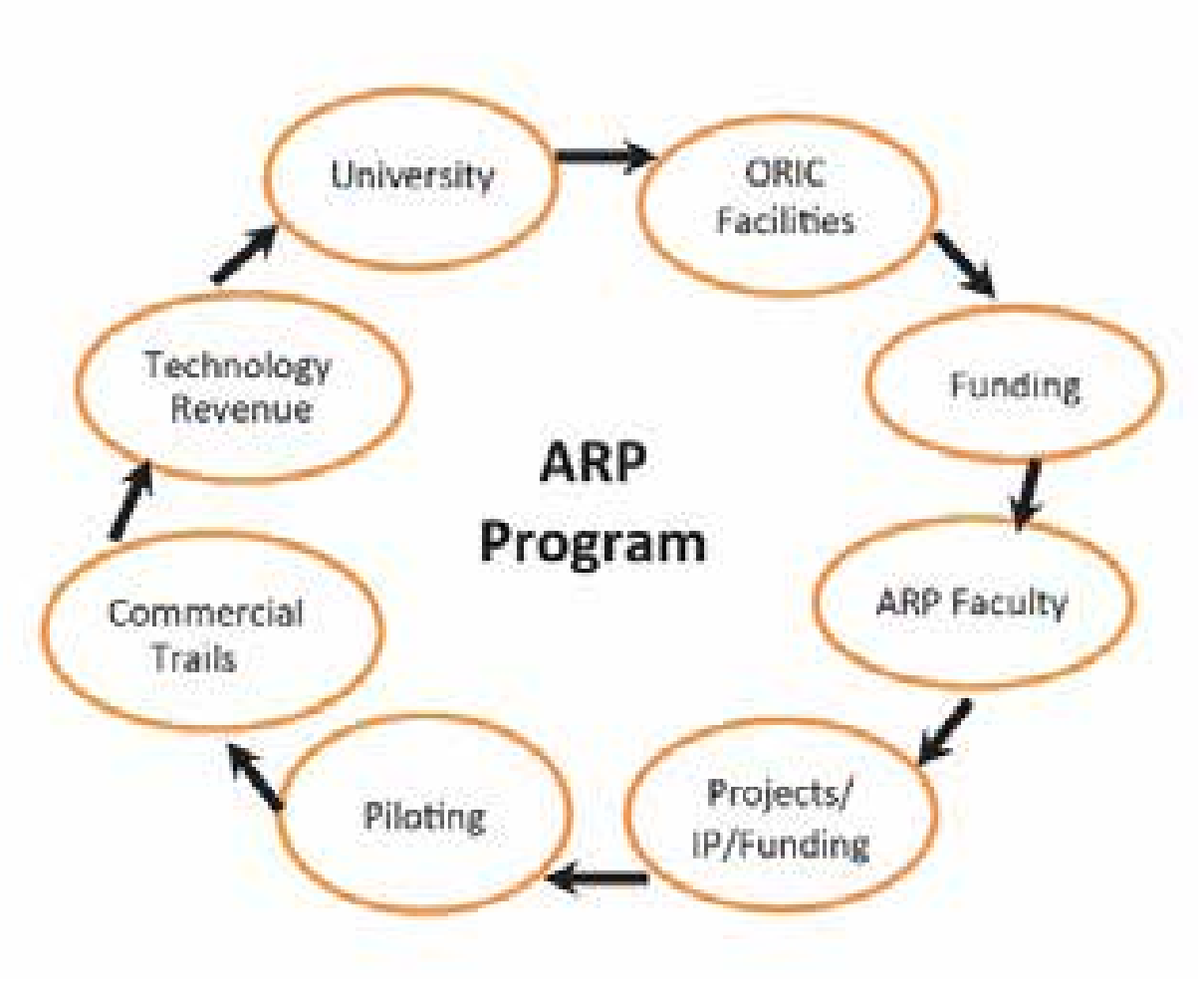

## 5.1 Policy Proposal 01 - Universities are Inspired to Adopt ARP

ARP is proposed to be parked in ORICs of universities of Pakistan. The faculty having interest in commercial research will join and register in the program called ARP Faculty. The university through the ORIC office will provide incentives and an enabling environment to faculty to deliver technologies to industry and society. ORICs will track the progress, make necessary policy revisions and provide support services for technology processes.

ORIC will help in technology planning, industry liaison, commercialization, and over all IP management and licensing. The universities may be encouraged by HEC to introduce this ARP program and provide necessary resources and policy inputs accordingly.

## 5.2 Policy Proposal 02 - Empowering ORIC to Run ARP

The strengthening of ORICs is proposed in the light of HEC policies and instructions to successfully run the ARP. Following are fundamental requirements in ORIC for ARP program:

- Internal committee and external industry committee of ORIC
- Dedicated marketing, management, and finance officers in ORIC
- Transportation to provide mobilization support to faculty and students.
- Operational budget to conduct various interaction activities between university and industry.
- Office structure for regular meetings and interactions

## 5.3 Policy Proposal 03 - Funding for ARP Faculty

The funding scheme for ARP faculty is different from the conventional funding designs. The scheme of funding has to support the path of technology development from idea to commercial trails and technology licensing. The project related to social interventions can also be funded. The university is proposed to provide a funding of 03 million to each of selected faculty members joining ARP with the following scheme.

- Funding amounts up to 03 million.
- 25% at baseline study phase /student thesis
- 50% for piloting in the university and product testing/analysis
- 25% for commercial trails, field survey, and market testing

ToRs for ARP Funded Project

- The project must be approved by end user, potential investor and ORIC.
- ORIC marketing officer must conduct independent analysis and endorse the project.
- Project duration should be 01 to 02 years.
- Faculty must have plan to assign multiple groups on the project.
- Faculty must spend good time in understanding problem, application, technology environment and end user perspective.
- The application must include its viability analysis in terms of market, price, investment, counter product, technology acceptance and user intentions.

The funding agencies are proposed to launch a smart grant program for ARP faculty providing 2-3 million for one project.

One project for ARP faculty must bring minimum three of following outputs:

- Industrial/society research contract
- Research grants and funding
- IP filing and patent
- Technology licensing to industry
- Revenue in terms of technology sale or licensing

## 5.4 Policy Proposal 04 - ARP Faculty – Incentives and Assessment

ARP faculty needs to be given active support by ORIC, initial funding and incentives to ensure successful technology development and delivery to industry and society. Following incentives to be given to ARP faculty

- No administration work for ARP faculty
- Up to 25% work load off.
- Flexible teaching schedule
- Points in annual appraisal for project initiation
- Points in annual appraisal for successful technology licensing and delivery
- 70% share in technology sale/ licensing revenue

ARP faculty needs to be assessed for technology development and delivery. The grant of next project and continuity of grant needs to be based on these technology outputs as.

- Successful industry collaboration and trustworthy linkages
- Increased industry/application orientation of faculty and students
- Winning research grant and funding
- IP filing (patent)
- Successful prototype development
- Successful commercial trails and product acceptance
- Technology licensing
- New industrial contract research

## 5.5 Policy Proposal 05- Academic Theses /Technology Selection

Academic these are very effective tools for technology identification, initial screening, initial trials and baseline study and experimentation.

The ARP faculty is supposed to.

1. Ensure all supervised theses are given by industry and society.
2. Develop understanding of technology needs and make technology development plan.

3. Assign topics to students and build their capacity to research according to need.
4. Ensure that thesis present good review of academic literature, patented technology, and problem to be solved.
5. Ensure that thesis present solution according to requirements and end user inputs.
6. Give remuneration/RA positions to students in funded and sponsored research.
7. Bring some support from industry and society for survey, chemicals, and consumables.
8. Try to file patent from the research work if possible.
9. Send these good projects for research grants and funding.

The department needs to involve ORIC in the review of the thesis supervised by ARP faculty. The above points need to be considered and given 50% weight in the review of proposal and thesis defense along other academic criteria.

Benefit of Revised Policy

- Students will get good training accruing to application.
- Students will have good job on higher salary.
- Industry will develop trust and may get some solutions from academia.
- Faculty will be able to develop solutions after series of problem-solving projects/theses.

## 5.6 Policy Proposal 06 - Piloting of Academic Research

Universities in Pakistan seriously need piloting of lab results to test their developed products and compare with standards, available products, and market requirements. ARP faculty is supposed to pilot their assigned projects, produce a big sample, and conduct testing for comparative analysis.

Each assigned project to ARP faculty is proposed to cover.

- Production in small number
- Consistency in production
- Comparative analysis with counter product

- Analysis with available ASTM standards
- Cost and market analysis
- Consumer testing

## 5.7 Policy Proposal 07 - Commercial Trails of Academic Research

There is very high-quality research in the universities of Pakistan. This research has never touched the phase of commercial trials where industry can pick it further and invest. ARP faculty needs to ensure their assigned research projects reach to commercial trials and consumer testing stages. Here are deliverables of this phase:

- Prior pilot trials confirm needs of commercial trials.
- Commercial trials are planned with investors in industry settings.
- Technoeconomic analysis is done and satisfying.
- Large number of products are produced and used by consumer.
- Consumer reports are ok for investors.
- Investors is ready to buy technology from the university/scientist.

Commercial trials are a pre final stage and based on small production. The success in this stage will lead to final large-scale plants and production set up by industry.

## 5.8 Policy Proposal 08 - Technology Sale and Licensing

The technology licensing has not started in Pakistan and needs to be planned and supported. ARP provides a window for academia to have an exclusive selected program to help faculty reach the technology licensing stage.

Here are outputs expected from ARP faculty in this phase.

- Investors and end users have been part of entire R&D process.
- Industry is satisfied to enter into licensing agreement.
- Legal expertise is involved in contract agreement.

- Contract draft is shared, well discussed, and negotiated.
- Terms are reviewed in detail.
- Royalty sharing, payments schedules and data sharing are planned.
- Other related matters

Technology will be owned and licensed out by the university and revenue will be distributed between university, scientists and any concerned third party. Faculty and students may create their own venture based on this technology. All matters will be treated according to IP policy of the institution**.**

Why should Universities Adopt ARP Program?

1. Development of faculty to contribute to socioeconomic development.
2. Training of graduate on real problems and assured employment.
3. Sale of technology and revenue for university and scientist.
4. Startup creation and promotion of entrepreneurship.
5. Increase in research grants and funded projects.

# Chapter 06

## 6. PESE Framework for Entrepreneurial University

PESE framework is a theoretical model with four constructs explaining theoretical relationships. The model is developed for entrepreneurial scientists who serve science and society through academic research. The framework includes four components as the personality, the environment, the scientific and the enterprising.

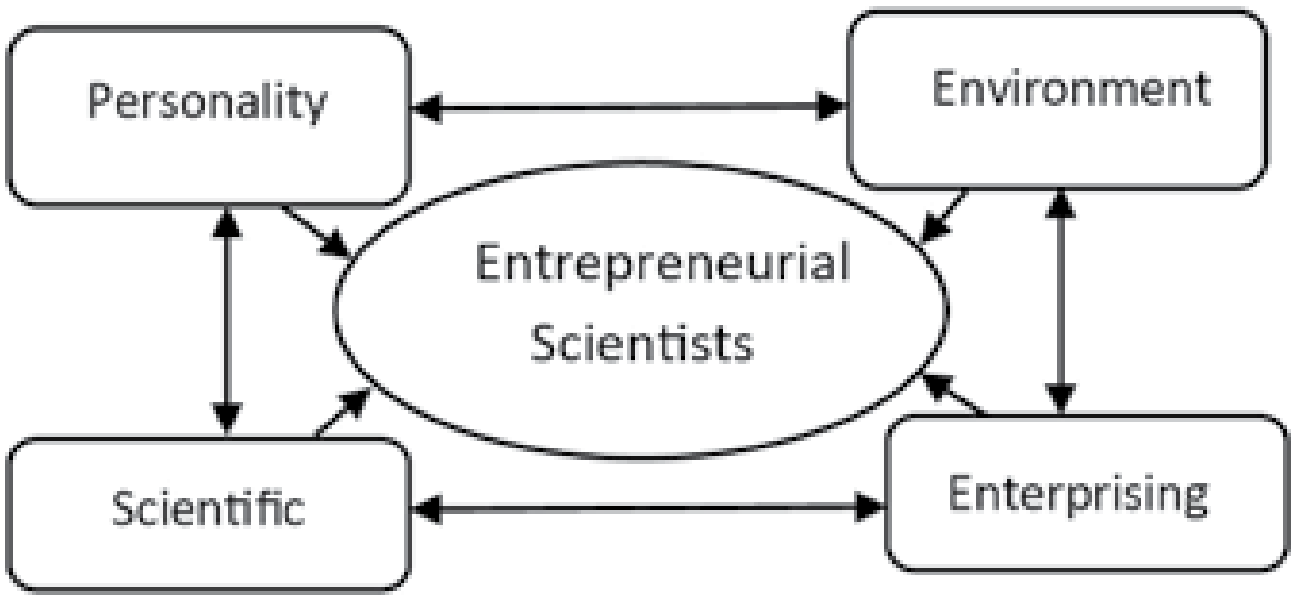

4: PESE Framework for Entrepreneurial University

The PESE framework guides university managers about developing individual scientists for the important role of entrepreneurship in academia. The scientists can be assessed on these indicators as how much potential they have for technology development and transfer. Further, the faculty of the universities can be assessed and gauged based on these components of the PESE framework.

**Table 5: PESE Framework and Related Studies**

| PESE Framework and Related Studies | |
|---|---|
| **Personality Component** | Louis, K. S., Blumenthal, D., Gluck, M. E., & Stoto, M. A. (1989). Entrepreneurs in academe: An exploration of behaviors among life scientists. *Administrative Science Quarterly*, 110-131.<br>Etzkowitz, H. (1998). The norms of entrepreneurial science: cognitive effects of the new university–industry linkages. *Research policy*, *27*(8), 823-833.<br>Moog, P., Werner, A., Houweling, S., & Backes-Gellner, U. (2012). The Impact of Balanced Skills, Working Time Allocation and Peer Effects on the Entrepreneurial Intentions of Scientists. |
| **Environment Component** | Degroof, J. J., & Roberts, E. B. (2004). Overcoming weak entrepreneurial infrastructures for academic. Sciences, 48(1), 24-43.<br>Roberts, E. B., & Malonet, D. E. (1996). Policies and structures for spinning off new companies from research and development organizations#. R&D Management, 26(1), 17-48.<br>Stuart, T. E., & Ding, W. W. (2006). When do scientists become entrepreneurs? The social structural antecedents of commercial activity in the academic life sciences. *American journal of sociology*, *112*(1), 97-144.<br>Gans, J. S., & Stern, S. (2003). The product market and the market for "ideas": commercialization strategies for technology entrepreneurs. *Research policy*, *32*(2), 333-350. |
| **Scientific Component** | Murray, F. (2004). The role of academic inventors in entrepreneurial firms: sharing the laboratory life. *Research policy*, *33*(4), 643-659.<br>Borgdorff, H. (2007). Artistic research and Pasteur's quadrant. *GRAY Magazine, Gerrit Rietveld Academy, Amsterdam, Netherlands*, 12-17. |
| **Enterprising Component** | Oliver, A. L. (2004). Biotechnology entrepreneurial scientists and their collaborations. *Research Policy*, *33*(4), 583-597.<br>Wright, M. (2014). Academic entrepreneurship, technology transfer and society: where next? *The journal of technology transfer*, *39*(3), 322-334.<br>Shane, S., & Stuart, T. (2002). Organizational endowments and the performance of university start-ups. *Management science*, *48*(1), 154-170. |

## 6.1 The Personality Component

The personality component is found major contributing in the concept of entrepreneurial scientists. The personality factor determines the person for having inbuilt qualities of being an entrepreneurial scientist. The model is in accordance with academic literature as the amount of research work advocates for the importance of personal factors in the development of the entrepreneurial scientist. The person with more ability to interact with the scientific community and carries a variety of personal skills presents higher likelihood to become the entrepreneurial scientist (Moog, P., et. al., 2015). Similarly, the scientist with self-belief of being a hybrid researcher makes him serve science and society together. This is termed as the role identity of the hybrid scientist by Jain, S., George, G., & Maltarich, M. (2009). The level of orientation about the entrepreneurial world also contributes to entrepreneurial scientific behavior as reported by Stuart, T. E., & Ding, W. W. (2006).

There may be two main reasons for the weak entrepreneurial progress of few countries. One is the lack of competent human resources and the lack adequate training for the potential entrepreneurs. The lack of awareness among people exists about mandatory characteristics for entrepreneurs. Lazear (2004) proposed jack of all trade's theory, explaining the utmost need of capital and expertise. According to Lazear's theory, to become a successful entrepreneur you do not need to be expert in any single skill, but all your success depends on how multi-talented you are and how effectively you deal with different aspects of your business.

The Lazear research guides the entrepreneurial scientists also in terms of developing the required personal skill to license out technology or initiate their own venture based on new technology.

## 6.2 The Environment Component

The workplace features also help in developing faculty as entrepreneurial scientists is the second presented component of the PESE theme endorsed by Stuart, T. E., & Ding, W. W. (2006). We name this theme as the environment contributes in the making of entrepreneurial scientists. Louis, K. S., et al., (1989) describes social norms related to local groups as an effective feature of the environment.

These concepts are supported by Moog, P., et. al., (2015) and Oliver, A. L. (2004) as they reported various aspects in the environment and policies of the state affect the birth and growth of entrepreneurial scientists. The entrepreneurial environment is a new challenge for society. In a country like Pakistan, people mostly prefer to have employment and secure a job. The entrepreneurial activities are at risk of declining if not backed by entrepreneurship from academia. Lee, Y. S. (1996) research shows that US academia is more eager to move towards entrepreneurship and commercialization in the current era as compared to the past. The large number of the individuals acknowledged the efforts made by their university for commercialization of new ideas. The study of Lee, Y. S. (1996) also reported denial by few respondents as their institutions are not much supportive in promoting startup ideas and commercializing them into the market through partnership with private industry.

It is not necessary that the innovator of an idea only benefits the most where the environment presents healthy market competition. Teece, D. J. (1986) presented a detailed study on factors that affect success of an innovator. He compared the success rate of innovators and their imitators and found that most of the time the one who copies someone's idea gets more profit as compared to the one who initiated it. Discussing the example of Electrical Musical Industries (EMI) Ltd., who developed the Computerized Axial Tomography CAT Scanners for the first time in the market in late 1960's. The company moved forward and introduced the technology to the whole US in the period of six months. The many imitators came forward and made a remarkable impact on EMI's market leadership. Although the technology was initiated by the EMI, but their following manufacturers became more successful within a short span of time. Another example is a small beverage company RC Cola, who introduced the canned packing of cold drinks. It was immediately adopted by Pepsi and Coca Cola; the two major beverage companies. This took away the advantage of being an innovator and initiator from RC Cola and the company went bankrupt.

These studies show that the environment gives an amount of opportunities to entrepreneurial scientists. The environment also presents challenges to be considered by entrepreneurial scientists.

## 6.3 The Scientific Component

The scientific capability in terms of specialization, professorship, in-depth knowledge and publishing in the same area is found complementing for entrepreneurial scientists. We present here both thesis and antithesis for the scientific capability as contributing to entrepreneurial scientists. Moog, P., et. al., (2015) found the generalized skills with diverse experience and exposure presents higher likelihood for the researcher to become the entrepreneurial scientist. The higher in publications means less in entrepreneurial activities. The study of Oliver, A. L. (2004) in the case of biotechnology reported that high rate of publications, academic supervision of PhDs, academic and industrial collaboration leads to high interest in academic entrepreneurship. Louis, K. S., et., al., (1989) supported the study thesis concluding that academic entrepreneurial activities stem from scholarly pursuits and scientific productivity.

## 6.4 The Enterprising Component

The academic scientists mostly qualify for the academic requirements in terms of teaching and research. The entrepreneurial scientists offer an added value to the world by contributing to public or industry life. That is their enterprising capability of using academic and scientific skills to solve the problems of industry and community. The enterprising capability is very much endorsed by other academic studies. According to Moog, P., et. al., (2015), these kinds of scientists interact more, make collaborations with third parties, present patents to the investors and are very much involved in the patent licensing activities. They do contract research and offer consultancy.

# Chapter 07

## 7. Technology Offices in the Entrepreneurial Universities

The technology offices or knowledge transfer offices emerged as holders of entrepreneurial university torch. The characteristics of the entrepreneurial university are built in the functions of technology offices. The technology offices assume a certain role and perform essential tasks as inputs measures for technology. The measures include IP management, exploring opportunities for IP exploitation, managing the technology transfer process, and developing policies and incentives to provide an enabling environment. Similarly, the output measures of technology offices include the number of patents, number of licenses, licensing revenue and number of a startup created from academic research. Both inputs and output measures of technology offices together develop the entrepreneurial university.

### 7.1 The Inputs Measures of Technology Office for Entrepreneurial University

The technology transfer offices were initiated as a result of the growing entrepreneurial university phenomenon in the USA. The university professor increased their engagement with industry and society through contract research, IP licensing and consultancy. The initial technology office history is traced in MIT when IP issues started emerging. Professor W. Bush refused to take royalty for his invention initially. After persuasion by the MIT committee, W. Bush agreed on only 25% to set an example and inspire others. MIT set up a dedicated office to manage technology and IP related issues. This model is replicated by other universities as technology outputs increased (Etzkowitz, 2002). The phenomena of the technology office spread to another world as universities assumed the entrepreneurial role by producing patents and commercializing them.

A technology office can significantly contribute to achieving the third mission of entrepreneurial universities. We term it as technical inputs and include support services, management services, facilitation, capacity in IP, contract and negotiation and other mobilization services can improve the process of technology transfer.

## 7.2 Managing Patenting Process

The scientists are in love with their inventions and fear of theft and loss. They are skeptical about disclosure and demand strict procedures and policies. They also need to have trust and confidence in technology office staff and their capacity to manage secrecy. The university needs to back the technology office with right disclosure policies, the system to protect invention and incentives for sharing with the technology office of the university. The technology office also needs its capacity to plan the right time and place of disclosure and guide the scientists accordingly. The public disclosure of any kind without filing provisional or full patent can kill the technology. Similarly, incorrect disclosure in a patent draft can miss the opportunity of a patent grant. The correct and well-planned disclosure with industry partners is also needed to secure the right partner and potential investment in technology. The disclosure with partners needs to proceed with a non-disclosure agreement.

The drafting of the patent is also very critical and errors in the drafting of claims can lead to patent rejection or much narrowed IP protection. The technology office needs to plan the right extent of the claim, the IP coverage and how to gain optimum protection coverage. This technology input of this section by technology office includes management of disclosure, right drafting of the patent claim, filing, and securing patent and IP policy and incentives.

## 7.3 Exploring Monetization Opportunities

The scientists are trained with a narrowed focus of going deep down and discovering something unknown. They are not trained to horizontally look at the opportunities around this discovery. The technology office assumes the role of finding opportunities around the technology. The technology office needs to surf around the potential market and explore the right partner to invest and commercialize the technology. The commercialization mostly requires the will of the partner, the resources with the partner and most important the capacity of the partner to optimize, produce and sell the technology-based product and services. The technology office needs to conduct seminars in business chambers, promote on social media, publish in newsletters, display in exhibitions, and share through technology partners. The technology office needs to reach out to maximum potential stakeholders who can license out the technology.

The technology inputs regarding exploring opportunities by technology office include sharing with a large number of prospects, demonstrating through prototype and working models, rightly assessing the technology worth and presenting in related meetings and conferences of the business community.

## 7.4 Managing Technology Transfer Process

The technology transfer process is very critical and involves multiple stakeholders. The process also demands a number of activities over different times and levels. Therefore, technology transfer needs to develop a good capacity of managing the technology transfer process. The process has the inbuilt risk of potential loss to inventor, institution, and industry partner too. The inability of scientists to deliver technology may lead to a break of contract and loss of earnings for both scientists and the institution. The ill-planning or inability of the industry to implement and commercialize technology can cause loss to the industry. The mismanagement, unrealistic expectations and delivery failure can damage the image and undermine a very good and high potential technology. The other aspect of technology transfer inputs includes strong negotiation for the right deal, accurate contract drafting, managing all legal aspects, and planning the right roles and timelines for implementation.

## 7.5 Technology Policy and Incentives

The policy provides the foundation for every phenomenon and enables an environment for every actor involved. The absence of policy creates confusion and distrust by all partners. The technology transfer and invention policy sometimes called IP policy are very crucial for healthy technology transfer culture. The policy must provide guidelines about invention rights, invention ownership, disclosure process, resource management and other aspects around invention and discovery process. Similarly, the reward for technology transfer must be planned and communicated. There are needs of incentives for inventors, sharing in income and responsibility of protecting IP from infringement. There should be recognition and appraisal for successful technology transfer. The general policy aspects include incentives for technology transfer, policy for IP, income sharing from earnings and policy for expenses and resources sharing.

# Chapter 08

## 8. Practicing Profeneurship through Applied Credit Hours

Profeneurship is the process in which the professors act as profeneurs to commercialize their academic works like an entrepreneur. The profeneurs involve themselves in activities beyond teaching and research to generate some socioeconomic impact. The profeneurs bring efficiency in converting science into humanly useful products and services when they are made part of economic benefits generated from the science commercialization. The practice of profeneurship can be supported by the introduction of applied credit hours in higher education. This profeneurship will lead to competency-based and outcome-based learning. The profeneurship can be systemized and institutionalized through applied credit hours to improve university-industry linkages and foster socio-economic progress in the country. The applied credit hours must be part of the academic governance system and backed by financial incentives, career incentives, and inspiration for all students, faculty, industry, and universities. This will promote the **Culture of Profeneurship.**

### 8.1 The Isolated Science World

From the inception of the organized academic science world around the 1500s till the mid1800s, science was considered a phenomenon of isolation. The scientists created the science and published it. The practitioners picked the science, tried it again, and developed technologies for human use.

William Rodger found this as a very inefficient system doing a lot of time and resource wastage. He founded MIT and asked professors to commercialize their academic skills and strengths. This was the birth of profeneurship when faculty started consultancy and selling patents (Etzkowitz, H., 2002). The profeneurs shifted from linear to a direct and non-linear model of academic commercialization, saving a lot of time and resource waste.

### 8.2 The Connected Science World- Profeneurship

The profeneurship developed a more connected, integrated, consulted, and efficient science world. The profeneurs are found more capable of commercializing their research works through creating their own ventures, transferring technologies or other modes of commercialization. The profeneurs produced more social and economic impact from state or industry-funded research grants. The profeneurs trained better graduates by sharing practical insight and involving students in real-life projects.

We have seen a new world of technologies and unicorns reshaping human life. This is mostly the result of profeneurship in the leading universities of the world.

## 8.3 Driving Profeneurship in Academia

The current systems of research grants, faculty hiring, faculty promotion, academic incentives and general business rules of universities favor only teaching, publications and administration. The academic business rules do not inspire faculty to become profeneurs and get involved in profeneurship. There is a strong old-fashioned concept in academia as a professor needs to create knowledge only and publish and not be supposed to commercialize. This concept has to be changed and professors need to be allowed to gain commercial benefits of academic research. The professor who creates the ideas is the best person to convert ideas into useful products in collaboration with commercial partners and stakeholders.

The new business rules are needed to inspire faculty to move from isolation science to profeneurship. The profeneurs can solve numerous problems of society and industry if trained and incentivized for profeneurship. The profeneurs can produce higher impact and returns on S&T funding.

## 8.4 The Applied Credit Hours

The two-century-old time-based measurement in higher education must be challenged. The academia needs a policy for a minimum of 25% learning through the application as Applied Credit Hours in the universities.

Laitinen, A. (2012) opened the academic debate of time-based education versus learning-based education by writing the “Cracking the credit hour”. The time measurement unit designed for faculty compensation and pension was adopted to measure students learning and awarding a degree in higher education back in the 19th Century. This time unit of education took very strong roots in the system and not changed for two centuries. The new scholarly discussion is challenging this time-based education measurement and inviting policymakers to consider competency-based measurement.

The revision in credit hours policy can be used to foster university-industry linkages. It will improve students' job placement, increase academic relevance with changing markets, make faculty more oriented to market needs and develop a culture of problem-solving research. This will lead to socio-economic development and economic returns for the country.

## 8.5 The Faculty Credit Hours in Industry and Community

The system of higher education measures faculty performance through the allocation of certain credit hours. In Pakistan, the faculty credit hours are dedicated to classroom teaching only. This contributes more than 70% to faculty performance. The faculty needs to be given certain credit hours for working on applied projects with industry and community.

The revised policy may advocate that 15%-25% of credit hours be offered for working in industry and community. This will demand little financial sacrifice from the universities but will bring higher economic returns in the long term. The faculty spending 3 credit hours in industry and community can sell this experience to win funded research projects, bring sponsorships from industry and improve applied research and education.

The new policy may encourage universities to spare 10% of faculty spend 3 credit hours in industry or community organizations per year.

## 8.6 The Students Credit Hours in Industry and Community

Currently, the students in Pakistani universities spend all credit hours on campus learning. The Pakistan Engineering Council has devised a policy for Outcome-based Learning that encourages students for outcome-based learning. The Higher Education Commission has started some initiatives in this regard. The students need to have the option of spending credit hours in industry and community. This option needs to be managed through strong governance to prevent fake industry letters and fake projects.

The policy in higher education must advocate for a minimum of 25% credit hours in field-based learning in industry, community, and other real-life settings.

## 8.7 Financial Voucher for Applied Credit Hours

The financial loss of the university may prevent sparing faculty for applied credit hours. Similarly, there will be no incentives for students to opt for applied credit hours and will go for easy escape. The industry will be reluctant to support and welcome students and faculty.

The Government of Pakistan needs to issue a financial voucher of applied credit hours given to the practitioners, industry and community organizations. The practitioners will hire faculty and students through this financial voucher and paid by the government.

Rs. 100,000/- may be given to faculty for spending 03 credit hours in industry and community and Rs. 30,000 to a student for 03 credit hours in industry and community.

## 8.8 The Tax Credit for Applied Credit Hours

Alternatively, the tax credit may be offered to the private sector for paying against applied credit hours of faculty and students. The paid amount be deducted from the payable taxes. This will inspire the private sector to increase engagement with academia and hire maximum faculty and students to get involved in practical life.

## 8.9 Policy for Academic Assessment for Applied Credit Hours

The academic assessment of faculty and students needs to have a good weight of applied credit hours. The faculty promotion, annual appraisal, and increments need to be linked with applied credit hours. Similarly, the students' final degree must be linked with a minimum of 25% applied credit hours.

# Chapter 09

## 9. Faculty Remuneration Policy under Entrepreneurial University

### 9.1 Summary

The universities operating on the conventional model of teaching and basic research are missing numerous revenue streams developed by the entrepreneurial universities. The faculty earn much more from entrepreneurial activities as compared to university paid salaries.

### 9.2 Conventional University Model

The salaries of faculty in Pakistan largely depend on public sector funding in government and on students' fees in private universities. This leads to a high financial burden on the state and students. The faculty is assigned maximum course load that leads to no time for quality research, industry engagements and other activities of socio-economic development. This is a conventional university model where teaching and basic research is focused.

### 9.3 Entrepreneurial University Model

The entrepreneurial universities developed other streams of revenue for institutions and faculty. The entrepreneurial faculty earnings include salary, research grants, industry funding, and licensing income from intellectual property. These non-salary streams of faculty remuneration generate a lot of other economic streams like new businesses, technology transfer, consulting, new products, and services. These streams bring a lot of money back to university, add in tax collection for the state, and generate a lot of employment, increase country export, and add in national GDP. Most importantly this entrepreneurial approach develops skillful students and enables them to start their own ventures.

These remuneration streams are still missed in Pakistan. The university faculty needs legal permission, incentives, recruitment criteria and promotional parameters to develop non-salary revenue streams like:

- Faculty earning from university salary.
- Faculty earning from research grants, entrepreneurial and industrial activity.
- Faculty earnings from IP licensing
- Faculty earnings from own startups and companies

## 9.4 The Economics of Entrepreneurial Universities

### 9.4.1 The Economic Output of Academia in EU, 2017

The Technology Transfer Offices (TTOs) in the USA and Knowledge Transfer Offices (KTOs) in the EU play the leading role for research commercialization and academic entrepreneurship. The TTOs and KTOs earn a lot of money from academic outputs and develop an enabling environment for faculty and students to engage in commercial activities. The below given two images show that an entrepreneurial university creates numerous economic activities and add significantly to national GDP. In the EU, around 5000 new companies are created by universities and research contracts worth 3.81 billion are made.

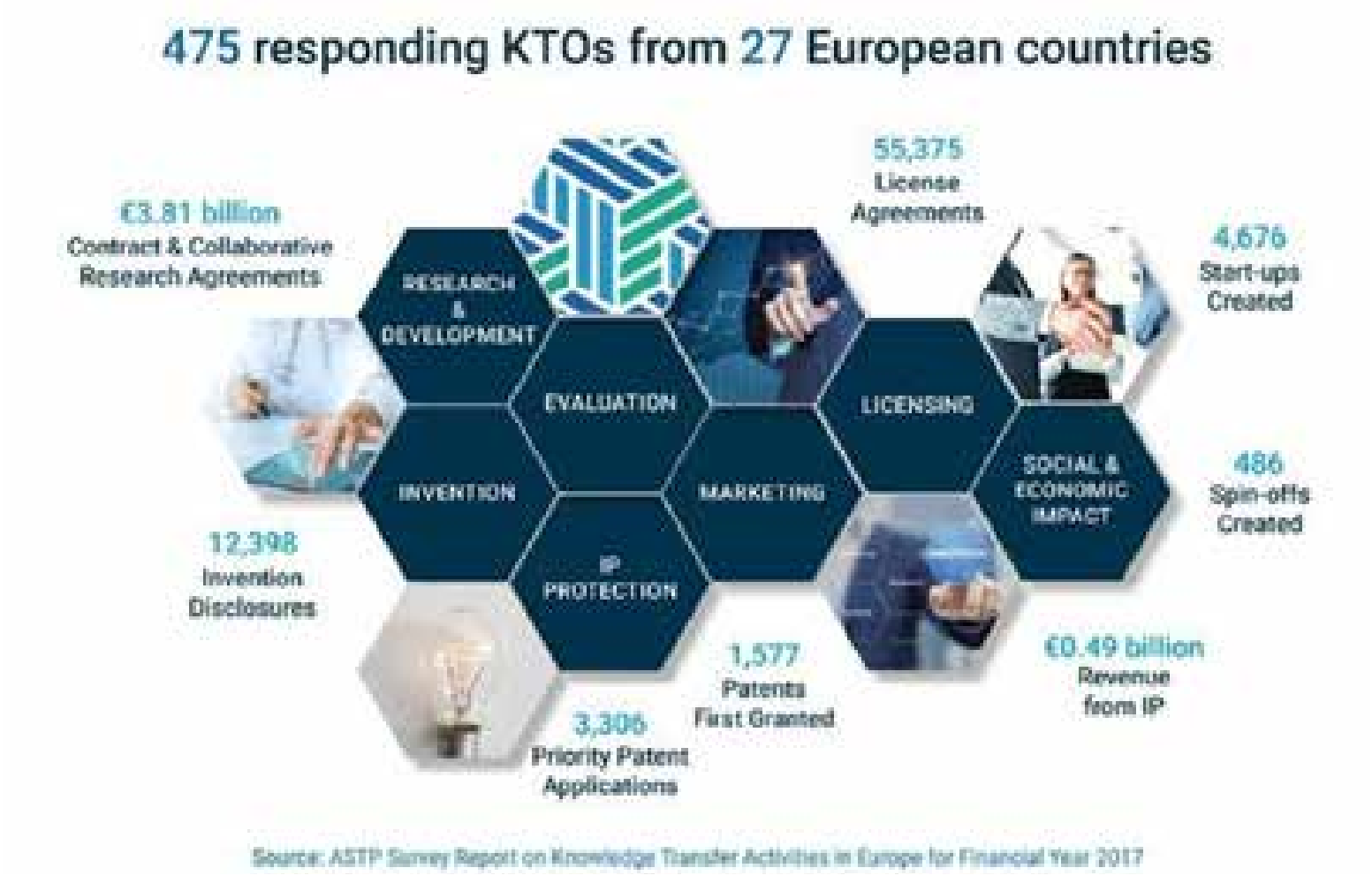

**Figure 5: KTOs from European Countries**

### 9.4.2 The Economic Output of Academia in the USA for 25 Years

According to the AUTM report 2017, the USA TTOs lead the world in research commercialization as USA universities frequently make engagements in entrepreneurial activities. The universities have created huge wealth through the commercialization of academic works in the last 25 years. The universities made an industrial impact of USD 1.3 trillion along with USD 591 billion in USA GDP and 4.3 million new employment opportunities. The 11,000 new companies are created by university faculty and students. The 200 news drugs and vaccines are introduced in the market generating a lot of economies and improving the health quality of USA people. Around 0.38 million new inventions are disclosed and 80000 are protected through patents for commercialization and revenue generation. This huge economic contribution is largely based on faculty works and developed culture of faculty involvement in applied research and commercialization activities.

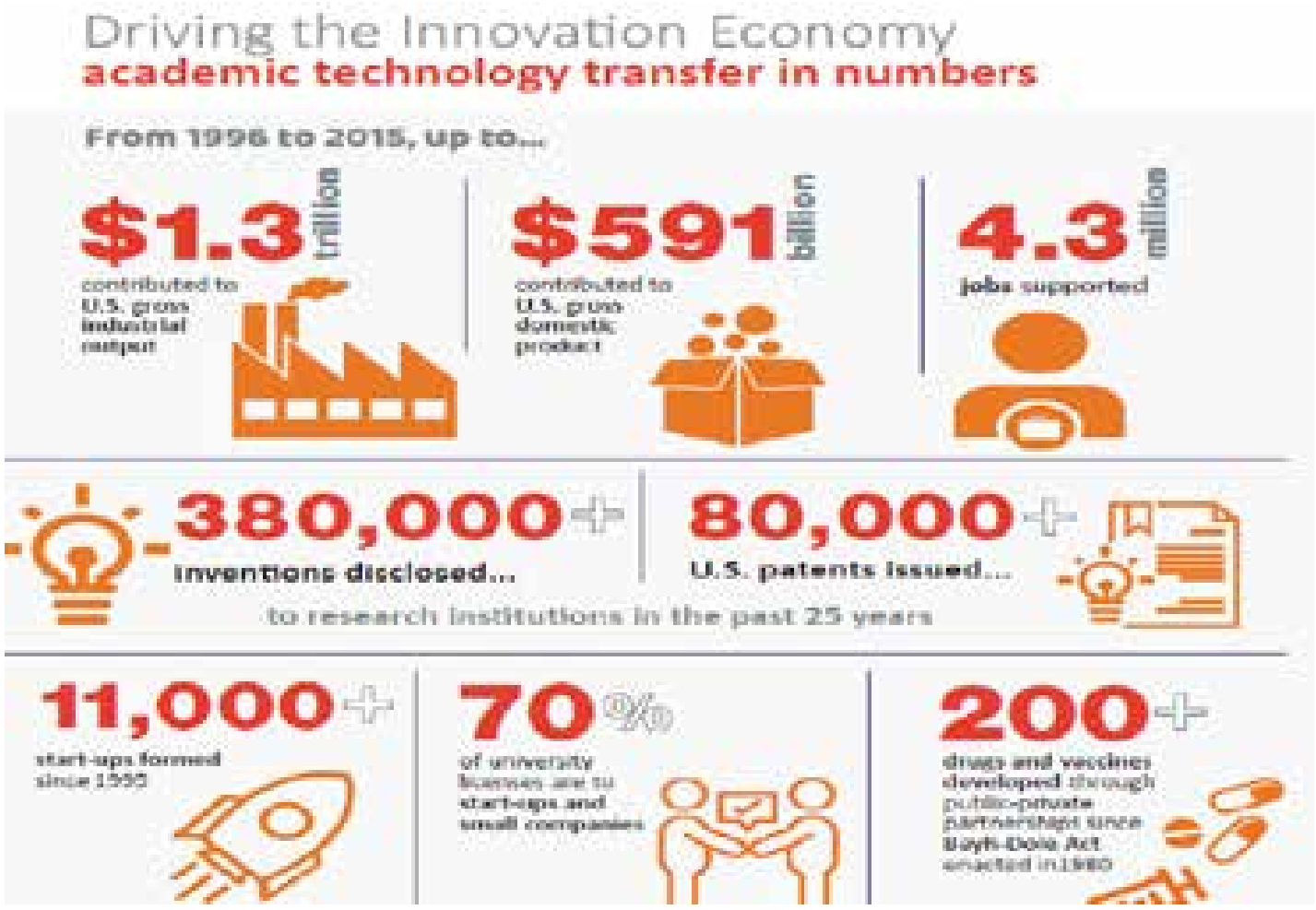

**Figure 6 Driving the Innovation Economy**

### 9.4.3 Faculty Earning from Industry and University Salary

The table 06 shows that faculty in Germany earns from industry yearly USD 200,000 followed by the USA, and China. The faculty in South Korea earns around USD 150,000 yearly from industry followed by the Netherlands, Turkey. The faculty members in Taiwan earn from industry around USD 100,000 and in India around USD 50000.

## Table 6: Faculty Earning from Industry and University Salary

| S. No | Country | Institution | Academic Salary ($US / Year) | Industry income per academic (PPP $US) | Salary % Ratio of Industry Income | Salary % Ratio of Total Earnings | WUR 2015-16 rank |
|---|---|---|---|---|---|---|---|
| 1. | USA | Johns Hopkins University | $102,402 | $249,900 | 41% | 29% | 11 |
| 2. | Germany | LMU Munich | $83,173 | $392,800 | 21% | 17% | 29 |
| 3. | Belgium | KU Leuven | $49,492 | $163,700 | 30% | 23% | 35 |
| 4. | Netherlands | Wageningen University | $97,230 | $242,500 | 40% | 28% | 47 |
| 6. | Sweden | Swedish University of Agri. | $75,828 | $144,200 | 52% | 34% | 201-250 |

*Data is altered into yearly average and USD.

**Sources:**

- https://www.timeshighereducation.com/world-university-rankings/funding-for-innovation-ranking-2016. (Accessed on: January 28, 2020)
- https://academicpositions.com/career-advice/professor-salaries-from-around-the-world. (Accessed on: January 28, 2020)

WUR Ranking: is an annual publication of university rankings by Times Higher Education magazine. https://www.timeshighereducation.com/world-university-rankings. (Accessed on: January 28, 2020)

The below-given table shows that faculty earn much more from industry sources as compared to university salary. A few selected universities are presented below and a detailed list is given in the appendix. The data shows that in USA Johns Hopkins University the university salary is only 41% of faculty earnings from his industry earnings and 29% of his total earnings. In the case of Germany LMU Munich, academic salary is 21% of his industry earnings and 17% of his total earnings. In the Netherlands, the faculty earn 40% of what they earn from industry and 28% of their total income. In Sweden, the faculty earn 52% of industry income from university salary and 34% of their total earnings from all sources.

### 9.4.4 Faculty Industry- Earnings – The Examples from the Countries

#### India

Indian Institutes of Management (IIM) lacks behind and yet to reach up to their global equivalents in terms of faculty earning policies but their corporate sector is gaining benefits from their expertise and making them richer. One of the prominent expertise and entrepreneurial activities of any faculty member could be the consulting activities. These consultancies help them earn more as compared to what they earn from their predefined salaries from institutes. The stats show the earning of IIM faculty on average as Rs. 12 lakhs from consulting fees and share the 45% of their earning with their home institute. This net amount is comparatively greater as on average a faculty member earns Rs. 8 lakhs as annual salary from the university.

**Source:***https://economictimes.indiatimes.com/industry/services/consultancy-/-audit/iim-professors-earn-more-from consultancy/articleshow/5122071.cms?from=mdr*

#### Malaysia

Malaysia is considered to be another exemplary country in the context of income generation activities among the faculty members other than the basic salary. The data generated from the interviews of researchers and analyzed by the interactive model reported that the researchers at higher education institutes of Malaysia acquire a significant amount of income from the research, consultancy and commercialization activities that contribute to the financial sustainability and advancement of a university. The following image summarizes some of the income-generating activities carried out by academic staff at HEIs in Malaysia.

**Table 7: Summary of Income generating activities carried out by academic staff at HEIs of Malaysia.**

| Author (s) | Income-generating activities |
|---|---|
| Albrecht $ Ziderman (1992) | Short ad-hoc vocationally oriented courses, applied contract research (industry) and consultancy. |
| Guang (2011) | Tuition fees (contract with private sector) and provision of service |
| Kassel (2011) | Teaching and training fee opportunities (presentation and workshops), conferences, meeting and full-length courses. |
| Kiamba (2004) | Pure consultancy, specialist-base production unit, general production unit, seminar, workshops and short courses. |
| Siswanto et al. (2013) | Education, research and community services. |
| Todericiu (2009) | Private companies fundraising through scientific research, sponsorship, education with tuition fees and funding from European projects |
| Wachter et al. (2012) | Contract with private partners, philanthropic funding, provision of service, consultancy facility-related services. |
| Wedgwood (2006) | Contract research, consultation R& D (linkage with industry). |
| Molas-Gallart et al. (2002) | Technology commercialization, entrepreneurial activities, advisory work and contracts, research, teaching and communication. |

Source: Types of income generating activities implemented in public universities in Malaysia (Source: Adopted from Asia (2012)) *https://www.researchgate.net/publication/277881266_Income_Generation_Activities_among_Academic_Staffs_at_Malaysian_Public_Universities*

## Jamaica

ORIC/TTO offices developed within the entrepreneurial university framework helps generate sustainable income streams that ultimately benefit the institutes and helps in the national development along with the financial stability of the faculty members. A study reported by the School of Graduate Studies, Research and Entrepreneurship and an Office of Development, University of Technology, Jamaica, exemplifies the identification of such independent expansion of revenue resources which included graduate programs, applied research and consultancy, endowments, and proficient training to industry.

**Table 8: Projection of Revenue from Income generating Streams over a period of five years in Jamaica**

| Income Generating Streams | Year | | | | |
|---|---|---|---|---|---|
| | 2007/8'000 ($) | 2008/9'000 ($) | 2008/10'000 ($) | 2008/11'000 ($) | 2008/12'000 ($) |
| Commission and Applied Research | 300,000 | 330,000 | 363,000 | 399,300 | 439,230 |
| Consultancy | 300,000 | 330,000 | 363,000 | 399,300 | 439,230 |
| Training | 300,000 | 330,000 | 363,000 | 399,300 | 439,230 |
| Grants | 300,000 | 330,000 | 363,000 | 399,300 | 439,230 |
| Projects | 300,000 | 330,000 | 363,000 | 399,300 | 439,230 |
| Total | 15000000 | 1650000 | 1815000 | 1996500 | 2196150 |

Source: *Oliver, G., Henry, M., Newman-Rose, C., & Streete, T. (2013). Strategies for Income Generation at the University of Technology, Jamaica. Latin American and Caribbean Journal of Engineering Education, 2(1).*

### 9.4.5 Summarized Concept

The entrepreneurial universities in the developed countries are promoting the collaborative approach of academia to connect with industry. They enable the professors to serve the society along with teaching and research activity. This entrepreneurial role needs to be supported by promising policy making, profitable incentives, preemptive ORIC/TTO and industry demand under a technology transfer drive.

The faculty needs to be classified into admin faculty, research faculty, teaching faculty and entrepreneurial faculty. The entrepreneurial faculty have keen interest to engage with society and industry to be further supported for commercialization activities. The faculty needs to be assessed and promoted based on their selected track as admin or teaching or basic research or entrepreneurial activities.

It is interesting to note that faculty in the advanced world hardly earns 30% of their income from academic salary and the rest 70% they generate from funding and commercial activities.

# Chapter 10

## 10. The Cooperative Education

Northeastern University in Boston, Massachusetts, started the cooperative education program for the first time in 1909 [1]. The cooperative education program at Northeastern University, also referred to as "co-op," is largely regarded as the country's first official cooperative education program. Since then, several universities, colleges, and organizations all around the world have embraced the cooperative education concept and provide variations of the curriculum in various fields, sectors, and locales.

Cooperative education, or "co-op," refers to a type of education that combines classroom instruction with real-world work experience. Cooperative education's origins can be found in the United States in the late 19th and early 20th centuries when it first appeared in response to the shifting demands of the industrial period (Finley, (2018). Rapid industrialization and technological development throughout the Industrial Revolution led to a demand for skilled laborers across a range of sectors, including manufacturing, agriculture, and engineering. However, at the time, the traditional educational system was not adequately preparing students for these new labour market demands. As a result, educators and employers began to realize that a more pragmatic approach to education was required to close the gap between academic learning and practical work experience. Herman Schneider, who established the "Cincinnati Cooperative School of Engineering" in 1906, was one of the early pioneers of cooperative education in the United States. Schneider thought that by mixing in-class instruction with real-world experience, students could learn more efficiently. For students to apply what they learned in the classroom to real-world job conditions. He envisioned a curriculum that would rotate between periods of classroom instruction and work terms. Following the success of Schneider's design, similar cooperative education initiatives spread throughout American educational institutions. The University of Cincinnati, Georgia Institute of Technology, Northeastern University, and Purdue University, among others, developed cooperative education programs in the 1910s and 1920s.

During the 1930s Great Depression, when there was a decline in the need for skilled labour and students had a difficult time finding job after graduation, cooperative education gained additional traction. Cooperative education programs gave students the chance to earn real-world work experience while still in school, increasing their employability in a competitive employment market. Cooperative education initiatives were extremely important in aiding the war effort during World War II. Many universities collaborated with businesses and governmental organizations to offer students who were hired for jobs related to the war practical training. This emphasized the need of cooperative learning in preparing students for the workforce even more. Cooperative education programs were growing throughout the country in the years following World War II, and many universities and colleges started including co-op as a fundamental component of their curricula. Cooperative education has changed and grown over time in response to the shifting demands of the labour market, and programs have been created in a variety of disciplines, including engineering, business, computer science, health sciences, and more.

Cooperative education programs are now a common feature in many American educational institutions and are highly acclaimed for their capacity to give students useful real-world work experience, improve their employability, and stimulate professional development (Lybeck,. (2004). Today, cooperative education programs are an integral part of many educational institutions in the United States and are highly regarded for their ability to provide students with valuable real-world work experience, enhance their employability, and foster their professional development. The origins of cooperative education in the USA can be traced to a desire to close the knowledge gap between the classroom and the real world and a response to the changing needs of the workforce (Bossert., & Yoder. (2001).

## 10.1 Institutions Offering Cooperative Education Programs in the United States

Northeastern University: Located in Boston, Massachusetts, Northeastern University is well-known for its extensive cooperative education program, or “co-op.” The co-op program at Northeastern combines classroom instruction with practical work experience, enabling students to complete paid internships in the fields they are studying. While pursuing their degrees, students typically alternate between academic semesters and co-op placements, gaining real-world work experience.

Georgia Institute of Technology: At Georgia Tech in Atlanta, Georgia, students can participate in a cooperative education program known as “co-op” that enables them to work full-time for employers during alternate semesters in their field of study. Students get paid, gain practical experience, and get academic credit for their work terms. The co-op program at Georgia Tech offers concentrations in engineering, computing, business, and the sciences.

Drexel University: In Philadelphia, Pennsylvania, Drexel University is well-known for its cooperative education program, or "co-op," which gives students a mix of in-class instruction and practical experience. Students at Drexel typically complete three six-month work terms in their major, giving them the opportunity to network with firms in that field and get practical experience.

Purdue University offers a cooperative education program called "co-op" that enables students to combine academic coursework with professional work experience. Purdue University is in West Lafayette, Indiana. The co-op program at Purdue offers concentrations in engineering, technology, and agriculture. Students often alternate between semesters of coursework and work terms to get real-world experience and insights into the business world.

University of Cincinnati: Cincinnati, Ohio's University of Cincinnati is renowned for its cooperative education program, which was among the nation's first and most significant co-op programs. The University of Cincinnati offers students the chance to take part in co-op positions in a variety of sectors, such as engineering, business, architecture, and applied sciences, where they can get practical experience and advance their careers.

## 10.2 Adoptions of Cooperative Education Programs in Asia

Asia has also embraced cooperative education programs, which combine academic study with real-world experience. Cooperative education initiatives have been implemented and adopted differently by each nation and institution throughout Asia, but the following are some broad observations:

In Japan, cooperative education programs, also referred to as "sandwich programs" or "sandwich courses," are very popular. Students in these programs often alternate between time spent learning in the classroom and time spent gaining experience working with business partners. Cooperative education programs are currently being used in Japan in a number of disciplines, including engineering, computer technology, business, and hospitality. These programs are frequently viewed as a tool to close the knowledge gap between the academy and business, as well to get students ready for the workforce.

**South Korea:** Cooperative education initiatives, sometimes known as "practical training" or "field practice," are now widely used in South Korea. Through internships or on-the-job training with business partners, these programs enable students to gain real-world experience in their majors or areas of study. In South

Korea, cooperative education programs are particularly well-liked in disciplines including business, computer science, and engineering.

**China:** Cooperative education initiatives are expanding there as well, especially in industries like engineering, business, and IT. Students in these programs frequently take part in internships or work placements with business partners, giving them the opportunity to learn practical skills and industry insights. Cooperative education programs in China are often seen to enhance students' employability and prepare them for the competitive job market.

**Singapore:** Singapore has also adopted "work-study programs," or cooperative education initiatives that let students combine academic work with professional experience. Students often rotate between semesters of coursework and work placements with industry partners in these programs, which are provided by a number of universities and polytechnics in Singapore. In Singapore, cooperative education programs are particularly well-liked in industries including engineering, business, and hospitality.

It's significant to highlight that country, institution, and subject of study may all influence how cooperative education initiatives are adopted and implemented in Asia. These programs might also change over time in response to shifting economic and educational environments.

## 10.3 Policy guidelines for Co-opt Programs for Higher Education

Cooperative education policy guidelines for higher education can ensure that these programs are successfully implemented and give students rich learning opportunities. The following are some important policy principles that institutions may take into account when creating or improving cooperative education programs:

**Clear Program Objectives:** Establishing specific goals for cooperative education programs that detail the intended learning outcomes, the program structure, and the anticipated advantages for students, businesses, and the institution is important. These goals should be clearly stated to all stakeholders and should be in line with the institution's overarching educational objectives.

**Quality Assurance:** To make sure that cooperative education programs adhere to accepted norms of academic rigor and professional relevance, institutions should use quality assurance techniques. This can entail defining standards for choosing and assessing business partners, keeping an eye on the calibre of the work assignments and mentorship given to students, and carrying out routine reviews and assessments of the program's efficacy.

**Curriculum Integration:** Programs for cooperative education should be incorporated into the institution's curriculum with distinct links between classroom instruction and practical application. This could entail creating curricula that smoothly juggle coursework and work terms, giving students chances to reflect on their professional experiences and relate them to their academic learning, and incorporating tests that evaluate both academic and workplace competencies.

**Student Support:** For students taking part in cooperative education programs, institutions should offer complete support services. This can entail helping students locate acceptable job placements, delivering pre- and post-placement support, providing resources for career development, and giving them chances to reflect on, get feedback from, and learn from their work experiences.

**Employer Engagement:** To create and preserve solid partnerships for cooperative education programs, institutions should actively engage with companies. In order to promote effective student placements and fruitful work experiences. This may need continuous communication and collaboration with businesses to understand their requirements, expectations, and feedback. It may also entail offering resources and help to companies.

**Equity and Inclusion:** Institutions should make sure that all students, regardless of their background, gender, color, or other characteristics, have access to and inclusion in cooperative education programs. In order to do this, it may be necessary to address potential participation barriers, offer assistance to students from underrepresented groups, encourage diversity among business partners and employers, and foster a welcoming and inclusive learning environment.

**Program Evaluation and Improvement:** Institutions should routinely assess the success of their cooperative education initiatives and make improvements based on feedback from students, companies, and other stakeholders. Making educated judgments on program modifications and upgrades may entail performing assessments, surveys, and focus groups.

It's crucial to remember that cooperative education policy guidelines may change depending on the unique setting and demands of a particular institution or nation. When creating policy guidelines for their programs, institutions should consult with the appropriate parties, such as faculty, students, employers, and policymakers, and take cooperative education best practices and research into account.

## 10.4 Assessment and Performance Evaluation

Cooperative education programs must include assessment and performance evaluation because they enable institutions to gauge the students' learning outcomes, competences, and performance during their work experiences. Here are some typical methods for performance review and assessment in cooperative education programs:

**Work Performance Evaluation:** During their cooperative education placements, students' work performance is typically evaluated by employers and/or supervisors in the workplace. This could entail evaluating how well students perform on the job, the caliber of their work, their capacity to meet deadlines, their problem-solving abilities, teamwork, communication, and other pertinent characteristics. Work performance evaluations, which can be carried out through routine feedback sessions, performance reviews, or written evaluations, can offer insightful information on how well students perform at work.

**Academic Assessment:** During their cooperative education placements, institutions may also evaluate students' academic progress. This may entail including academic assessments into the cooperative education program, such as case studies, research papers, presentations, or reflective journals, to aid students in connecting their classroom learning and professional experiences. Academic exams can help determine how well students can use their theoretical knowledge in practical settings and show what they have learned.

**Self-Assessment:** It may be recommended for students to evaluate their own performance during their co-op education placements. In order to do this, students may reflect on their educational experiences, recognize their strengths and areas for development, and make goals for their futures both personally and professionally. Self-assessment can help students take charge of their learning and growth by fostering metacognition and self-awareness.

**Peer Assessment:** Cooperative education programs can also make use of peer assessment, where students evaluate and comment on their peers' work. This can be accomplished by having students evaluate their peers' work performance, teamwork contributions, and other pertinent qualities through structured peer evaluations or peer feedback sessions. Peer assessment can enhance communication and teamwork among students and offer a variety of viewpoints on their achievement.

**Program Evaluation:** To determine the overall success of their cooperative education programs, institutions may also carry out program assessments. Surveys, focus groups, or interviews may be used to gather input from students, employers, teachers, and other stakeholders. Program evaluations can shed light on the program's advantages and disadvantages, point out areas that need improvement, and guide the development of new programs in the future.

Cooperative education programs should use assessment and performance evaluation techniques that are legitimate, trustworthy, fair, and in line with the program's learning objectives and skills. Establishing clear evaluation standards, providing sufficient training on assessment methods for faculty, staff, and students, and ensuring that feedback is timely and constructive are all responsibilities of educational institutions. Cooperative education programs can benefit from ongoing improvement and refining of assessment and performance evaluation techniques based on input and best practices.

## 10.5 Effective Governance of Co-opt Programs

Cooperative education program management and implementation success depend on effective governance. Here are some essential components of efficient governance for higher education's cooperative education programs:

**Clear Program Goals and Objectives:** Programs for cooperative education should have clear objectives that complement the institution's overall mission and top strategic priorities. All stakeholders, including students, instructors, employers, and other partners, should be made aware of these goals and objectives in order to ensure a common understanding of the program's purpose and desired results.

**Established Policies and Procedures:** Cooperative education programs need to have set policies and procedures that regulate program operations, such as student eligibility requirements, employer partnerships, placement procedures, academic requirements, assessment and evaluation methodologies, and program evaluation procedures. To ensure uniformity and fairness in the execution of the program, these policies and procedures should be written down and periodically reviewed and modified.

**Collaborative Decision-Making:** Cooperative decision-making mechanisms with input from many stakeholders, including academics, employers, students, and other pertinent parties, are necessary for the effective governance of cooperative education programs. To encourage discussion, feedback, and decision-making on program-related issues, regular communication channels such as program advisory boards, faculty committees, and student feedback mechanisms should be developed.

**Resource Allocation:** To enable the successful operation of cooperative education initiatives, significant resources, including financial, human, and infrastructure support, are required. The administration of programs, employer relations, student support, assessment and evaluation, and other program-related activities should be supported by the appropriate resources, which institutions should make sure are allocated.

**Quality Assurance:** Cooperative education programs should be of a high standard, and quality assurance procedures should be in place to assure this. This could involve benchmarking against cooperative education best practices, frequent program assessments, stakeholder input systems, monitoring of program outcomes and student performance. Program improvement activities should be guided by the results of quality assurance measures.

**Professional Development:** In order to improve their knowledge, abilities, and competencies in program administration, employer relations, student support, assessment and evaluation, and other pertinent areas, faculty and staff involved in cooperative education programs should have access to appropriate professional development opportunities. In doing so, you may make sure that the program personnel have the knowledge and skills essential to administer and conduct cooperative education programs.

**Partnerships and Engagement:** Strong collaborations with employers, community organizations, and other stakeholders are frequently essential to cooperative education programs. In order to ensure the relevance and longevity of the program, effective governance of cooperative education programs include creating and maintaining these partnerships, involving employers in program planning and implementation, and promoting continuing collaboration.

**Program Evaluation and Continuous Improvement:** To determine the cooperative education program's efficacy, pinpoint areas that need improvement, and guide program improvement activities, regular program evaluation should be carried out. It is important to leverage feedback from students, employers, staff, and other stakeholders to guide efforts at continuous development and make sure the program is still applicable and in line with the institution's and the workforce's evolving demands.

Institutions may make sure that their cooperative education programs are well-managed, in line with institutional objectives, and produce favorable results for students, businesses, and other stakeholders by including these components of successful governance. Cooperative education initiatives in higher education can be made more successful and sustainable through regular review, evaluation, and governance reform.

## 10.6. Impact of Co-opt Programs

Programs for cooperative education have been proved to benefit students, companies, and institutions of higher learning. Following are some instances of the effects of cooperative education programs, along with corroborating information and sources:

**Enhanced Student Learning and Career Outcomes:** Cooperative education programs give students the chance to apply what they have learned in the classroom to actual workplace situations, which can improve their learning and skill development. According to studies, students who take part in cooperative education programs report being more prepared for the workforce, are more satisfied at work, and have better employment outcomes than their peers who do not (Sung, Kim, & Woo. (2018). For instance, according to a study by the National Association of Colleges and Employers (NACE), students who successfully completed a cooperative education program had higher average starting salaries, higher rates of job offer, and lower unemployment rates than those who did not (NACE 2018).

**Improved Employer Engagement and Workforce Readiness:** Strong ties between employers and higher education institutions are fostered by cooperative education initiatives, and this can lead to an increase in employer engagement and workforce preparation. Access to a pool of capable and enthusiastic students for internships, co-op positions, and other work experiences is advantageous to employers. Additionally, they can spot and hire potential employees early in the academic careers of the students, improving recruitment outcomes and lowering hiring costs (Singh & Weckman. (2018). Additionally, compared to graduates from traditional programs, employers frequently express greater levels of satisfaction with the skills and level of job readiness of cooperative education program graduates (NACE 2018).

**Enhanced Institutional Reputation and Community Engagement:** By demonstrating their dedication to experiential learning, career development, and community engagement, higher education institutions can improve their reputation through cooperative education initiatives. By forming solid alliances with neighborhood companies and organizations, attending to community needs, and acting as a talent pipeline for the local workforce, these initiatives can help promote good town-gown relations (Green & Hammer. (2019). By generating graduates who are prepared for the workforce and have the skills required by regional businesses, cooperative education initiatives can also assist institutions in meeting regional workforce needs and supporting economic growth.

**Increased Diversity, Equity, and Inclusion:** By giving minority students access and opportunities, cooperative education programs have the potential to enhance diversity, equity, and inclusion in the workforce. Students from a variety of backgrounds can benefit from these programs by building their professional networks, gaining valuable job experience, and enhancing their career chances. Cooperative education programs can also enhance diversity in the workplace by encouraging firms to use inclusive hiring practices and by building an inclusive workplace for students from different backgrounds.

## 10.7. Economic Outcome of Co-opt Programs

Cooperative education programs can have economic outcomes and impacts at various levels. Here are some examples:

**Economic Impact on Students:** By giving students paid work experiences, cooperative education programs can benefit students financially by covering their educational expenses and lowering their debt from student loans. Through co-ops, internships, and other work opportunities, students can make money to pay for their education, pay for living expenses, and gain valuable work experience without having to rely solely on loans or other forms of financial aid (Meisenbach. (2016). Students may benefit from having better financial literacy, budgeting abilities, and long-term financial stability as a result of this.

**Economic Impact on Employers:** Employers might gain from cooperative education programs by giving them a low-cost talent acquisition method. Employers have access to a pool of motivated, capable, and eager-to-learn students who can benefit their businesses. Employers can find and hire prospective talent early in students' academic careers, saving time and resources used for conventional recruitment approaches, which can lead to lower recruitment expenses (Harris & Fiechter. (2018). Additionally, cooperative education programs enable employers to shape students to meet their unique workforce needs, resulting in increased productivity and long-term cost savings.

**Economic Impact on Higher Education Institutions:** Programs for cooperative education can help higher education institutions remain financially viable. Institutions can attract more students, improve enrolment, and keep them by offering co-op positions, internships, or other work experiences by supplying beneficial career development chances. This may lead to higher graduation rates, better student retention rates, and more tuition income (Daugherty & Phillips. (2017). Cooperative education initiatives can also bring in money through employer collaborations, sponsorships, or contributions, supporting institutions' financial stability.

**Economic Impact on Communities and Regions:** By creating a workforce that is prepared for employment and in line with regional economic demands, cooperative education initiatives can have a positive economic influence on local communities and regions. Graduates of cooperative education programs frequently possess more relevant skills, are more prepared for the workforce, and have professional networks that can help local employers and support regional economic growth (Hill,. (2019). When cooperative education initiatives are used, this can lead to less skill gaps, higher productivity levels, and enhanced economic competitiveness for the local communities and regions.

# 11. The Conclusion

## 11.1 From Passive to Entrepreneurial

Professor Langer's lab at MIT has an annual budget of USD 17.3 million, initiated 40 companies worth USD 23 billion, including contribution in Moderna (Vaccine), licensed 1100 patents to more than 300 companies, impacting the lives of 4.7 billion people, and published 1500 papers with 36000 citations (Prokesch, S., 2017). Pakistan is still a passive country in 2023 exporting raw materials and offering its natural resources to foreign companies for exploitation. Pakistan imports are increasing, and exports are stagnant inspite of huge investment in science and technology. Pakistan has 60% youth and hardly exports IT services worth USD 02 billion. Pakistan foreign debt is increasing and a very low level of reserves putting it on the edge of default. The entire nation of Pakistan needs to get out of passive sleeping mode and become an entrepreneurial nation to lead a proactive life. The mindset shift is needed from consuming the resources to generating the resources. The entrepreneurial orientation of Pakistani people will enable them to create high value and commercialize this created value to generate wealth. The institutions like education sector, media, chambers of commerce and public sector needs to act decisively to promote entrepreneurial orientation in the society.

## 11.2. Towards Entrepreneurial Higher Education

An entrepreneurial university is an institution of higher education that actively promotes and fosters an entrepreneurial mindset among its students, faculty, and staff. It goes beyond traditional academic activities and embraces an innovative and enterprising culture. Here are some features commonly associated with entrepreneurial universities: a) Entrepreneurial Mindset: An entrepreneurial university encourages and cultivates an entrepreneurial mindset among its stakeholders. This mindset includes attributes such as creativity, problem-solving, risk-taking, adaptability, and a willingness to explore new opportunities, b) Innovation and Research: Entrepreneurial universities prioritize innovation and research, focusing on creating new knowledge, technologies, and solutions that have practical applications. They often have dedicated research centers or incubators that support the development of startups and commercialization of research outcomes, c) Entrepreneurial Education: These universities offer specialized programs and courses that integrate entrepreneurship into various disciplines, enabling students to develop entrepreneurial skills and knowledge. They provide opportunities for experiential learning, business plan competitions, and internships with startups and established companies. d) Industry Collaboration: Entrepreneurial universities foster close collaborations with industry partners, creating opportunities for students and faculty to engage in real-world projects, internships, and research collaborations. This collaboration helps bridge the gap between academia and industry, enhancing the practical relevance of education and research. e) Technology Transfer and Commercialization: These universities actively promote the transfer of technology and the commercialization of intellectual property developed by their faculty and students. They provide support for patenting, licensing, and startup creation, facilitating the transition of innovative ideas and research outcomes into viable businesses, e) Entrepreneurial Support Ecosystem: Entrepreneurial universities establish a supportive ecosystem that includes incubators, accelerators, and entrepreneurship centers. These resources provide mentoring, funding, networking opportunities, and infrastructure to nurture and support entrepreneurial ventures. f) Entrepreneurial Faculty: The faculty members at entrepreneurial universities often have a mix of academic qualifications and practical industry experience. They play an active role in promoting entrepreneurship, mentoring students, and engaging in research with commercial potential, f) Community Engagement: These universities actively engage with the local community and contribute to regional economic development. They collaborate with local businesses, governments, and community organizations to foster entrepreneurship, create jobs, and stimulate economic growth, g) Entrepreneurial Alumni Network: Entrepreneurial universities maintain strong connections with their alumni, who often become successful entrepreneurs, industry leaders, or mentors. The alumni network provides a valuable resource for current students and serves as an inspiration for aspiring entrepreneurs, h) Global Orientation: Entrepreneurial universities have a global outlook and encourage international collaborations and partnerships. They attract students and faculty from diverse backgrounds and provide opportunities for international experiences, such as study abroad programs and global entrepreneurship initiatives.

# About The Authors

**Rahmat Ullah**

Executive Director, IRP, CEO, The Souvenir Company, Director ORIC, International Institute of Science, Arts, and Technology, Gujranwala - rahmat@irp.edu.pk

**Dr. Rashid Aftab**

Director, Riphah Institute of Public Policy, Islamabad - rashid.aftab@riphah.edu.pk

**Saeed Siyal**

Research Fellow, School of Economics and Management, BUCT Beijing China. saeed@mail.ustc.edu.cn

**Kashif Zaheer**

Faculty Member, Riphah Institute of Public Policy, Islamabad kashif.zaheer@riphah.edu.pk

The authors present practical reflections and draw lessons from the last 17 years of working with more than 500 hundred universities, research labs, industries, S&T organizations, and policy institutes for research commercialization. The authors have been involved in more than 100 academic projects for funding, development, and commercialization. The authors have conducted training on applied research and have been involved in policy training and dialogues. This localized experience along with active participation in the triple helix platform at the international level helped the authors understand the dynamic of research commercialization, social impediments, and governance of technology development and diffusion in society.

***_____“The End”_____***

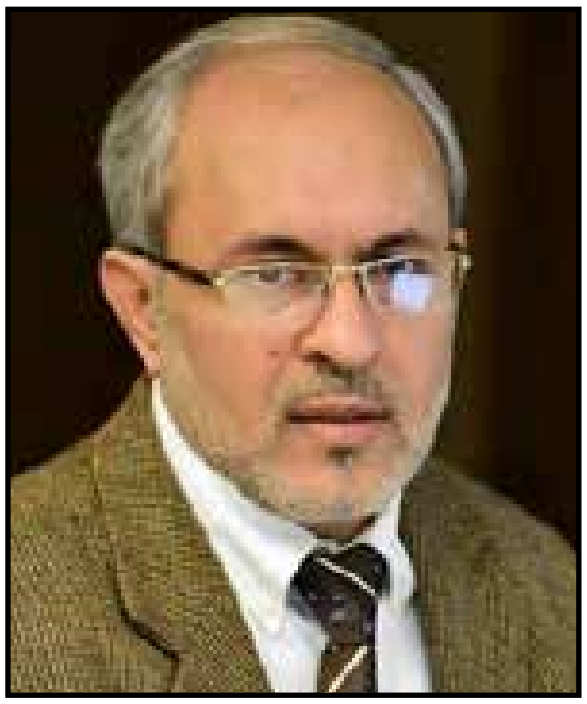

# Message

This book aims to ignite a transformative spark within the realm of academia, urging universities worldwide to embrace the spirit of entrepreneurship and integrate wealth creation into their mission, practices, teaching, and scientific endeavours. By presenting comprehensive policies and operational execution frameworks, the book inspires a new era of innovative and prosperous higher education.

This book serves as a guiding light, illuminating the path for universities to reimagine their role in fostering wealth creation. We now must delve into the transformative power of entrepreneurial education, offering insights on how universities can foster an ecosystem that nurtures creativity, innovation, and problem-solving skills. This book explores various case studies that highlight the success stories of universities that have already embarked on this journey, sharing valuable lessons and best practices.

I have been part of IRP foundation days and journey of the last 15 years. I see this book as the very right step to provide literature to current and next generation on university roles in technology transfer and creation of economic value. This book will surely provide the foundation for our academia to grow and emerge as a center for economic development.
The time for change is now. Let us seize this opportunity to shape a brighter future for higher education—one that is entrepreneurial, inclusive, and truly impactful.

**Prof. Dr. Khalid Mehmood**
Pro Vice Chancellor
The University of Punjab

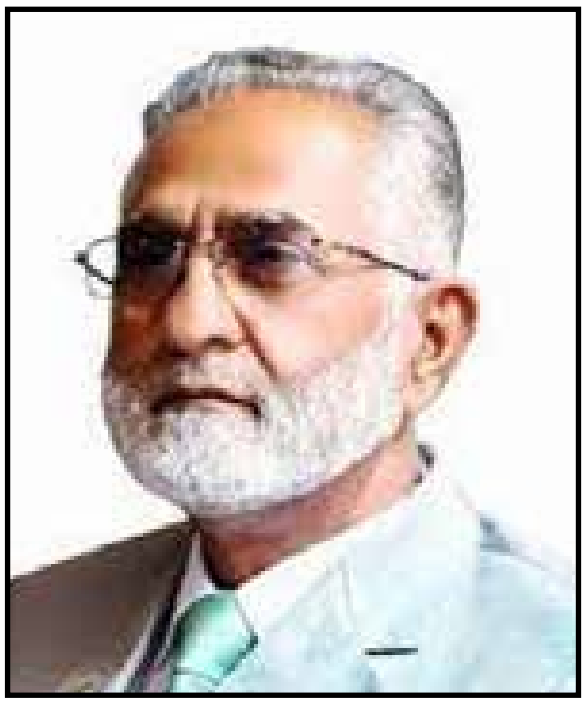

# Message

"Entrepreneurial Higher Education" is a groundbreaking research book that sheds light on the critical issue of the academia-industry gap. With meticulous research and insightful analysis, the book explores the indispensable need for industry linkages and specialized skills in higher education, emphasizing their vital role in the survival and growth of both academia and the industry.

The author presents a compelling argument, supported by real-world examples that traditional higher education must adapt to the evolving demands of the job market. The recent report by The State Bank of Pakistan, revealing that a mere 10% of Information Technology Graduates possess employable skills, underscores the urgency of this issue. Through comprehensive research, the book offers a roadmap to align educational programs with industry needs, equipping graduates with the practical skills and knowledge required for successful entrepreneurship. What sets this book apart is its balanced approach, advocating for collaboration between academia and industry, rather than pitting them against each other. The author highlights the importance of fostering strong partnerships, encouraging academic institutions to actively engage with industry experts, entrepreneurs, and employers. This holistic approach ensures graduates are equipped with the skills, knowledge, and entrepreneurial mindset necessary to thrive in a dynamic job market. The writing style is engaging, making complex concepts accessible to a wide range of readers. The author's deep understanding of both academia and industry enables them to present a compelling case for change, backed by solid research and practical insights.

"Entrepreneurial Higher Education" is an essential read for educators, policymakers, industry leaders, and anyone interested in the future of higher education. It serves as a wake-up call, urging stakeholders to come together, bridge the academia-industry gap, and shape a future where graduates possess the employable skills needed to drive innovation, economic growth, and entrepreneurship.

**Muhammad Bashir Malik**
Social Entrepreneur
Founder & Chairman Bin Qutab Group

Suite No 7, Ground Floor, Evacuee Trust Complex,
Agha Khan Road, Sector F-5/1, Islamabad.
Phone: +92 (51) 843 8370-7